\documentclass[twoside,11pt]{article}

\usepackage[preprint,nohyperref]{jmlr2e}

\graphicspath{{figures/}}
\usepackage{amsmath,bm}
\usepackage{mathtools}
\mathtoolsset{showonlyrefs}
\providecommand{\mathscr}[1]{\mathcal{#1}} 
\usepackage{enumerate}
\let\ExpOriginalEnumerate\enumerate
\renewcommand{\enumerate}[1][]{%
  \def\ExpRequestedLabel{#1}%
  \def\ExpAlphaLabel{label=(\alph*)}%
  \def\ExpRomanLabel{label=(\roman*)}%
  \ifx\ExpRequestedLabel\ExpAlphaLabel
    \ExpOriginalEnumerate[(a)]%
  \else\ifx\ExpRequestedLabel\ExpRomanLabel
    \ExpOriginalEnumerate[(i)]%
  \else\ExpOriginalEnumerate\fi\fi}

\usepackage{longtable,booktabs,array}
\usepackage{algorithm}
\usepackage{algorithmic}
\usepackage{comment}

\newcommand{\R}{\mathbb{R}}					
\newcommand{\pr}{\mathbb{P}}				
\newcommand{\E}{\mathbb{E}}					
\newcommand{\argmin}{\operatornamewithlimits{arg\,min}}
\newcommand{\supp}{\operatorname{supp}}
\newcommand{\abs}[2][]{#1\lvert #2 #1\rvert}
\newcommand{\norm}[2][]{#1\lVert #2 #1\rVert}
\newcommand{\inprod}[3][]{#1\langle #2,#3 #1\rangle}
\newcommand{\tr}{\operatorname{tr}}
\newcommand{\op}{\mathrm{op}}
\newcommand{\HS}{\mathrm{HS}}

\usepackage{xcolor}
\usepackage[colorlinks=false,allbordercolors={1 1 1}]{hyperref}

\AtEndDocument{\label{LastPage}} 

\ShortHeadings{Transfer Learning for Operators}{Li, Li and Tan}
\firstpageno{1}

\begin{document}

\title{Instrumental Transfer Learning for Solution Operators}

\author{\name Tongyu Li\thanks{The authors contributed equally and are listed in alphabetical order.} \email tongyuli@nus.edu.sg \\
\addr Department of Statistics and Data Science\\
National University of Singapore, Singapore 117546, Singapore
\AND
\name Zhuoyuan Li\footnotemark[1] \email zy.li@nus.edu.sg \\
\addr Institute for Functional Intelligent Materials\\
National University of Singapore, Singapore 117544, Singapore
\AND
\name Ruoxu Tan\footnotemark[1] \email ruoxut@tongji.edu.cn \\
\addr School of Mathematical Sciences and School of Economics and Management\\
Tongji University, Shanghai 200092, China
}

\editor{}

\maketitle 

\begin{abstract}
Data-driven learning for the solution operator of a differential equation has broad scientific importance, but observations from a target system are often scarce and costly to collect. Related source systems may provide abundant observations under similar physical conditions, whereas even small changes in the governing equation can induce complex changes in the solution operator, making naive data pooling ineffective. We propose a structured transfer learning method that accounts for unknown sparse differences in the governing equations from a known library of candidate terms. Using the resolvent identity and a whitened instrumental variable approach, we construct unbiased source estimating equations for the target operator. Building on these equations, we develop a three-stage estimator that consists of a target-only ridge pilot, source-wise sparse calibration, and a projected refit designed to remove the leading calibration error inherited from the pilot. An iterative recalibration scheme further reduces the remaining error. We establish non-asymptotic prediction bounds showing when transfer achieves the pooled all-source rate, together with minimax lower bounds that characterize both rate optimality and fundamental identifiability limits under common shifts. Simulations and an electronic band-pass filter application demonstrate clear gains over target-only learning, naive pooling and existing transfer methods.
\end{abstract}


\section{Introduction}\label{sec:introduction}
Learning solution operators of partial differential equations (PDEs) is central to many problems in science and engineering. Representative applications include simulating fluid flows in aerodynamics \citep{li2021fourier,kovachki2023neural}, forecasting atmospheric dynamics in weather prediction \citep{bi2023accurate,lam2023learning}, predicting temperature or stress fields in continuum mechanics \citep{goswami2022deep,long2026deep}, and modeling wave propagation in seismic imaging \citep{yang2021seismic,yang2023rapid}. In these settings, both the input and the output are functions over space or time, such as an external forcing profile, boundary heat flux, or wave excitation, which gives rise to the need for studying operators between function spaces.

Recently, operator learning and function-on-function regression have provided a data-driven framework for learning mappings between function spaces from paired functional observations \citep{morris2015functional,subedi2026operator}. The object of interest is often the solution operator, which maps an input function, such as a forcing term, boundary condition, or control signal, to the resulting system state over a continuous domain. Once learned, a solution operator acts as a fast surrogate that predicts system responses to new inputs without repeated physical experiments or numerical solves. This fast evaluation is particularly valuable for many-query and real-time tasks, including PDE-constrained design optimization, real-time optimal control, uncertainty quantification, and Bayesian inverse problems \citep{stuart2010inverse,peherstorfer2018survey,nelsen2024random}.

However, accurate operator learning usually requires observations that cover a sufficiently rich range of system responses \citep{dehoop2023convergence}. For a newly configured system, such observations are often scarce because experiments and high-fidelity simulations are costly \citep{goswami2022deep}. Related systems, by contrast, may provide abundant historical data collected under different physical parameters or operating conditions. These systems share physical mechanisms with the system of interest, while their input-to-output relations may differ. Their data offer an important source of information, provided that this heterogeneity is properly modeled.

We refer to a newly deployed regime as the \emph{target} system, which affords only a small sample of input-output observations, and to related auxiliary systems with abundant historical data as \emph{sources}. Discarding the sources wastes valuable physical information, whereas naively pooling all data under a single shared operator introduces severe misspecification bias, particularly when the source and target systems obey different governing equations. Transfer learning \citep{zhuang2020comprehensive} seeks to leverage the source observations to improve estimation for the target system, taking into account such heterogeneity. Nevertheless, since differential operator inversion is globally coupled, even a localized change in a governing equation alters its inverse solution operator globally across input directions. This makes prevailing statistical transfer learning methods that pool the source data first and then correct a small contrast on the target ineffective here \citep{li2022transfer,lin2024hypothesis,hu2025functional}.

To address this challenge, we notice that related systems often share a natural structural link through their governing equations. Variations in physical properties, such as thermal conductivity, transport velocity, or external modulation, typically modify only a few terms in the governing equation \citep{rozza2007reduced,lataire2015benchmark}. Similarly, sparse equation-discovery methods represent governing laws using a few active terms from a prescribed library \citep{rudy2017data,rudy2019data}. Motivated by these examples, we consider related systems whose governing operators differ by only a few terms from a known collection of candidate differential operators. Because these sparse differences occur in the governing equations prior to inversion, they do not translate into simple sparse shifts between the solution operators. This paper aims to exploit unknown, sparse perturbations in the governing equations to improve target solution-operator learning from noisy data when target observations are limited.

\subsection{Our contributions}
We develop a transfer learning framework that turns partial knowledge of how governing laws change into information for estimating a target solution operator. The framework models differences among the governing equations as sparse combinations of terms from a known candidate library, while allowing the corresponding solution operators to differ globally and in complex ways. Our main contributions are as follows.

\begin{itemize} 
\item \emph{A principled route from governing-law similarity to solution-operator transfer.}
We use the resolvent relation to express each source system through the target solution operator, and construct unbiased estimating equations using regularized, whitened source inputs as instruments. This removes the endogeneity introduced by the response-dependent input adjustment. The resulting framework enables information sharing without first identifying the entire target equation or assuming a sparse additive contrast between solution operators.

\item \emph{Transfer that can improve beyond the accuracy of the initial target estimate.}
We construct a projection of the source moments that, under the stated calibration coverage condition, removes the leading error inherited from estimating the unknown task differences at a target pilot. The remaining calibration contribution is a product of the pilot and coefficient-estimation errors. This reduction is essential for converting large source samples into an improvement over target-only learning. A target ridge pilot, sparse source calibration, and projected refitting implement the construction through convex optimizations. We also propose an iterative recalibration scheme that uses the improved target estimate to reduce the remaining error further. Our analysis accounts for the reuse of observations across these stages without sample splitting.

\item \emph{Conditions for attaining the all-source rate and limits imposed by identifiability.}
We establish non-asymptotic prediction bounds showing that, when the projected sources jointly retain enough information about the target, our one-step estimator achieves the prediction error rate 
\begin{equation}
\norm{\widehat A-A^{(0)}}_{\Gamma^{(0)}}^2
\lesssim
\Bigl(\frac{t}{N}\Bigr)^{\frac{2\beta}{2\beta+1}}
+\Bigl(\frac{t}{n_0}\Bigr)^{\frac{4\beta}{2\beta+1}},
\end{equation}
where $n_0$ and $N$ are the target and total sample sizes, respectively, $\beta>1/2$ describes the decay of the target prediction spectrum, and $t$ controls the failure probability. The bound separates the all-source estimation rate from a higher-order cost of learning the task differences. It improves on the target-only rate $(t/n_0)^{2\beta/(2\beta+1)}$ when $N\gg n_0$, and each recalibration round further shrinks the second term. An all-source minimax lower bound in the general closed-operator model establishes rate optimality at fixed confidence in the corresponding regimes. A separate common-shift lower bound of order $n_0^{-1}$ shows that source observations cannot resolve a target change that can be absorbed into every source perturbation. These results explain both the value of source information retained after projection and the need for conditions ensuring that this information identifies the target.

\item \emph{Evidence that mechanistic calibration yields useful gains with limited target data.}
Through simulations on functional regression and advection--diffusion equations, we examine the effects of calibration, projection, and recalibration. On the electronic band-pass filter benchmark of \citet{lataire2015benchmark}, the projected estimator reduces mean full-band spectral prediction error by approximately $36.3\%$ relative to target-only learning across 24 held-out evaluations, each using one target training record. The application demonstrates that physically structured calibration can extract useful information from operating regimes for which naive pooling is strongly biased.
\end{itemize}

\subsection{Related work}

\paragraph{Functional regression and operator learning.}
Function-on-function regression provides a classical statistical framework for learning linear maps between function spaces; see \citet{morris2015functional,wang2016functional} for comprehensive reviews, and \citet{kadri2016operator,li2025function,yan2026functional} for recent advances. In scientific computing, operator learning has emerged as a parallel paradigm for approximating solution maps of partial differential equations, with recent surveys synthesizing neural and statistical perspectives \citep{kovachki2023neural,subedi2026operator}. 
In particular, statistical theory for linear operator estimation has established convergence rates under noisy observations \citep{dehoop2023convergence,mollenhauer2022linear} and kernel-based representations \citep{stepaniants2023learning}. 
While these works study a single operator from homogeneous data, we exploit the structural link that arises when multiple solution operators are governed by differential equations that differ by unknown, sparse perturbations.

\paragraph{Statistical transfer learning.}
Statistical transfer learning quantifies the benefit of auxiliary source data under explicit assumptions on task relatedness \citep{zhu2025recent,bi2026heterogeneity}, such as sparse coefficient contrasts in high-dimensional linear and generalized linear models \citep{li2022transfer,tian2023transfer} or bounded discrepancies in nonparametric and kernel regression \citep{cai2024nonparametric,wang2025kerneltransfer}. 
In functional linear regression, existing methods measure source--target similarity directly on the forward regression mappings: \citet{lin2024hypothesis} penalize RKHS norm contrasts, \citet{hu2025functional} model sparse basis contrasts under posterior drift, and \citet{jiao2025coefficient} transfer a shared functional shape across tasks. In contrast to the methods that assume small additive contrasts or proportional shapes directly between the target and source mappings, our formulation places sparsity on the differences between the underlying governing differential operators. The resulting notion of task relatedness therefore does not require a sparse contrast between solution operators. 

\paragraph{Transfer and multi-tasks learning for physical systems.}
In scientific machine learning, knowledge transfer across physical regimes is commonly achieved by fine-tuning neural operators under conditional shift \citep{goswami2022deep}, leveraging multifidelity simulations \citep{lu2022multifidelity}, or incorporating shared physical invariants \citep{ma2026physics}. More recently, large-scale PDE foundation models \citep{herde2024poseidon} and in-context operator networks \citep{li2026context} have enabled retrain-free adaptation across diverse physical systems. The key distinction here is that we develop a rigorous statistical framework that exploits source estimating equations, which allows us to characterize when these equations provide useful target information despite unknown task differences. 

\paragraph{Instrumental variables and orthogonal estimation.}
Functional instrumental-variable regression \citep{florens2015instrumental}, nonparametric conditional-moment estimation \citep{chen2012estimation}, and kernel or minimax instrumental-variable methods \citep{singh2019kerneliv,dikkala2020minimax} provide tools for estimation with endogenous regressors. In our setting, endogeneity is created by the physical correction of the source input, and the original forcing supplies an instrument under the conditional centering assumptions. Our projection also connects to the principle of reducing sensitivity to nuisance estimation in double/debiased machine learning, locally robust estimation, and orthogonal statistical learning \citep{chernozhukov2018double,chernozhukov2022locally,foster2023orthogonal}. Unlike classical semiparametric problems targeting a low-dimensional parameter, our challenge is to remove first-order pilot bias while retaining sufficient information to recover an infinite-dimensional target operator. We formalize an algebraic coverage condition that reduces the calibration bias to a second-order error product and link the retained source subspace to the resulting minimax prediction rates.

\subsection{Organization and notation}
\paragraph{Organization.}
Section~\ref{sec:method} defines the model and estimator. Section~\ref{sec:theory} presents the theoretical results. Section~\ref{sec:experiment} reports the numerical studies, including the electronic-filter application in Section~\ref{sec:real_ltv}. Section~\ref{sec:discussion} discusses the findings and limitations. Proofs and further details are in the appendices.

\paragraph{Notation.}
Positive constants $c$ and $C$ may change between appearances, with dependence specified in each result. For nonnegative $a,b$, write $a\lesssim b$ if $a\le Cb$, and $a\asymp b$ if both $a\lesssim b$ and $b\lesssim a$. Also, $a\wedge b=\min(a,b)$ and $a\vee b=\max(a,b)$.

For a real random variable $X$, denote $\norm{X}_{\psi_2}=\inf\{c>0:\E\{\exp(X^2/c^2)\}\le2\}$.
For a vector $v$, denote $\norm{v}_{\ell^q}=(\sum_k|v_k|^q)^{1/q}$ for $q=1,2$, $\norm{v}_{\ell^\infty}=\norm{v}_\infty=\max_k|v_k|$, and $\supp(v)=\{k:v_k\ne0\}$. For an index set $S$ contained in $\{1,\dots,p\}$, let $|S|$ be its size, $S^\complement$ its complement, and $v_S$ the corresponding subvector of $v$. For a matrix $M$, write $M[S,T]$ for the submatrix with rows in $S$ and columns in $T$, and $\norm{M}_\infty=\max_k\sum_l|M_{kl}|$.

For a Hilbert space $H$, write $\inprod{\cdot}{\cdot}_H$, $\norm{\cdot}_H$, and $I_H$ for the inner product, norm, and identity. If $\Pi$ is an orthogonal projection on $H$, write $\Pi_\perp=I_H-\Pi$. For an operator $T$, let $T^*$, $\operatorname{dom}(T)$, $\operatorname{ran}(T)$, and $\tr(T)$ denote its adjoint, domain, range, and trace, when defined. For bounded $T$, $\norm{T}_{\op}=\sup_{\norm{x}\le1}\norm{Tx}$. For bounded self-adjoint operators, either $A\preceq B$ or $B\succeq A$ means $\inprod{(B-A)x}{x}\ge0$ for all $x$, and $\asymp$ inherits this order for nonnegative operators. For $\Gamma\succeq0$, write $\norm{x}_\Gamma^2=\inprod{\Gamma x}{x}$ for the weighted seminorm.

We identify $H_1\otimes H_2$ with the Hilbert--Schmidt operators from $H_2$ to $H_1$, with $\norm{T}_{\HS}^2=\norm{T}_{H_1\otimes H_2}^2=\sum_k\norm{Te_k}_{H_1}^2$ for any orthonormal basis $\{e_k\}$ of $H_2$. The rank-one operator is $(v_1\otimes v_2)w=\inprod{v_2}{w}_{H_2}v_1$. For bounded $B:H_2\to H_2$, it holds that $(I_{H_1}\otimes B)T=TB^*$.

\section{Model and methodology}\label{sec:method}

We specify the functional operator model in Section~\ref{sec:model}, construct the three-stage transfer estimator in Section~\ref{sec:estimator}, and introduce an iterative recalibration scheme for further refinement in Section~\ref{sec:recalibration}.  

\subsection{Functional operator model}\label{sec:model}
Let $V$ and $W$ be separable Hilbert spaces representing the spaces of functional responses (states) and predictors (forcings), respectively, and let $V_0$ be a dense linear subspace of $V$ encoding prescribed boundary and regularity conditions.
We consider $m+1$ independent but structurally related systems. The system indexed by $j=0$ is the target system of primary interest, for which data are limited, while the remaining $m$ systems ($j=1,\dots,m$) serve as auxiliary source systems. 
Although the source systems need not coincide with the target, they are assumed to share sufficient physical structure to provide complementary information. Our goal is to leverage source observations to improve estimation of the target system while explicitly correcting for the structural differences across tasks.

Specifically, for each task $j=0,1,\dots,m$, let $L^{(j)}:V_0\subset V\to W$ be a closed, bijective linear operator that characterizes the governing law of the corresponding physical system. In PDE applications, $L^{(j)}$ is typically an unbounded differential operator. Its inverse is the associated solution operator, $A^{(j)}=\{L^{(j)}\}^{-1}:W\to V_0\subset V$, which maps an input to the resulting system state. Rather than observing the differential operator $L^{(j)}$ directly, we observe input--output pairs generated by its inverse $A^{(j)}$. For task $j$, we collect independent and identically distributed observations $(U_i^{(j)},F_i^{(j)})$, $i=1,\dots,n_j$, following the functional linear regression model
\begin{equation}\label{eq:reg_model}
U^{(j)}=A^{(j)}F^{(j)}+\varepsilon^{(j)}.
\end{equation}
Here, $F^{(j)}\in W$ represents a random predictor or forcing term satisfying $\E\{F^{(j)}\}=0$ and $\E\{\norm{F^{(j)}}_W^2\}<\infty$, and $U^{(j)}\in V_0$ is the observed functional response contaminated by random noise $\varepsilon^{(j)}\in V_0$.

A central challenge in operator transfer learning is that even a simple, localized change in a differential operator $L^{(j)}$ generally induces a global, non-sparse change in its inverse $A^{(j)}$. Consequently, we impose structure on the governing operators rather than directly on the solution operators. We assume that each source governing operator differs from the target governing operator through a linear combination of known candidate mechanisms \citep{rudy2019data}. Specifically, on $V_0$, we assume that $L^{(j)}=L^{(0)}+\theta^{(j)}\cdot D$ for $j=1,\dots,m$, where $\theta^{(j)}\cdot D=\sum_{l=1}^p\theta_l^{(j)}D_l$. Here $D_l:\operatorname{dom}(D_l)\subset V\to W$, $1\le l\le p$, are known closed linear operators such that $V_0\subset\cap_{l=1}^p\operatorname{dom}(D_l)$ and $D_lA^{(j)}:W\to W$ is bounded for every $j$ and $l$. These regularity conditions ensure that the perturbation operators are well defined on the admissible state space and that their composition with each solution operator remains stable. The unknown vector $\theta^{(j)} =(\theta^{(j)}_1,\dots,\theta^{(j)}_p)^\top\in\R^p$  
quantifies how source task $j$ deviates from the target task.
Since real-world system changes are typically driven by only a few physical mechanisms, we assume $\theta^{(j)}$ to be sparse in the sense that $S_j=\supp(\theta^{(j)})$ has size $s_j=|S_j|$ bounded uniformly in $j$. This is also implied by our assumptions in Section~\ref{sec:theory}.

\begin{example}[\citealp{rozza2007reduced}]
\label{ex:elliptic}
On the two-dimensional torus $\mathbb T^2=(\mathbb R/2\pi\mathbb Z)^2$,
take $V=W=L^2(\mathbb T^2)$ and $V_0=H^2_{\rm per}(\mathbb T^2)$.
The periodic Sobolev space encodes both the regularity and the periodic boundary conditions imposed on the states. For a fixed $\kappa>0$, consider the target operator $L^{(0)}=\kappa^2-\Delta$ and source operators $L^{(j)}=\kappa^2-\Delta+\theta_1^{(j)}\partial_1+\theta_2^{(j)}\partial_2$.
The target system is governed by an elliptic diffusion--reaction operator, while each source system additionally contains a constant advection component. 
Here $D_1=\partial_1$ and $D_2=\partial_2$, so the perturbation parameters $\theta^{(j)}$ describe the change in horizontal and vertical advection between source task $j$ and the target task.
The Fourier symbol of $L^{(j)}$ is
$\ell_j(k)=\kappa^2+|k|^2+i\theta^{(j)}\cdot k$ for $k\in\mathbb Z^2$.
Its positive real part makes $L^{(j)}$ closed and bijective, with bounded inverse $A^{(j)}:L^2\to H^2_{\rm per}$ and bounded $\partial_lA^{(j)}$.
\end{example}

To perform nonparametric estimation of the target inverse, we assume that $A^{(0)}$ belongs to an operator reproducing kernel Hilbert space $H$ \citep{kadri2016operator,stepaniants2023learning}. The error term $\varepsilon^{(j)}\in V_0$ in \eqref{eq:reg_model} is assumed to satisfy
\begin{equation}\label{eq:conditional_centering}
\E\{\varepsilon^{(j)}\mid F^{(j)}\}=0,
\qquad
\E\{D_l\varepsilon^{(j)}\mid F^{(j)}\}=0,
\quad l=1,\dots,p.
\end{equation}
The first condition rules out systematic response-level bias conditional on the predictor. The second imposes the analogous centering condition after applying each operator $D_l$.

\subsection{Transfer estimator via instrumental moments}\label{sec:estimator}
Standard transfer learning methods typically assume that source and target parameters differ by a simple, sparse shift. However, a sparse perturbation $\theta^{(j)}\cdot D$ in the governing differential operator $L^{(j)}$ induces a complex, non-sparse change in its inverse $A^{(j)}=\{L^{(j)}\}^{-1}$, making direct transfer in the solution-operator space ineffective. A core contribution of our framework is to bypass this obstacle by exploiting the algebraic link between the governing operators and their inverses. The following resolvent identity re-expresses each source regression model directly in terms of the target solution operator $A^{(0)}$.   

\begin{lemma}\label{lem:resolvent}
For each $j=1,\ldots,m$,
\begin{equation}\label{eq:resolvent}
A^{(j)}=A^{(0)}\{I_W-\theta^{(j)}\cdot DA^{(j)}\},
\end{equation}
and thus, 
\begin{equation}\label{eq:transformed_regression}
U^{(j)}-A^{(0)}\{F^{(j)}-\theta^{(j)}\cdot DU^{(j)}\}
=\{I_V+A^{(0)}\theta^{(j)}\cdot D\}\varepsilon^{(j)}.
\end{equation}
\end{lemma}

Lemma~\ref{lem:resolvent} shows that if we adjust the source forcing $F^{(j)}$ by subtracting the state-dependent perturbation $\theta^{(j)}\cdot DU^{(j)}$, the source response $U^{(j)}$ follows a linear regression model governed by the exact target operator $A^{(0)}$. However, this identity introduces a new statistical challenge: Because $U^{(j)}$ contains the observation noise $\varepsilon^{(j)}$, the adjusted predictor $F^{(j)}-\theta^{(j)}\cdot DU^{(j)}$ is correlated with the transformed regression error $\{I_V+A^{(0)}\theta^{(j)}\cdot D\}\varepsilon^{(j)}$. Consequently, naive least-squares regression on \eqref{eq:transformed_regression} would suffer from endogeneity bias even if $\theta^{(j)}$ were known.

To eliminate this bias, we develop an infinite-dimensional instrumental variable formulation \citep{florens2015instrumental} using the exogenous forcing $F^{(j)}$. 
Write $\Sigma^{(j)}=\E\{F^{(j)}\otimes F^{(j)}\}$ and $\widehat\Sigma^{(j)}=n_j^{-1}\sum_{i=1}^{n_j}F_i^{(j)}\otimes F_i^{(j)}$ for each task $j$. For $\eta>0$, we introduce the regularized whitened instruments,
\begin{equation}\label{eq:instrument-variable}
Z^{(j)}(\eta)
=\{\Sigma^{(j)}+\eta I_W\}^{-1/2}F^{(j)},\qquad
\widehat Z_i^{(j)}(\eta)
=\{\widehat\Sigma^{(j)}+\eta I_W\}^{-1/2}F_i^{(j)}.
\end{equation}
Whitening rescales the forcing directions according to their variability. The regularization parameter $\eta$ stabilizes the inverse square root in directions where the forcing covariance is small or nearly singular, a commonly used technique in infinite-dimensional problems. For $A\in H$ and $\theta\in\R^p$, we define the population and empirical residual-instrument cross-covariance moments by, respectively,
\begin{equation}\label{eq:emp_moment}
\begin{aligned}
M_\eta^{(j)}(A,\theta)
&=\E\Bigl(
\bigl[U^{(j)}-A\{F^{(j)}-\theta\cdot D U^{(j)}\}\bigr]
\otimes Z^{(j)}(\eta)\Bigr),
\\
\widehat M_\eta^{(j)}(A,\theta)
&=\frac1{n_j}\sum_{i=1}^{n_j}
\bigl[U_i^{(j)}-A\{F_i^{(j)}-\theta\cdot D U_i^{(j)}\}\bigr]
\otimes\widehat Z_i^{(j)}(\eta).
\end{aligned}
\end{equation}
Because $Z^{(j)}(\eta)$ depends only on $F^{(j)}$, the conditional centering conditions in \eqref{eq:conditional_centering} together with \eqref{eq:transformed_regression} guarantee that the population moment vanishes at the true parameters: 
\begin{equation}\label{eq:EE}
\begin{aligned}
M_\eta^{(j)}(A^{(0)},\theta^{(j)})=0 .
\end{aligned}
\end{equation}
Equation~\eqref{eq:EE} provides an unbiased estimating equation that links the source observations directly to $A^{(0)}$ and $\theta^{(j)}$ without requiring a shared input--output law across tasks. 

In practice, however, $\theta^{(j)}$ is unknown and must be calibrated using a preliminary target-only pilot estimator $\widetilde A$. A second major novelty of our approach is a moment-projection mechanism that insulates the final transfer estimator from this pilot error. To see how estimation errors propagate, we define the empirical moment maps
\begin{equation}
\widehat\Phi_l^{(j)}(\eta)T
=\frac1{n_j}\sum_{i=1}^{n_j}
TD_lU_i^{(j)}\otimes\widehat Z_i^{(j)}(\eta),\qquad
\widehat O^{(j)}(\eta)T
=\frac1{n_j}\sum_{i=1}^{n_j}
TF_i^{(j)}\otimes\widehat Z_i^{(j)}(\eta),
\end{equation}
for any operator $T\in H$. By linearity, for $A\in H$ and $\theta\in\R^p$, $\widehat M_\eta^{(j)}(A,\theta)$ admits the exact decomposition
\begin{equation}\label{eq:emp_moment_expansion}
\begin{aligned}
\widehat M_\eta^{(j)}(A,\theta)
={}&\widehat M_\eta^{(j)}(A^{(0)},\theta^{(j)})
-\{\widehat O^{(j)}(\eta)
-\theta^{(j)}\cdot\widehat\Phi^{(j)}(\eta)\}
\{A-A^{(0)}\}\\
&+\{\theta-\theta^{(j)}\}\cdot\widehat\Phi^{(j)}(\eta)A ,
\end{aligned}
\end{equation}
which captures the random sampling noise of the source moment at the true parameters, the signal measuring how far a candidate operator $A$ is from the true target operator $A^{(0)}$, and the calibration bias caused by replacing the unknown true perturbation $\theta^{(j)}$ with an estimate $\theta$, respectively. This third term creates a bottleneck for naive plug-in transfer. Because $\theta^{(j)}$ is unknown, we must estimate it by $\widetilde\theta^{(j)}$ using a preliminary pilot $\widetilde A$ fitted only on the small target sample of size $n_0$. Consequently, the calibration error $\widetilde\theta^{(j)}-\theta^{(j)}$ inherits the estimation error of $\widetilde A$ and does not vanish even as the source sample size $n_j$ grows.

To break this bottleneck, we rewrite the calibration bias at $A^{(0)}$ around the pilot $\widetilde A$:
\begin{equation}\{\widetilde\theta^{(j)}-\theta^{(j)}\}\cdot\widehat\Phi^{(j)}(\eta)A^{(0)} =\{\widetilde\theta^{(j)}-\theta^{(j)}\}\cdot\widehat\Phi^{(j)}(\eta)\widetilde A +\{\widetilde\theta^{(j)}-\theta^{(j)}\}\cdot\widehat\Phi^{(j)}(\eta)\{A^{(0)}-\widetilde A\}.\end{equation}
Notice that the leading first-order error, $\{\widetilde\theta^{(j)}-\theta^{(j)}\}\cdot\widehat\Phi^{(j)}(\eta)\widetilde A$, lies entirely within the subspace spanned by the known operators $\widehat\Phi_l^{(j)}(\eta)\widetilde A$. By projecting the empirical source moment onto the orthogonal complement of the active span of $\widehat\Phi_l^{(j)}(\eta)\widetilde A$, we completely eliminate this first-order calibration error.

Motivated by \eqref{eq:EE} and  \eqref{eq:emp_moment_expansion}, we propose a three-stage transfer estimation procedure as follows, with tuning parameters $\lambda_0, \eta, \lambda > 0$ and $\{\lambda_j>0 : 1\le j\le m\}$. 
\begin{enumerate}[label=(\roman*)]
\item Target-only pilot estimation:
\begin{equation}
\widetilde A=\argmin_{A\in H}\biggl\{
\frac1{n_0}\sum_{i=1}^{n_0}\norm{U_i^{(0)}-AF_i^{(0)}}_V^2
+\lambda_0\norm A_H^2\biggr\};
\end{equation}
\item Source-wise sparse calibration:
\begin{equation}\label{eq:calibration}
\widetilde\theta^{(j)}
=\argmin_{\theta\in\R^p}
\bigl\{\norm{\widehat M_\eta^{(j)}(\widetilde A,\theta)}_{V\otimes W}^2
+\lambda_j\norm\theta_{\ell^1}\bigr\},
\qquad 1\le j\le m;
\end{equation}
\item Calibrated transfer refitting:
\begin{equation}
\begin{aligned}
\widehat A=\argmin_{A\in H}\biggl[&
\frac1{n_0}\sum_{i=1}^{n_0}
\norm{U_i^{(0)}-AF_i^{(0)}}_V^2+\lambda\norm A_H^2\\
&+\sum_{j=1}^m\rho_j
\norm{\widetilde\Pi_\perp^{(j)}(\eta)
\widehat M_\eta^{(j)}(A,\widetilde\theta^{(j)})}_{V\otimes W}^2\biggr],
\qquad \rho_j=\frac{n_j}{n_0}.
\end{aligned}
\label{eq:refit}
\end{equation}
Here $\widetilde\Pi^{(j)}(\eta)$ denotes the orthogonal projection onto the empirical calibration span $\{\theta\cdot\widehat\Phi^{(j)}(\eta)\widetilde A:\supp(\theta)\subseteq\supp(\widetilde\theta^{(j)})\}$.
\end{enumerate}
 
Each stage serves a distinct purpose. Stage~(i) yields a consistent baseline $\widetilde A$ of the target operator. Stage~(ii) substitutes $\widetilde A$ into the source moments to identify which physical mechanisms distinguish source $j$ from the target. Stage~(iii) pools information across all tasks while using $\widetilde\Pi_\perp^{(j)}(\eta)$ to orthogonalize the source moments against nuisance calibration errors (see Proposition~\ref{prop:revision-coverage}). We also evaluate the use of $\widetilde\Pi_\perp^{(j)}(\eta)$ in experiments in Section~\ref{sec:experiment}.

Our formulation is also computationally tractable. Because $\widehat M_\eta^{(j)}(A,\theta)$ is affine in $\theta$ for fixed $A$ and affine in $A$ for fixed $\theta$, our three-stage decomposition avoids nonconvex joint optimization over $(A,\theta^{(1)},\dots,\theta^{(m)})$. Instead, Stage~(ii) reduces to $m$ parallel $p$-dimensional Lasso-type problems, and Stage~(iii) reduces to a single convex quadratic program in $H$.

\subsection{Recalibrated improvement}\label{sec:recalibration}

Although the final refit uses all source data, its calibrations are constructed from the target-only pilot $\widetilde A$. Thus, the three-stage estimator is limited by the accuracy of the pilot estimator.  If the source samples make the first refit $\widehat A$ more accurate, it is natural to perform subsequent calibration using $\widehat A$ as a new pilot. By alternating between recalibrating the source perturbations and refitting the target operator, we expect to progressively shrink the remaining pilot bias.

Specifically, we retain the active supports $\widehat S_j=\supp(\widetilde\theta^{(j)})$ selected in Stage~(ii) so that we do not need to repeat $\ell^1$ variable selection. Set $\widehat A^{[0]}=\widetilde A,\widehat\theta^{(j)[0]}=\widetilde\theta^{(j)}$, and $\widehat A^{[1]}=\widehat A$. For $k=1,2,\dots$, we recalibrate each source on $\widehat S_j$ to obtain
\begin{equation}\label{eq:iterated_calibration}
\widehat\theta^{(j)[k]}
=\argmin_{\theta:\,\supp(\theta)\subset\widehat S_j}
\norm{\widehat M_\eta^{(j)}(\widehat A^{[k]},\theta)}_{V\otimes W}^2.
\end{equation}
The restricted optimization makes the corrected source forcing compatible with the current target estimate without repeating variable selection.
Then let $\widehat\Pi^{(j)[k]}(\eta)$ be the orthogonal projection onto $\{\theta\cdot\widehat\Phi^{(j)}(\eta)\widehat A^{[k]}:
\supp(\theta)\subset\widehat S_j\}$. Using the updated calibration and projection, the next refit is
\begin{equation}\label{eq:iterated_refit}
\begin{aligned}
\widehat A^{[k+1]}=\argmin_{A\in H}\biggl[&
\frac1{n_0}\sum_{i=1}^{n_0}
\norm{U_i^{(0)}-AF_i^{(0)}}_V^2+\lambda\norm{A}_H^2\\
&+\sum_{j=1}^m\rho_j
\norm{\widehat\Pi_\perp^{(j)[k]}(\eta)
\widehat M_\eta^{(j)}(A,\widehat\theta^{(j)[k]})}_{V\otimes W}^2\biggr].
\end{aligned}
\end{equation}

The intuition behind this iterative refinement is straightforward. Whenever $\widehat S_j$ contains the true support $S_j$, the updated projection $\widehat\Pi_\perp^{(j)[k]}(\eta)$ removes the first-order calibration error evaluated at the current iterate $\widehat A^{[k]}$. Consequently, at the true operator $A^{(0)}$, the residual calibration bias is proportional to the product of the updated coefficient error $\widehat\theta^{(j)[k]}-\theta^{(j)}$ and the current operator error $\widehat A^{[k]}-A^{(0)}$. Because $\widehat\theta^{(j)[k]}$ is itself calibrated using $\widehat A^{[k]}$, its error scales with $\widehat A^{[k]}-A^{(0)}$, making the product error quadratic in the accuracy of $\widehat A^{[k]}$. Each refit step therefore squares the pilot-dependent remainder from the previous step, successively driving it down until the estimator is limited only by the irreducible fluctuation across all $N$ observations; see Theorem~\ref{thm:iterated_refit}.

\section{Theoretical results}
\label{sec:theory}
We first list and explain the assumptions
that allow sources to improve target estimation in Section~\ref{sec:assumptions}. Then Section~\ref{sec:upper_bounds} gives the one-step upper bound, and Section~\ref{sec:iterative_theory} studies its improvement through repeated refitting and the role of calibration coverage. In Section~\ref{sec:minimax}, we derive two lower bounds describing two limits on prediction accuracy: one determined by the total number of observations, and the other determined by the number of target observations when source data cannot distinguish changes in the target operator.

\subsection{Assumptions}\label{sec:assumptions} 
We assume that, for every $j$ and $l$, the map $T\mapsto T\circ D_l\circ A^{(j)}$ is bounded from $H$ to $H$. This ensures that the population moment operators defined below are bounded linear maps from $H$ into the tensor-product space $V\otimes W$. The following gives the population counterparts of the empirical quantities defined in Section~\ref{sec:estimator}.
Define the target curvature $\Gamma^{(0)}:H\to H$ by
$\inprod{\Gamma^{(0)}T_1}{T_2}_H=\E\{\inprod{T_1F^{(0)}}{T_2F^{(0)}}_V\}$. For $j=1,\dots,m$, define the population operators associated with the regularized whitened instrument by
\begin{equation}
\begin{aligned}
O^{(j)}(\eta)T
&=\E\{TF^{(j)}\otimes Z^{(j)}(\eta)\},\\
\Phi_l^{(j)}(\eta)T
&=\E\{TD_lU^{(j)}\otimes Z^{(j)}(\eta)\}
=O^{(j)}(\eta)\{T\circ D_l\circ A^{(j)}\}.
\end{aligned}
\end{equation}
Denote by $\Pi^{(j)}(\eta)$ the orthogonal projection in $V\otimes W$ onto $\{\theta\cdot\Phi^{(j)}(\eta)A^{(0)}: \supp(\theta)\subset S_j\}$.
Define 
$\Psi^{(j)}(\eta) =\Pi_\perp^{(j)}(\eta) \{O^{(j)}(\eta)-\theta^{(j)}\cdot\Phi^{(j)}(\eta)\}$,
and the projected source curvature operator $\Gamma^{(j)}(\eta):H\to H$ by $\inprod{\Gamma^{(j)}(\eta)T_1}{T_2}_H =\inprod{\Psi^{(j)}(\eta)T_1} {\Psi^{(j)}(\eta)T_2}_{V\otimes W}$.

Throughout the theoretical analysis, we use the following shorthands:
\begin{equation}
\begin{aligned}
n_S&=\sum_{j=1}^mn_j,\qquad
N=n_0+n_S,\qquad
w_j=\frac{n_j}{N}\ (0\le j\le m),\qquad
\rho=\frac{n_S}{n_0}, \\
\bar\lambda&=\frac{\lambda}{1+\rho} =\frac{n_0\lambda}{N},\qquad
\mathcal G=\Gamma^{(0)}+ \sum_{j=1}^m\rho_j\Gamma^{(j)}(\eta)+\lambda I_H.
\end{aligned}
\end{equation}
For $a>0$, define the effective dimension
$\mathcal N(a)=\tr[\Gamma^{(0)}\{\Gamma^{(0)}+aI_H\}^{-1}]$.

With these definitions in place, we state four assumptions governing the moment geometry, sparse calibration, source informativeness, and concentration tails.

\begin{assumption}[Moment-map compatibility]\label{asm:nonp}
For all $j=1,\ldots,m$, there are finite constants $b_j,\phi_j>0$ such that, for all $0<\eta\le1$ and $T\in H$,
\begin{enumerate}[label=(\alph*)]
\item\label{asm:nonp:feat}
$\norm{\Pi^{(j)}(\eta) \{O^{(j)}(\eta)-\theta^{(j)}\cdot\Phi^{(j)}(\eta)\}T}_{V\otimes W} \le b_j \norm{T}_{\Gamma^{(0)}}$;

\item\label{asm:nonp:perturb}
$\max_{1\le l\le p}\norm{\Phi_l^{(j)}(\eta)T}_{V\otimes W} \le \phi_j \norm{T}_{\Gamma^{(0)}}$.

\end{enumerate}
\end{assumption}

For $T=\widetilde A-A^{(0)}$, Assumption~\ref{asm:nonp}(\ref{asm:nonp:feat}) bounds the component of the source-moment error selected by $\Pi^{(j)}(\eta)$, which is the component used to estimate $\theta^{(j)}$. Write $g_l^{(j)}(\eta)=\Phi_l^{(j)}(\eta)A^{(0)}$. Part~(\ref{asm:nonp:perturb}) bounds the change in each $g_l^{(j)}(\eta)$ when $A^{(0)}$ is replaced by $\widetilde A$: $\norm{\Phi_l^{(j)}(\eta)\widetilde A-g_l^{(j)}(\eta)}_{V\otimes W}\le\phi_j\norm{\widetilde A-A^{(0)}}_{\Gamma^{(0)}}$.
Thus, a small target prediction error also gives small errors in the quantities used for source calibration. Similar assumptions are used in conditional moment estimation \citep{chen2012estimation}.

Define the Gram matrix $G_j(\eta)$ by its entries \begin{equation}G_j(\eta)_{kl}=\inprod{g_k^{(j)}(\eta)}{g_l^{(j)}(\eta)}_{V\otimes W}, \quad 1\le k,l\le p.\end{equation}
Write $S_j^\complement=\{1,\ldots,p\}\setminus S_j$ for the inactive indices. 
\begin{assumption}[Sparse calibration geometry]\label{asm:para}
For all $j=1,\ldots,m$, there are constants
$\kappa_j,\alpha_j,\tau_j>0$ such that, for all $0<\eta\le1$,
\begin{enumerate}[label=(\alph*)]
\item\label{asm:para:eigen}
$\lambda_{\min}(G_j(\eta)[S_j,S_j]) \ge \kappa_j^2$;
\item\label{asm:para:irrep}
$\norm{G_j(\eta)[S_j^\complement,S_j]
G_j(\eta)[S_j,S_j]^{-1}}_\infty \le 1-\alpha_j$; 
\item\label{asm:para:min} $\min_{l\in S_j}|\theta_l^{(j)}|\ge \tau_j$.
\end{enumerate}
\end{assumption} 

Assumption~\ref{asm:para} adapts familiar conditions for Lasso support recovery to the calibration vectors $g_l^{(j)}(\eta)$ \citep{zhao2006model}. Part~(\ref{asm:para:eigen}) prevents cancellation among the active vectors $g_l^{(j)}(\eta)$, $l\in S_j$, so different active coefficient vectors produce different moments. In part~(\ref{asm:para:irrep}), each row of $G_j(\eta)[S_j^\complement,S_j]G_j(\eta)[S_j,S_j]^{-1}$ gives the coefficients for projecting an inactive $g_l^{(j)}(\eta)$ onto the span of the active vectors; the bound limits their total absolute magnitude. Part~(\ref{asm:para:min}) requires every nonzero coefficient to have magnitude at least $\tau_j$. The later sample-size and tuning conditions make estimation errors small enough to distinguish these coefficients from zero. For $S_j=\varnothing$, the conditions on the active coefficients are vacuous and $\Pi^{(j)}(\eta)=0$, while control of false selection is still required; see Appendix~\ref{app:empty_supports}.

\begin{assumption}[Spectral regularity and source richness]\label{asm:spectrum}
There are constants $0<c_\Gamma\le C_\Gamma<\infty$ and $\beta>1/2$ such that the eigenvalues $\nu_k$ of $\Gamma^{(0)}$ satisfy $c_\Gamma k^{-2\beta}\le\nu_k\le C_\Gamma k^{-2\beta}$, $k\ge1$.
For every $1\le j\le m$, $0<\eta\le1$, and $T\in H$,
\begin{equation}
c_\Gamma\norm{T}_{\Gamma^{(0)}}^2
\le \E\{\norm{TF^{(j)}}_V^2\}
\le C_\Gamma\norm{T}_{\Gamma^{(0)}}^2,
\qquad
\Gamma^{(j)}(\eta)
\preceq
C_\Gamma\{\Gamma^{(0)}\}^2
\{\Gamma^{(0)}+\eta I_H\}^{-1}.
\end{equation}
Moreover, $\tr\bigl[ \Sigma^{(j)} \{\Sigma^{(j)}+\eta I_W\}^{-1} \bigr] \le C_\Gamma\mathcal N(\eta)$.
All sources jointly satisfy
\begin{equation}\sum_{j=1}^m\rho_j\Gamma^{(j)}(\eta) \succeq c_\Gamma\rho \{\Gamma^{(0)}\}^2\{\Gamma^{(0)}+\eta I_H\}^{-1}.\end{equation}
\end{assumption} 

Assumption~\ref{asm:spectrum} has three roles. 
First, the polynomial decay $\nu_k\asymp k^{-2\beta}$ follows the spectral framework used in functional prediction \citep{cai2012minimax} and gives $\mathcal N(a)\asymp a^{-1/(2\beta)}$. 
Second, $\E\{\norm{TF^{(j)}}_V^2\}\asymp\norm{T}_{\Gamma^{(0)}}^2$ ensures that the source and target forcing distributions measure operator errors on comparable scales. It is analogous to the bounded covariance eigenvalue conditions used for heterogeneous designs in transfer learning \citep{li2022transfer}.
The trace condition ensures $\E\{\norm{Z^{(j)}(\eta)}_W^2\}\le C_\Gamma\mathcal N(\eta)$, so the second moments of the source instruments are controlled by the same complexity measure.
Third, the lower bound requires the projected source moments, taken together, to control every direction with nonzero target prediction error. Individual sources may lose such directions through projection, provided other sources retain them. 

\begin{assumption}[Tail conditions]\label{asm:bernstein}
There is a constant $C>0$ such that
\begin{equation}\label{eq:subgaussian_forcing}
\bigl\|\norm{TF^{(0)}}_V\bigr\|_{\psi_2}
\le C\norm{T}_{\Gamma^{(0)}},\qquad T\in H.
\end{equation}
For every $1\le j\le m$ and $w\in W$,
\begin{equation}\label{eq:source_subgaussian_forcing}
\norm{\inprod{F^{(j)}}{w}_W}_{\psi_2}
\le C\inprod{\Sigma^{(j)}w}{w}_W^{1/2}.
\end{equation}
Moreover, for all $0<a,\eta\le1$, $1\le j\le m$, and integers $q\ge2$, each nonnegative random quantity $G$ in the table below satisfies $\E(G^q)\le {q!}v^2b^{q-2}/2$, where $v^2$ and $b$ are given in the corresponding row.
\begin{center}
\normalfont\small
\renewcommand{\arraystretch}{1.5}
\begin{tabular}{@{}ccc@{}}
\toprule
Random quantity $G$ & $v^2$ & $b$ \\
\midrule
$\displaystyle \sup_{\norm{T}_H\le1}\bigl\lvert\inprod{\varepsilon^{(0)}}{\bigl[\{\Gamma^{(0)}+aI_H\}^{-1/2}T\bigr]F^{(0)}}_V\bigr\rvert$
& $C^2\mathcal N(a)$ & $C/a^{1/2}$ \\
$\displaystyle \bigl\lVert\{I_V+A^{(0)}\theta^{(j)}\cdot D\}\varepsilon^{(j)}\otimes Z^{(j)}(\eta)\bigr\rVert_{V\otimes W}$
& $C^2\mathcal N(\eta)$ & $C/\eta^{1/2}$ \\
$\bigl\lVert\bigl[T\mapsto TY\otimes Z^{(j)}(\eta)-\E\{TY\otimes Z^{(j)}(\eta)\}\bigr]\{\Gamma^{(0)}+aI_H\}^{-1/2}\bigr\rVert_{\HS}$
& $C^2\mathcal N(\eta)\mathcal N(a)$ & $C/(\eta a)^{1/2}$ \\
\bottomrule
\end{tabular}
\end{center}
In the last row, the bound holds separately for $Y=F^{(j)}$ and for $Y=D_\ell U^{(j)}$, $1\le\ell\le p$.
\end{assumption}

The first two inequalities in Assumption~\ref{asm:bernstein} bound the tails of $\norm{TF^{(0)}}_V$ and $\inprod{F^{(j)}}{w}_W$. The common Bernstein bound controls the target score, the transformed source noise, and the centered source moment maps, allowing concentration bounds for sums of Hilbert-space-valued random variables \citep{pinelis1994optimum}.

\begin{example}[Assumptions~\ref{asm:nonp}--\ref{asm:bernstein}]
\label{ex:elliptic_theory}
Consider the model in Example~\ref{ex:elliptic}. Let $H$ be the Fourier-diagonal operator space $T_ae_k=a_ke_k$, with
$\|T_a\|_H^2=\sum_k|a_k|^2$. Then $A^{(0)}\in H$, and right composition by $\partial_lA^{(j)}$ preserves $H$.
Use the same forcing spectrum $q_k=c_s(1+|k|^2)^{-s}$, $s>1$, in every task, with independent Rademacher coefficients in a real Fourier basis. Take independent, centered Rademacher noise with amplitudes $|\sigma_k|\le C(1+|k|^2)^{-(s+3)/2}$.

Suppose further that two sources have complementary supports $S_1=\{1\}$ and $S_2=\{2\}$, with $0<\tau\le|\theta_1^{(1)}|,|\theta_2^{(2)}|\le\epsilon$ for a sufficiently small fixed $\epsilon>0$. Provided each of these two sources accounts for at least a fixed positive fraction of the total source sample size $n_S$, Assumptions~\ref{asm:nonp}--\ref{asm:bernstein} all hold. In particular, the ordered target eigenvalues satisfy $\nu_r\asymp r^{-s}$, so $\beta=s/2$, and the complementary sources jointly retain a fixed fraction of the full source information. Appendix~\ref{app:elliptic_verification} specifies the real Fourier convention and proves these claims.
\end{example}

\subsection{Non-asymptotic upper bounds}\label{sec:upper_bounds}
We bound $\norm{\widehat A-A^{(0)}}_{\Gamma^{(0)}}^2$ by accounting for sampling errors in the target and source data, together with errors in $\widetilde A$ and $\widetilde\theta^{(j)}$. Throughout this subsection, we assume that, for every $1\le j\le m$,
\begin{equation}
b_j+\phi_j+\alpha_j^{-1}+\kappa_j^{-1}+\tau_j^{-1}+C_\Gamma+\norm{A^{(0)}}_H+\norm{\theta^{(j)}}_{\ell^1}\le C,
\end{equation}
for a constant $C$ that is independent of the sample sizes, $m$, and $p$. In particular, $s_j\tau_j\le\norm{\theta^{(j)}}_{\ell^1}$ implies that $s_j$ is uniformly bounded.
The parameter $t\ge1$ determines the failure probability $C(m+1)\mathrm{e}^{-t}$; taking $t=\log\{C(m+1)/\delta\}$ gives probability at least $1-\delta$.

For any regularization level $a>0$, the empirical covariance and score fluctuations on the target sample ($j=0$, at $a\in\{\lambda_0,\bar\lambda\}$) and source samples ($1\le j\le m$, at $a=\eta$) are governed by the concentration radii
\begin{equation}
\begin{aligned}
r_\Sigma^{(j)}(a)&=\Bigl\{\frac{\mathcal N(a)+t}{n_j}\Bigr\}^{1/2}+\frac{\mathcal N(a)+t}{n_j}, \qquad 0\le j\le m,\\
	r_M^{(j)}(a)&=\Bigl\{\frac{t\mathcal N(a)}{n_j}\Bigr\}^{1/2}+\frac{t}{n_ja^{1/2}}, \qquad 0\le j\le m,\\
	r_{\op}^{(j)}(\eta,a)&=r_\Sigma^{(j)}(\eta)+\Bigl\{\frac{(t+\log p)\mathcal N(\eta)\mathcal N(a)}{n_j}\Bigr\}^{1/2}+\frac{t+\log p}{n_j(\eta a)^{1/2}}, \qquad 1\le j\le m.
\end{aligned}
\end{equation}
Here $r_\Sigma^{(j)}$ bounds empirical covariance deviations, $r_M^{(j)}$ bounds regression and moment noise at the true parameters, and $r_{\op}^{(j)}(\eta,a)$ bounds errors in the empirical moment operators $\widehat O^{(j)}(\eta)$ and $\widehat\Phi_l^{(j)}(\eta)$ (incorporating $r_\Sigma^{(j)}(\eta)$ because the instruments $\widehat Z_i^{(j)}(\eta)$ are whitened using the same source sample).

In Stage~(i), Proposition~\ref{prop:pilot_rate} guarantees that the target-only pilot error in the norm of $\mathcal G_0=\Gamma^{(0)}+\lambda_0 I_H$ is bounded by $e_0=r_M^{(0)}(\lambda_0)+\lambda_0^{1/2}\norm{A^{(0)}}_H$. Because Stage~(ii) calibrates $\theta^{(j)}$ at $\widetilde A$, the penalty $\lambda_j$ must dominate both the source moment noise $r_M^{(j)}(\eta)$ and the inherited pilot error $e_0$:
\begin{equation}\label{eq:lambda_j_choice}
	\lambda_j\ge\frac C{\alpha_j}\bigl(r_M^{(j)}(\eta)
	+[1+\{1+\norm{\theta^{(j)}}_{\ell^1}\}
	r_{\op}^{(j)}(\eta,\lambda_0)]e_0\bigr).
\end{equation}
Under \eqref{eq:lambda_j_choice}, Proposition~\ref{prop:calibration_rate} recovers the exact support $\supp(\widetilde\theta^{(j)})=S_j$ and yields $\norm{\widetilde\theta^{(j)}-\theta^{(j)}}_{\ell^1}\le Cq_j$ with high probability, where $q_j=s_j\lambda_j/\kappa_j^2$.

The theorem below also requires the estimated quantities to remain sufficiently close to their population values. Thus, we further write
\begin{equation}
\begin{aligned}
d_j&=r_{\op}^{(j)}(\eta,\lambda_0)+e_0,
	\qquad
	\varepsilon_{\mathrm{curv}}=
	\frac{n_0}{N}r_\Sigma^{(0)}(\bar\lambda)
	+\sum_{j=1}^m\rho_j\delta_j^2,\\
	\delta_j&=
	\biggl\{b_j\Bigl(\frac{n_0}{N}\Bigr)^{1/2}
	+\Bigl(\frac{n_0}{n_j}\Bigr)^{1/2}\biggr\}
	\Bigl\{r_\Sigma^{(j)}(\eta)
	+\frac{s_j^{1/2}}{\kappa_j}d_j\Bigr\}\\
	&\quad+\Bigl(\frac{n_0}{N}\Bigr)^{1/2}
	\bigl[\{1+\norm{\theta^{(j)}}_{\ell^1}+q_j\}
	r_{\op}^{(j)}(\eta,\bar\lambda)+q_j\bigr].
\end{aligned}
\end{equation}
The quantity $d_j$ combines the errors in estimating the source operators and the target pilot. The quantities $\delta_j$ and $\varepsilon_{\mathrm{curv}}$ are used to ensure that the quadratic term in \eqref{eq:refit} remains bounded below by a constant multiple of $\norm{T}_{\mathcal G}^2$.
\begin{theorem}\label{thm:upper_bound}
Let $0<\lambda_0\le1$ and $0<\eta\le\bar\lambda\le1$.
Choose $\lambda_j$ according to \eqref{eq:lambda_j_choice}.
If
\begin{equation}\label{eq:upper_admissibility}
r_\Sigma^{(0)}(\lambda_0)
+\max_{1\le j\le m}(d_j+\lambda_j)
+\varepsilon_{\mathrm{curv}}\le c
\end{equation}
for a sufficiently small constant $c>0$, then, with probability at least $1-C(m+1)\mathrm{e}^{-t}$,
\begin{equation}\label{eq:upper_rate_explicit}
\norm{\widehat A-A^{(0)}}_{\Gamma^{(0)}}^2
\le C\biggl[
w_0^2
r_M^{(0)}(\bar\lambda)^2
+\bar\lambda\norm{A^{(0)}}_H^2
+\sum_{j=1}^m w_j\bigl\{r_M^{(j)}(\eta)+q_je_0\bigr\}^2
\biggr].
\end{equation}
\end{theorem}

In \eqref{eq:upper_rate_explicit}, $w_0^2r_M^{(0)}(\bar\lambda)^2$ is the contribution from target sampling error, and $\bar\lambda\norm{A^{(0)}}_H^2$ is the error due to regularization in this bound. For source $j$, $r_M^{(j)}(\eta)$ accounts for sampling error, while $q_je_0$ accounts for errors in $\widetilde\theta^{(j)}$ and $\widetilde A$. The product $q_je_0$ arises because the projection removes $\{\widetilde\theta^{(j)}-\theta^{(j)}\}\cdot\widehat\Phi^{(j)}(\eta)\widetilde A$; the remaining term also contains $\widetilde A-A^{(0)}$. Thus, improving either estimate reduces this contribution.
The proof of this theorem and those of the two corollaries below are given in Appendix~\ref{app:upper_proofs}.

The next corollary chooses $\lambda_0$ and $\lambda_j$ to simplify the bound. With $\lambda_0=(t/n_0)^{2\beta/(2\beta+1)}$, the pilot bound gives $e_0\le C(t/n_0)^{\beta/(2\beta+1)}$, so we take
\begin{equation}\label{eq:calibration_rate_simplified}
\lambda_j=C\biggl\{r_M^{(j)}(\eta)
+\Bigl(\frac{t}{n_0}\Bigr)^{\frac{\beta}{2\beta+1}}\biggr\}.
\end{equation}
To state the resulting bound, we set
\begin{equation}
\begin{aligned}
v(t)=\frac{t}{N}(m+w_0),
\quad
u(t)=\Bigl(\frac{t}{N}\Bigr)^2\biggl(1+\sum_{j=1}^m\frac1{w_j}\biggr),\quad
c_\theta=\sum_{j=1}^mw_j\frac{s_j^2}{\alpha_j^2\kappa_j^4}.
\end{aligned}
\end{equation}

\begin{corollary}
\label{cor:transparent_rate}
Under the conditions of Theorem~\ref{thm:upper_bound}, take $\lambda_0=(t/n_0)^{2\beta/(2\beta+1)}$ and $\eta=\bar\lambda$, and choose $\lambda_j$ as in \eqref{eq:calibration_rate_simplified}. Then, with probability at least $1-C(m+1)\mathrm{e}^{-t}$,
\begin{equation}\label{eq:readable_upper_rate}
\norm{\widehat A-A^{(0)}}_{\Gamma^{(0)}}^2
\le C\biggl\{\bar\lambda\norm{A^{(0)}}_H^2
+v(t)\bar\lambda^{-1/(2\beta)}+u(t)\bar\lambda^{-1}
+c_\theta\Bigl(\frac{t}{n_0}\Bigr)^{\frac{4\beta}{2\beta+1}}\biggr\}.
\end{equation}
The four terms are, in order, regularization bias, effective-dimension variance, the Bernstein remainder, and the one-step calibration remainder. 
\end{corollary}

In \eqref{eq:readable_upper_rate}, $\bar\lambda\norm{A^{(0)}}_H^2$ decreases when $\bar\lambda$ decreases, whereas $v(t)\bar\lambda^{-1/(2\beta)}+u(t)\bar\lambda^{-1}$ increases. Thus, choosing $\bar\lambda$ balances the error due to regularization against sampling errors. The term $c_\theta(t/n_0)^{4\beta/(2\beta+1)}$ does not depend on $\bar\lambda$ and comes from estimating the source coefficients using the target pilot.

\begin{corollary}\label{cor:transfer_improvement}
Let $m$ be fixed. As $n_0\to\infty$, assume $1\le t=o(n_0), n_0=o(N),n_j\asymp N$ for $1\le j\le m$, and
\begin{equation}
t+\log p=o\Bigl(N^{\frac{2\beta-1}{2\beta+1}}t^{\frac{2}{2\beta+1}}\wedge N^{\frac{1}{2\beta+1}}t^{\frac{2\beta}{2\beta+1}}\Bigr).
\end{equation}
Choose
\begin{equation}
\lambda_0=\Bigl(\frac{t}{n_0}\Bigr)^{\frac{2\beta}{2\beta+1}},\qquad \bar\lambda=\eta=\Bigl(\frac{t}{N}\Bigr)^{\frac{2\beta}{2\beta+1}},\qquad \lambda_j=C\Bigl(\frac{t}{n_0}\Bigr)^{\frac{\beta}{2\beta+1}},
\end{equation}
with a sufficiently large constant in $\lambda_j$. Then, for all sufficiently large $n_0$, with probability at least $1-C\mathrm{e}^{-t}$,
\begin{equation}\label{eq:transfer_improvement}
\norm{\widehat A-A^{(0)}}_{\Gamma^{(0)}}^2\le C\biggl\{\Bigl(\frac{t}{N}\Bigr)^{\frac{2\beta}{2\beta+1}}+\Bigl(\frac{t}{n_0}\Bigr)^{\frac{4\beta}{2\beta+1}}\biggr\}.
\end{equation}
In particular, if $tN\le Cn_0^2$, then $\norm{\widehat A-A^{(0)}}_{\Gamma^{(0)}}^2\le C(t/N)^{2\beta/(2\beta+1)}$.
\end{corollary}

Corollary~\ref{cor:transfer_improvement} separates two sources of error. The term $(t/N)^{2\beta/(2\beta+1)}$ reflects information from all observations. The term $(t/n_0)^{4\beta/(2\beta+1)}$ comes from using the target pilot in one-step calibration. Both improve on the target-only bound $(t/n_0)^{2\beta/(2\beta+1)}$ when $N/n_0\to\infty$ and $t/n_0\to0$; see Proposition~\ref{prop:pilot_rate} in Appendix~\ref{app:pilot_proof}. For $N\lesssim n_0^2/t$, the all-source term determines the bound. For larger $N$, the calibration remainder determines this one-step bound.  

\subsection{Iterative refinement and calibration coverage}\label{sec:iterative_theory}
The one-step bound in Corollary~\ref{cor:transfer_improvement} contains the term $(t/n_0)^{4\beta/(2\beta+1)}$ from calibration at the target pilot. The recalibration in Section~\ref{sec:recalibration} uses each refit as a new pilot to reduce this term in the bound. We first give the rate after a fixed number of refits, then state exactly which calibration error the projection must remove.

\begin{theorem}\label{thm:iterated_refit}
Under the conditions and tuning choices of Corollary~\ref{cor:transfer_improvement}, let $\widehat A^{[1]}=\widehat A$ and define later iterates by \eqref{eq:iterated_calibration}--\eqref{eq:iterated_refit}. For each fixed integer $K\ge1$ and all sufficiently large $n_0$, with probability at least $1-C\mathrm{e}^{-t}$,
\begin{equation}\label{eq:iterated_rate}
\norm{\widehat A^{[K]}-A^{(0)}}_{\Gamma^{(0)}}^2\le C_K\biggl\{\Bigl(\frac{t}{N}\Bigr)^{\frac{2\beta}{2\beta+1}}+\Bigl(\frac{t}{n_0}\Bigr)^{\frac{2^{K+1}\beta}{2\beta+1}}\biggr\},
\end{equation}
where $C_K$ depends on $K$. In particular, if $N/t\le C(n_0/t)^{2^K}$, then $\norm{\widehat A^{[K]}-A^{(0)}}_{\Gamma^{(0)}}^2\le C_K(t/N)^{2\beta/(2\beta+1)}$.
\end{theorem}

The proof of Theorem~\ref{thm:iterated_refit} is in Appendix~\ref{app:iterative_proof}. In \eqref{eq:iterated_rate}, each additional refit doubles the exponent of $t/n_0$ in the calibration term, while leaving $(t/N)^{2\beta/(2\beta+1)}$ unchanged. For example, the condition for the first term to determine the bound changes from $N\le Cn_0^2/t$ at $K=1$ to $N\le Cn_0^4/t^3$ at $K=2$. 

The proof of Theorem~\ref{thm:iterated_refit} uses $\widehat S_j=S_j$ to ensure that each projection removes the calibration error evaluated at the current target estimate. The next proposition states the requirement for arbitrary calibration vectors and projections.

Fix $\eta>0$, write $\widehat\Phi^{(j)}=\widehat\Phi^{(j)}(\eta)$, and let $\widetilde A\in H$ be any pilot, including a later iterate. For source $j$, let $\widetilde\theta^{(j)}\in\R^p$ and let $\Pi_j$ be an arbitrary orthogonal projection on $V\otimes W$, with $\Pi_{\perp,j}=I-\Pi_j$. These quantities are held fixed during the refit. Set $e_j=\widetilde\theta^{(j)}-\theta^{(j)}$ and define
\begin{equation}
\begin{aligned}
r_{\Pi,j}&=\Pi_{\perp,j}\widehat M_\eta^{(j)}(A^{(0)},\widetilde\theta^{(j)}), & z_{\Pi,j}&=\Pi_{\perp,j}\widehat M_\eta^{(j)}(A^{(0)},\theta^{(j)}),\\
u_{\Pi,j}&=\Pi_{\perp,j}\{e_j\cdot\widehat\Phi^{(j)}\}\widetilde A, & p_{\Pi,j}&=\Pi_{\perp,j}\{e_j\cdot\widehat\Phi^{(j)}\}\{A^{(0)}-\widetilde A\}.
\end{aligned}
\end{equation}
Linearity in the coefficients gives $r_{\Pi,j}=z_{\Pi,j}+u_{\Pi,j}+p_{\Pi,j}$. Here $z_{\Pi,j}$ is the projected moment noise at the true parameters, $u_{\Pi,j}$ is the coefficient error evaluated at the pilot that remains after projection, and $p_{\Pi,j}$ combines coefficient and pilot errors. 

\begin{proposition}\label{prop:revision-coverage}
If 
\begin{equation}
	\{e_j\cdot\widehat\Phi^{(j)}\}\widetilde A\in\operatorname{ran}(\Pi_j),
	\label{eq:revision-vector-coverage}
\end{equation}
 then $u_{\Pi,j}=0$ and
\begin{equation}\label{eq:revision-product-residual}
\|r_{\Pi,j}\|\le\|z_{\Pi,j}\|+\|e_j\|_{\ell^1}\max_{1\le l\le p}\|\widehat\Phi_l^{(j)}\{A^{(0)}-\widetilde A\}\|.
\end{equation}
\end{proposition} 

Proposition~\ref{prop:revision-coverage} explains why \eqref{eq:revision-vector-coverage} removes the calibration error evaluated at the pilot. Its proof is in Appendix~\ref{app:coverage_proof}. The proposition is easiest to understand when $\widetilde A=A^{(0)}$. Under \eqref{eq:revision-vector-coverage}, projection then removes the calibration error completely, even if $\widetilde\theta^{(j)}$ is inaccurate. When $\widetilde A\neq A^{(0)}$, the remaining calibration error is bounded by a product of the coefficient error and the pilot error, as shown in \eqref{eq:revision-product-residual}. A better pilot therefore allows the same coefficient error to have a smaller effect on the source moments.

Removing calibration error is only part of the task. For example, choosing $\Pi_j=I$ makes $r_{\Pi,j}=0$, but also removes every source term from the refit objective. Thus, the projection need remove calibration error while keeping source information useful for estimating $A^{(0)}$.

\subsection{Minimax lower bounds}\label{sec:minimax}
We establish two lower bounds for prediction of $A^{(0)}$. The first is $N^{-2\beta/(2\beta+1)}$ on a class satisfying Assumptions~\ref{asm:nonp}--\ref{asm:bernstein}. The second is $n_0^{-1}$ on a class in which the source laws remain fixed as $A^{(0)}$ changes.
Both constructions belong to the closed-operator model in Section~\ref{sec:model}; applying these bounds to a specified PDE coefficient class requires that it contains the corresponding construction.

For a data law $P$, write $\pr_P$ for probability under $P$ and let $\ell_P(\widehat A)=\|\widehat A-A_P^{(0)}\|_{\Gamma_P^{(0)}}^2$.
For a parameter class $\mathcal P$, define
\begin{equation}
R_\mathrm{e}^*(\mathcal P)=\inf_{\widehat A}\sup_{P\in\mathcal P}\E_P\{\ell_P(\widehat A)\},
\qquad
R_\gamma^*(\mathcal P)=\inf_{\widehat A}\inf\{r>0:
\sup_{P\in\mathcal P}\pr_P\{\ell_P(\widehat A)>r\}\le\gamma\}.
\end{equation}
The infima range over all estimators using all target and source observations. The quantity $R_\mathrm{e}^*$ measures expected prediction error, whereas $R_\gamma^*$ measures the error threshold guaranteed uniformly over $\mathcal P$ with probability at least $1-\gamma$. We use $R_\gamma^*$ when comparing the lower bounds with the high-probability upper bounds proved above.

For the first result, we construct a class in which all $N$ observations have the same distribution. A lower bound on this class shows that estimation remains difficult even when every source is identical to the target.
\begin{example}\label{exp:3}
Let $V=W=\mathbb R\oplus\ell^2$ and let $\Lambda:\ell^2\to\ell^2$ be defined by $(\Lambda a)_k=\mu_ka_k$, $k\ge1$, where $\mu_k>0$ and $\sum_k\mu_k^2<\infty$. Set $V_0=\mathbb R\oplus\operatorname{ran}(\Lambda)$. For $a\in\ell^2$, define $A_a:W\to V_0$ and $L_a:V_0\to W$ by
\begin{equation}\label{eq:block_operator}
A_a(x,f)=(x+\langle a,f\rangle,\Lambda f),\qquad L_a(y,g)=(y-\langle a,\Lambda^{-1}g\rangle,\Lambda^{-1}g).
\end{equation}
Let
\begin{equation}\label{eq:exp-3-H-def}
H=\{T_{c,a}:(x,f)\mapsto(cx+\langle a,f\rangle,c\Lambda f)\mid(c,a)\in V\},\qquad\|T_{c,a}\|_H^2=c^2+\|a\|_{\ell^2}^2,
\end{equation}
so $A_a\in H$. Fix a nonzero $h_0\in\ell^2$, take $p=1$, and use the known bounded operator $D(y,g)=(-\langle h_0,g\rangle,0)$. For a fixed $\beta>1/2$, take $\nu_k=c_\beta k^{-2\beta}$ and observe in every task
\begin{equation}\label{eq:block_observation}
F=(0,(\nu_k^{1/2}\xi_k)_{k\ge1}),\qquad U=A_aF+(\sigma\zeta,0),
\end{equation}
where the $\xi_k$ are independent Rademacher variables, $\zeta\sim N(0,1)$ is independent of them, and $\sigma>0$. All tasks have the same $a$, and $\theta^{(j)}=0$. Denote the resulting family with $\|a\|_{\ell^2}\le R$, where $R>0$, by $\mathcal P_{\rm anc}(R)$.
\end{example}
Proposition~\ref{prop:anchored_submodel} in Appendix~\ref{app:anchored_model_verification} verifies that $\mathcal P_{\rm anc}(R)$ consists of valid observations from closed operators and satisfies Assumptions~\ref{asm:nonp}--\ref{asm:bernstein} uniformly.

\begin{theorem}\label{thm:minimax_source}
There are constants $c>0$ and $\gamma_0\in(0,1/2)$, depending only on $R,\beta,\sigma$, and $\Lambda$, such that, for all sufficiently large $N$, $R_\mathrm{e}^*(\mathcal P_{\rm anc}(R))\ge cN^{-2\beta/(2\beta+1)}$ and $R_{\gamma_0}^*(\mathcal P_{\rm anc}(R))\ge cN^{-2\beta/(2\beta+1)}$.
\end{theorem}

Theorem~\ref{thm:minimax_source} shows that even with $\theta^{(j)}=0$ for every source, no estimator can improve on $N^{-2\beta/(2\beta+1)}$ uniformly over $\mathcal P_{\rm anc}(R)$. The result uses all $N$ functional observations and remains valid under the sampling design $n_j\ge n_0$ for every source. For fixed $K$ and confidence level, Theorem~\ref{thm:iterated_refit} attains the same rate when $N\lesssim n_0^{2^K}$ and its other conditions hold. Thus the upper and lower bounds agree in that range for fixed-confidence risk. An expected-risk upper bound for the proposed estimator would require a separate argument. The lower-bound proof is in Appendix~\ref{app:all_source_lower_proof}.

The next result considers a common perturbation that can be transferred between the target operator and all source coefficient vectors without changing any source law. Source observations then provide no information about this change, and the resulting prediction error is limited by the target sample size.

For a law $P_0$ and $h\in\mathbb R^p\setminus\{0\}$, let $P_s$ denote the law obtained by setting $L_s^{(0)}=L^{(0)}+sh\cdot D$, $A_s^{(0)}=\{L_s^{(0)}\}^{-1}$, and $\theta_s^{(j)}=\theta^{(j)}-sh$, while keeping the forcing and noise laws fixed. Write $T_h=A^{(0)}(h\cdot D)A^{(0)}$ and let $P_s^{(0)}$ denote the law of one target observation.

\begin{theorem}\label{thm:minimax_shift}
Assume that $(h\cdot D)A^{(0)}$ is bounded, $\supp(h)\subseteq\bigcap_{j=1}^m S_j$, $T_h\in H$ and $\norm{T_h}_{\Gamma^{(0)}}>0$. Suppose that, for all sufficiently small $|s|$, $P_s\in\mathcal P$ and $\operatorname{KL}(P_s^{(0)},P_0^{(0)})\le Cs^2$. Then the source laws do not depend on $s$, and, for all sufficiently large $n_0$, $R_\mathrm{e}^*(\mathcal P)\ge c/n_0$ and $R_{\gamma_1}^*(\mathcal P)\ge c/n_0$, where $c>0$ depends only on the fixed path and $\gamma_1\in(0,1/4)$ is numerical.
\end{theorem}

Along the path in Theorem~\ref{thm:minimax_shift}, $L_s^{(0)}+\theta_s^{(j)}\cdot D=L^{(j)}$, so source observations do not help distinguish different values of $s$. The condition $\norm{T_h}_{\Gamma^{(0)}}>0$ ensures that these values give different target predictions. Only the $n_0$ target observations provide information about this change, which yields the $n_0^{-1}$ lower bound regardless of the source sample sizes. Its proof is in Appendix~\ref{app:common_shift_lower_proof}.

To illustrate Theorem~\ref{thm:minimax_shift}, we use the block model in Example~\ref{exp:3}. Source $j$ depends on $a$ and $\theta^{(j)}$ only through $a+\theta^{(j)}\Lambda h_0$. With source coefficients bounded away from zero, the shifts $a\mapsto a+s\Lambda h_0$ and $\theta^{(j)}\mapsto\theta^{(j)}-s$ leave all source laws unchanged for small $|s|$, but change the target prediction at $F=(0,f)$ by $(s\langle\Lambda h_0,f\rangle,0)$. Thus, only target observations can identify $s$. Appendix~\ref{app:block_shift_verification} verifies the theorem's conditions.
 
For a class containing both constructions, Theorems~\ref{thm:minimax_source} and \ref{thm:minimax_shift} give a lower bound proportional to $N^{-2\beta/(2\beta+1)}\vee n_0^{-1}$, equivalently the sum of these terms up to constants. This holds for expected risk and for fixed-confidence risk with $\gamma\le\min(\gamma_0,\gamma_1)$.

\section{Numerical studies}\label{sec:experiment}
We evaluate the finite-sample performance of the proposed transfer estimators through two simulation studies and a real-world engineering application. The three studies are designed to examine complementary aspects of our methodology and theory:
\begin{enumerate}
 \item Section~\ref{sec:functional-per-n0} considers a structured functional regression model (Example~\ref{exp:3}) to illustrate how governing-operator calibration and source-specific projection harness complementary sources
 in regression-coefficient space.
 \item Section~\ref{sec:exp-diffusion} studies a periodic advection--diffusion equation (Example~\ref{ex:elliptic}) across varying noise levels and target sample sizes, testing whether projection eliminates first-order pilot error (Proposition~\ref{prop:revision-coverage}) and whether iterative recalibration further refines the remaining higher-order error (Theorem~\ref{thm:iterated_refit}).
 \item Section~\ref{sec:real_ltv} applies our framework to experimental measurements from a linear time-varying electronic band-pass filter, demonstrating practical transfer gains in a data-scarce regime.
\end{enumerate}

\subsection{Competing methods and evaluation metrics}
\label{sec:baselines}

We evaluate the three-stage estimator from Section~\ref{sec:estimator}, referred to as \emph{Projected} ($\widehat A_{\mathrm P}$), alongside its unprojected counterpart \emph{Unprojected} ($\widehat A_{\mathrm U}$), which sets $\widetilde\Pi_\perp^{(j)}(\eta)\equiv I_{V\otimes W}$ in \eqref{eq:refit}. Where indicated, we also examine the iterative recalibration refinements from Section~\ref{sec:recalibration}.

Our estimators are compared against five baselines. The first two represent the extremes of no transfer and uncalibrated data integration:
\begin{itemize}
 \item \emph{Target-only} ($\widehat A_{\mathrm T}$), which fits the target observations alone via ridge regression:
 \begin{equation}
  \widehat A_{\mathrm T}
  =\underset{A\in H}{\operatorname{arg\,min}}
  \biggl\{\frac{1}{n_0}\sum_{i=1}^{n_0}
  \|U_i^{(0)}-AF_i^{(0)}\|_V^2+\lambda\|A\|_H^2\biggr\};
 \end{equation}
 \item \emph{Pooling} ($\widehat A_{\mathrm{pool}}$), which fits a single shared operator across all tasks without adjusting for structural shifts:
 \begin{equation}
  \widehat A_{\mathrm{pool}}
  =\underset{A\in H}{\operatorname{arg\,min}}
  \biggl\{\frac{1}{N}\sum_{j=0}^{m}\sum_{i=1}^{n_j}
  \|U_i^{(j)}-AF_i^{(j)}\|_V^2+\lambda\|A\|_H^2\biggr\}.
 \end{equation}
\end{itemize}
Here the regularization parameter $\lambda$ is tuned separately for each estimator.

The remaining three baselines are recent statistical transfer methods developed for functional linear regression (FLR):
\begin{itemize}
 \item \emph{Jiao--McKeague} ($\widehat A_{\mathrm{JM}}$; \citealt{jiao2025coefficient}), a coefficient-shape transfer estimator that averages sign-aligned target and source least-squares fits in empirical principal-component coordinates and rescales the resulting common shape on the target sample;
 \item \emph{Lin--Reimherr} ($\widehat A_{\mathrm{LR}}$; \citealt{lin2024hypothesis}), an adaptive hypothesis-transfer estimator that pairs a pooled-source ridge fit with a ridge-penalized target offset across nested subsets of ranked sources and aggregates the candidates on held-out target data;
 \item \emph{Hu--Lin} ($\widehat A_{\mathrm{HL}}$; \citealt{hu2025functional}), an adaptive transfer estimator that combines pooled-source principal-component regression with an $\ell^1$-penalized target offset and aggregates candidates over ranked source sets.
\end{itemize}

Although JM, LR, and HL were originally formulated for scalar-response FLR, their estimation, source-ranking, and aggregation steps depend on the observations only through the quadratic empirical prediction loss. In our experiments, each candidate operator $A\in H$ is represented in a finite-dimensional basis of the operator space $H$. Because the functional-response squared error $\|U_i^{(j)}-AF_i^{(j)}\|_V^2$ (or the output prediction error in Section~\ref{sec:real_ltv}) is a quadratic form in these operator coordinates, it induces an empirical normal (Gram) operator and predictor--response cross-covariance directly on $H$. We therefore implement JM, LR, and HL in $H$ by treating the operator coordinates as the regression parameter vector, extracting empirical principal directions from the task-specific normal operators, and applying their methods without supplying the true informative source sets. Full adaptation and tuning details are provided in Appendix~\ref{app:literature-baselines}.

In the simulation studies, estimation accuracy is quantified by the out-of-sample prediction risk $R(\widehat A;A^{(0)},\Gamma^{(0)})=\|\widehat A-A^{(0)}\|_{\Gamma^{(0)}}^2$. We denote its Monte Carlo average across independent repetitions by $\overline R_q$ for each estimator $\widehat A_q$, $q\in\{\mathrm U,\mathrm P,\mathrm T,\mathrm{pool},\mathrm{JM},\mathrm{LR},\mathrm{HL}\}$. Within every repetition, all estimators are fitted and evaluated on the exact same data realizations. The real-data application is evaluated by held-out spectral prediction error as defined in Section~\ref{sec:real_ltv}. Complete experimental protocols are deferred to Appendix~\ref{app:experiments}.

\subsection{Functional regression with block operator structure}
\label{sec:functional-per-n0}

We first consider a finite-dimensional realization of the block operator model in Example~\ref{exp:3} with $V=W=\mathbb R\oplus\mathbb R^K$ ($K=512$) and $m=2$ source tasks. Following the notation of Example~\ref{exp:3}, we set $\beta=\sigma=1$, $\mu_k=k^{-1}$, and $\nu_k=k^{-2}$ for $k=1,\ldots,K$. The target operator is $A^{(0)}=T_{1,a^{(0)}}$, where the slope vector $a^{(0)}=(a_k)_{k=1}^K$ satisfies $a_k\propto k^{-3/4}$ and $\|a^{(0)}\|_2=1$. Task differences are governed by a two-element dictionary $D_l(y,g)=(-\langle h_l,g\rangle,0)$, $l=1,2$, where $h_1$ and $h_2$ are unit vectors with disjoint supports on the odd and even indices among the first 16 coordinates. The two sources have disjoint active supports $S_1=\{1\}$ and $S_2=\{2\}$, so each source perturbs the target governing operator along a distinct dictionary direction. While Target-only, Unprojected, and Projected estimate the full operator parameter $(c,a)$ in \eqref{eq:exp-3-H-def-app}, Pooling and the three FLR baselines are supplied with the exact scalar $c=1$
and estimate only the scalar-response slope vector $a\in\mathbb R^K$.

\begin{figure}[t]
 \centering
 \includegraphics[width=\textwidth]{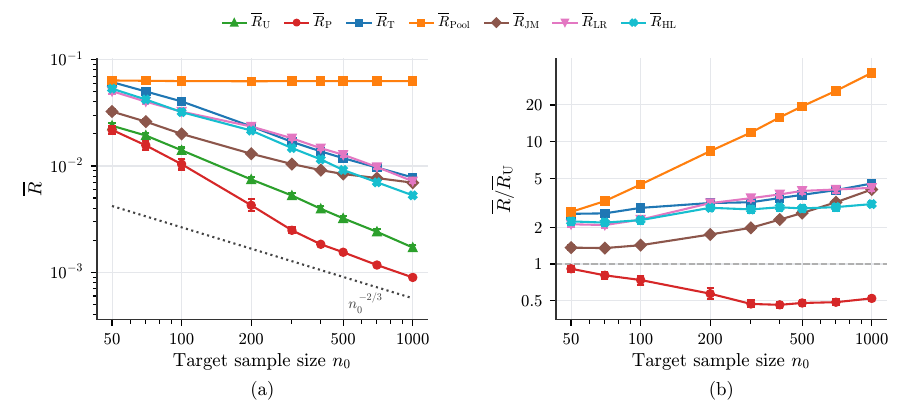}
 \caption{Functional regression across target sample sizes $n_0$ with fixed task ratio $N=20n_0$ ($n_1=n_2=9.5n_0$). (a) Mean prediction risks $\overline R_q$ over 1,000 independent sample-size trajectories, with $n_0^{-2/3}$ shown as a reference rate. (b) Mean prediction risks relative to Unprojected ($\overline R_q/\overline R_{\mathrm U}$).
 Error bars represent pointwise $95\%$ confidence intervals from 5,000 bootstrap samples.}
 \label{fig:functional-per-n0}
\end{figure}

We vary the target sample size $n_0$ while maintaining a fixed total-to-target sample ratio $N=20n_0$ ($n_1=n_2=9.5n_0$). At each $n_0$, regularization and truncation hyperparameters are tuned on independent development datasets and held fixed during evaluation. Figure~\ref{fig:functional-per-n0} reports the mean prediction risks $\overline R_q$ and the relative risks $\overline R_q/\overline R_{\mathrm U}$ for $n_0\in[50,1000]$. For comparison, panel~(a) includes a reference line with rate $n_0^{-2/3}$. Corollary~\ref{cor:transfer_improvement} provides a related all-source bound under its stated assumptions.

As shown in Figure~\ref{fig:functional-per-n0}(a), both Unprojected and Projected substantially outperform all five baselines across the entire sample-size range and closely track the $n_0^{-2/3}$ reference line. In contrast, the risk of Pooling remains nearly flat as $n_0$ grows, demonstrating that abundant source data cannot compensate for uncalibrated structural misspecification ($A^{(j)}\ne A^{(0)}$). Figure~\ref{fig:functional-per-n0}(b) further shows that Projected consistently improves upon Unprojected, cutting the prediction risk by roughly half at moderate-to-large sample sizes.

A closer inspection of the literature baselines highlights the fundamental difference between parameter-space transfer and governing-operator calibration. Although both sources differ from the target by a one-sparse shift in $L^{(j)}$, the induced regression-coefficient contrast is $a^{(j)}-a^{(0)}=\theta_1^{(j)}\Lambda h_1+\theta_2^{(j)}\Lambda h_2$. Because $\mu_k$ and $\nu_k$ decay with $k$, equalizing the calibration signal strength across the two dictionary directions requires a larger perturbation magnitude on the second source, yielding $\|a^{(2)}-a^{(0)}\|_2\approx 3.73$ compared with $\|a^{(1)}-a^{(0)}\|_2\approx 0.39$. Consequently, parameter-space transfer methods view Source~2 as severely misaligned: JM selects only Source~1 at every sample size, and LR and HL consistently rank Source~1 first while heavily penalizing or downweighting Source~2. By calibrating the sparse perturbation in the governing operator $L^{(j)}$ rather than penalizing the dense shift in $A^{(j)}$, Unprojected and Projected effectively leverage both sources.

Finally, the block structure of Example~\ref{exp:3} makes the role of moment projection transparent. As shown in Appendix~\ref{app:fr-estimation}, for every $A\in H$ the empirical source moment satisfies
\begin{equation}
 \widetilde\Pi_\perp^{(j)}(\eta)
 \widehat M_\eta^{(j)}(A,\widetilde\theta^{(j)})
 =
 \widetilde\Pi_\perp^{(j)}(\eta)
 \widehat M_\eta^{(j)}(A,0).
\end{equation}
Thus, once the active supports are identified, the projected refit objective is completely invariant to estimation error in the continuous perturbation magnitudes $\widetilde\theta^{(j)}$. 
Because the active calibration directions for $S_1=\{1\}$ and $S_2=\{2\}$ are mutually orthogonal, the direction removed from Source~1 is preserved by Source~2 and vice versa, directly illustrating how source-specific projections satisfy the complementary source richness condition in Assumption~\ref{asm:spectrum}.

\subsection{Periodic advection--diffusion PDE}
\label{sec:exp-diffusion}

We next investigate moment projection and iterative recalibration in the two-dimensional periodic advection--diffusion model of Example~\ref{ex:elliptic}. On the torus $\mathbb T^2=(\mathbb R/2\pi\mathbb Z)^2$ with $V=W=L^2(\mathbb T^2)$ and $V_0=H^2_{\mathrm{per}}(\mathbb T^2)$, we consider the governing operators
\begin{equation}
 L^{(0)}=I-\Delta,\qquad
 L^{(1)}=L^{(0)}+\tau\partial_1,\qquad
 L^{(2)}=L^{(0)}+\tau\partial_2,
\end{equation}
corresponding to dictionary elements $D_l=\partial_l$ ($l=1,2$) and sparse advection shifts $\theta^{(1)}=(\tau,0)$ and $\theta^{(2)}=(0,\tau)$ with $\tau=1$. We estimate the target solution operator $A^{(0)}=\{L^{(0)}\}^{-1}$ in the real Hilbert space $H$ of Hilbert--Schmidt Fourier multipliers with conjugate-symmetric symbols. Retaining Fourier modes $\|k\|_\infty\le 12$ yields a 625-dimensional parameter space. The forcing and observation noise follow power-law spectral decay in the real Fourier basis with smoothness exponent $\beta=1$ and noise scale $\sigma>0$, and the source sample sizes are fixed at $n_1=n_2=16384$ (see Appendix~\ref{app:exp-diffusion} for full specifications).

\begin{figure}[t]
 \centering
 \includegraphics[width=\textwidth]{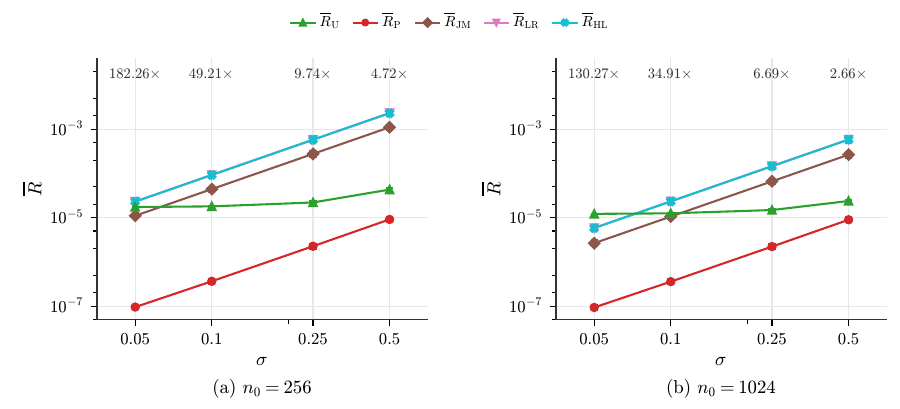}
 \caption{Mean prediction risks in the advection--diffusion model across noise levels $\sigma$ with $\tau=1$ and $n_1=n_2=16384$: (a) $n_0=256$; (b) $n_0=1024$. Both axes are on a logarithmic scale. Annotations above the curves report the risk ratio $\overline R_{\mathrm U}/\overline R_{\mathrm P}$. The LR and HL curves nearly coincide. Error bars (nearly invisible) indicate pointwise $95\%$ percentile intervals from 20,000 bootstrap samples over 300 repetitions.}
 \label{fig:exp-projection}
\end{figure}

\paragraph{Effect of projection across noise levels.}
We first vary the noise level $\sigma\in\{0.05,0.1,0.25,0.5\}$ at two target sample sizes, $n_0=256$ and $n_0=1024$. Figure~\ref{fig:exp-projection} compares Unprojected and Projected against the three FLR transfer baselines. Across all settings, Projected achieves the lowest prediction risk, and its margin over Unprojected widens dramatically as $\sigma$ decreases—reaching a $182$-fold risk reduction at $n_0=256$ and $\sigma=0.05$. Notably, as $\sigma$ decreases, the risk of Projected scales down steadily, whereas the risk of Unprojected hits a floor and stagnates at low noise levels, allowing even the literature baselines to overtake Unprojected at $\sigma=0.05$.

This stagnation is not caused by variable-selection failure: at $\sigma=0.05$ and $0.1$, both Unprojected and Projected recover the exact supports $S_1=\{1\}$ and $S_2=\{2\}$ in $100\%$ of the repetitions. Rather, it stems from continuous pilot bias.  Even when $\sigma$ is small, the target-only ridge pilot $\widetilde A$ carries regularization bias, inducing a Stage~(ii) calibration error $\widetilde\theta^{(j)}-\theta^{(j)}$. Without projection, the resulting first-order bias $\{\widetilde\theta^{(j)}-\theta^{(j)}\}\cdot\widehat\Phi^{(j)}(\eta)\widetilde A$ enters Stage~(iii) directly and creates an error floor that does not vanish as $\sigma\to 0$.  By Proposition~\ref{prop:revision-coverage}, the projection $\widetilde\Pi_\perp^{(j)}(\eta)$ eliminates this leading term,
\begin{equation}
 \widetilde\Pi_\perp^{(j)}(\eta)
 \bigl[\{\widetilde\theta^{(j)}-\theta^{(j)}\}
        \cdot\widehat\Phi^{(j)}(\eta)\bigr]\widetilde A=0,
\end{equation}
leaving only the second-order product residual \eqref{eq:revision-product-residual} and thus the estimator can fully benefit from low-noise data. 

\begin{figure}[t]
 \centering
 \includegraphics[width=\textwidth]{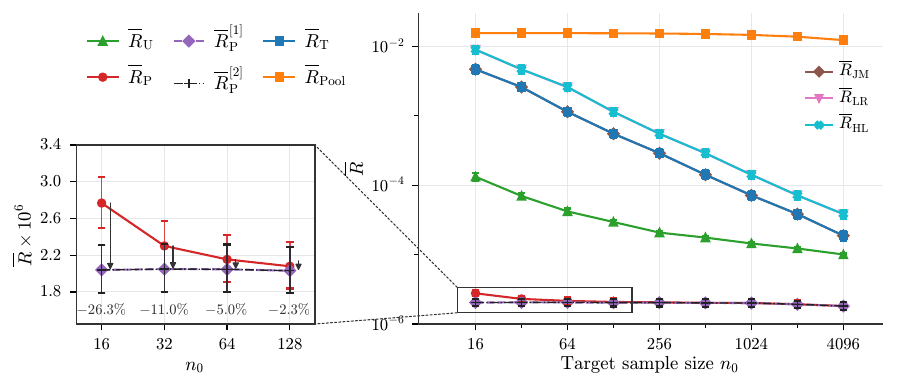}
 \caption{Mean prediction risks in the advection--diffusion model across target sample sizes $n_0$, with $\sigma=0.25$, $\tau=1$, and $n_1=n_2=16384$. The left inset magnifies the small-$n_0$ regime for Projected ($\overline R_{\mathrm P}$) and its first two recalibration iterates ($\overline R_{\mathrm P}^{[1]},\overline R_{\mathrm P}^{[2]}$), with annotations indicating the percentage risk reduction from the first recalibration step. JM nearly coincides with Target-only, and LR with HL. Error bars denote pointwise $95\%$ bootstrap intervals over 300 repetitions.}
 \label{fig:exp-recalibration}
\end{figure}

\paragraph{Effect of target sample size and iterative recalibration.}
Although projection removes the first-order calibration error, the remaining second-order product error still depends on the accuracy of the initial pilot $\widetilde A$ when $n_0$ is very small. To evaluate how iterative recalibration (Section~\ref{sec:recalibration}) mitigates this remainder, we fix $\sigma=0.25$ and vary $n_0$ from $16$ to $4096$. Figure~\ref{fig:exp-recalibration} compares the seven estimators alongside one and two additional recalibration rounds, whose mean risks are denoted by $\overline R_{\mathrm P}^{[1]}$ (for $\widehat A^{[2]}$) and $\overline R_{\mathrm P}^{[2]}$ (for $\widehat A^{[3]}$), respectively.

Across the entire range of $n_0$, Unprojected and Projected achieve orders-of-magnitude lower risk than all five baselines. While Target-only improves steadily with $n_0$ and Pooling remains dominated by misspecification bias, JM, LR, and HL fail to extract useful transfer gains: JM always selects the empty source set and coincides with Target-only, while LR and HL perform similarly to a target-only fit on the half-sample used prior to their aggregation step. The spectral diagnostics in Appendix~\ref{app:exp-diffusion} explain why parameter-space transfer fails here. In the Fourier basis, the constant mode ($k=0$) is unaffected by advection ($\mathrm{i}\theta^{(j)}\cdot 0=0$). It is therefore shared across all tasks and accounts for nearly $90\%$ of the target risk. However, advection induces discrepancies at low nonzero frequencies, so $A^{(j)}$ is not a scalar multiple of $A^{(0)}$. To accommodate these low-frequency shifts, LR and HL must select small offset penalties, which inadvertently allow the shared constant mode to be re-estimated from the noisy target sample rather than locked to the precise source estimate. By modeling the differential shift in $L^{(j)}$, our estimators correct the low-frequency advection while preserving the shared constant mode.

The left inset of Figure~\ref{fig:exp-recalibration} isolates the benefit of iterative recalibration. When target data are scarce ($n_0=16$), a single recalibration round ($\overline R_{\mathrm P}^{[1]}$) reduces the mean prediction risk of Projected by a further $26.3\%$. As $n_0$ increases ($n_0=32,64,128$), the initial pilot $\widetilde A$ becomes more accurate and the incremental gain from recalibration tapers off ($11.0\%$, $5.0\%$, and $2.3\%$). Because Projected already recovers the true supports $S_1=\{1\}$ and $S_2=\{2\}$ in every repetition, this improvement stems entirely from re-estimating the continuous perturbation coefficients and updating the projection subspaces at the refined operator $\widehat A_{\mathrm P}$. Moreover, the second recalibration round ($\overline R_{\mathrm P}^{[2]}$) is virtually indistinguishable from the first, suggesting that an additional round yields little further improvement in this experiment.

\subsection{Application to a linear time-varying electronic filter}
\label{sec:real_ltv}

Finally, we evaluate all seven methods on experimental measurements from the electronic band-pass filter benchmark of \citet{lataire2015benchmark}. In this system, an external periodic scheduling voltage modulates the internal circuit parameters over time, turning the circuit into a linear time-varying (LTV) system. Each operating regime (defined by the harmonic content and amplitude of the scheduling signal) constitutes a task: the predictor $F^{(j)}$ is the measured input voltage, and the state $U^{(j)}$ is the time-integrated output voltage, so that the measured output voltage is the time derivative of $U^{(j)}$.

This benchmark poses two practical challenges for operator transfer. First, whereas linear time-invariant systems act diagonally in the frequency domain, periodic time variation induces inter-frequency coupling that scatters input energy from directly excited frequencies into surrounding harmonic sidebands. All methods therefore estimate a non-diagonal harmonic transfer operator in a shared B-spline operator space $H$. Second, target data are extremely scarce: each of the eight target regimes has only three experimental records with distinct multisine input realizations. Cyclically assigning these three records to training, validation, and testing yields 24 evaluation runs, each with only a single functional target training record ($n_0=1$). To aid estimation, we construct a candidate bank of 12 source tasks (30 records in total) from related constant and time-varying scheduling regimes, and represent inter-regime physical differences via $L^{(j)}-L^{(0)}=\theta^{(j)}\cdot D$, where $D$ comprises products of scheduling-voltage harmonics and time derivatives. Within each evaluation run, Unprojected and Projected select sources and their common weight multiplier using the single target validation record. The candidate source bank is constructed from input and scheduling channels, and calibration uses source records and a pilot fitted on the target training record; see Appendix~\ref{app:real_ltv} for details.

\begin{table}[t]
 \centering
 \caption{Mean normalized spectral prediction error $E_{\mathcal K}$ across the 24 held-out evaluations on the electronic band-pass filter benchmark. Boldface and underlining mark the lowest and second-lowest mean errors in each frequency band (evaluated prior to rounding).}
 \label{tab:real_ltv}
 \begin{tabular}{lccc}
 \toprule
 Method & Full band & Forced lines & Sidebands\\
 \midrule
 Target-only
 & $4.17\times10^{-5}$
 & $1.02\times10^{-6}$
 & $3.47\times10^{-4}$ \\
 Pooling
 & $1.12\times10^{-1}$
 & $1.87\times10^{-2}$
 & $7.01\times10^{-1}$ \\
 \midrule
 Jiao--McKeague
 & $1.93\times10^{-3}$
 & $1.71\times10^{-3}$
 & $4.83\times10^{-3}$ \\
 Lin--Reimherr
 & $4.15\times10^{-5}$
 & $8.06\times10^{-7}$
 & $3.48\times10^{-4}$ \\
 Hu--Lin
 & $5.53\times10^{-5}$
 & $1.27\times10^{-5}$
 & $3.83\times10^{-4}$ \\
 \midrule
 Unprojected
 & $\underline{2.67\times10^{-5}}$
 & $\underline{6.42\times10^{-7}}$
 & $\underline{2.06\times10^{-4}}$ \\
 Projected
 & $\boldsymbol{2.65\times10^{-5}}$
 & $\boldsymbol{5.98\times10^{-7}}$
 & $\boldsymbol{2.06\times10^{-4}}$ \\
 \bottomrule
 \end{tabular}
\end{table}

For each held-out test record, let $F_k$ and $Y_k$ denote the measured input and output Fourier coefficients at angular frequency $\omega_k$, and let $\widehat Y_k=\mathrm{i}\omega_k(\widehat AF)_k$ be the predicted output spectrum. We evaluate the normalized spectral prediction error
\begin{equation}
 E_{\mathcal K}(\widehat A)
 =\frac{\sum_{k\in\mathcal K}|Y_k-\widehat Y_k|^2}
       {\sum_{k\in\mathcal K}|Y_k|^2},
 \label{eq:real-ltv-error}
\end{equation}
across three frequency sets $\mathcal K$: the full $0.5$--$20\text{ kHz}$ evaluation band, the subset of directly excited input frequencies (\emph{Forced lines}), and the remaining unexcited frequencies (\emph{Sidebands}).

Table~\ref{tab:real_ltv} summarizes the mean normalized prediction errors across the 24 evaluations. Both Unprojected and Projected outperform all five baselines across every frequency set. For Projected, the reductions in mean prediction error relative to Target-only are $36.3\%$ on the full band, $41.5\%$ on the forced lines, and $40.7\%$ on the sidebands. Moreover, both proposed estimators improve upon Target-only in all 24 evaluations on the full band and sidebands, and in 15 of the 24 evaluations on the forced lines. The substantial reduction in sideband error confirms that calibrated operator transfer successfully extracts cross-frequency coupling dynamics from auxiliary regimes rather than merely smoothing the directly excited lines.

By contrast, naive Pooling increases prediction error by more than three orders of magnitude relative to Target-only, as ignoring scheduling-induced shifts severely distorts the filter's resonance profile. Among the FLR transfer baselines, LR matches Target-only on the full band and sidebands while achieving a moderate gain on the forced lines, whereas JM and HL incur higher errors across all three bands. In particular, JM selects no sources in any evaluation; its higher error relative to Target-only arises because hard principal-component truncation discards poorly conditioned spline directions that Target-only and LR retain through smooth regularization (see Appendix~\ref{app:real_ltv}). 

In this application, Unprojected and Projected perform very similarly.
With only $n_0=1$ training record and hyperparameter selection driven by a single validation record, finite-sample variance and source-weight selection dominate the second-order projection refinement. Overall, these results demonstrate that governing-operator calibration combined with validation-based source selection reliably transfers complex time-varying dynamics using a single target training record and a separate target validation record, where neither naive pooling nor direct parameter-shift transfer succeeds.

\section{Discussion}\label{sec:discussion}
In this work, we have developed an instrumental transfer framework that exploits sparse differences between governing equations to better estimate the target solution operator.
A central implication is that knowledge of how systems differ can support prediction without identifying their complete governing equations.
Several limitations point to promising directions for future research. 

In terms of theoretical investigation, our analysis treats the forcings $F_i^{(j)}$ and state responses $U_i^{(j)}$ as fully observed functions in Hilbert spaces. In many physical applications, however, states and inputs are recorded only on discrete spatial or temporal grids with errors and must be pre-smoothed or numerically differentiated before applying $D_l$. Extending our non-asymptotic guarantees to discretized observations requires quantifying how grid resolution and reconstruction errors propagate into the pilot, whitened instruments, and projected source moments. In addition, for finite time-series records, preprocessing steps such as time-domain windowing prior to Fourier transformation can alter the optimal linear predictor relative to the true physical operator, warranting a unified treatment of signal processing and operator transfer. Moreover, while our main rate theorems rely on exact support recovery and analyze a fixed number of recalibration rounds $K$, developing data-driven rules for selecting candidate projections $\Pi_j$ and stopping the recalibration loop remains an important open problem.

In terms of methodological extension, several directions could broaden the scope of operator transfer learning. In particular, our current formulation assumes that all source tasks are informative once calibrated and share a common function-space domain with the target. Extending projected instrumental transfer to automatically detect and safeguard against uninformative or misspecified sources, as well as accommodating changes in nonlinear governing equations, is a compelling avenue for future work.


\section*{Data availability}
All the code for data generation, model learning, performance evaluation and visualization is publicly available at \url{https://github.com/zylipku/OpTrans}.

\newpage
\appendix
\begin{center}
\Large\textbf{Appendices}
\end{center} 

Appendix~\ref{sec:proof} proves the main results. Appendix~\ref{sec:supporting_bounds} gives the auxiliary results including  concentration, pilot, calibration, and projection bounds. Appendix~\ref{sec:additional_examples} verifies the periodic elliptic and block models. Appendix~\ref{app:experiments} contains the experimental details.

\section{Proofs of the main results}\label{sec:proof}

Throughout Appendices~\ref{sec:proof} and~\ref{sec:supporting_bounds}, we use the following error quantities.

Define
\begin{equation}
\begin{aligned}
&R_{\Sigma,\Gamma}^{(0)} = \sup_{T\in H:\norm{T}_{\Gamma}\le1} \Biggl\lvert\frac{1}{n_0}\sum_{i=1}^{n_0}\norm{TF_i^{(0)}}_V^2-\norm{T}_{\Gamma^{(0)}}^2\Biggr\rvert,\enspace R_{M,\Gamma}^{(0)} = \sup_{T\in H:\norm{T}_{\Gamma}\le1} \Biggl\lvert\frac{1}{n_0}\sum_{i=1}^{n_0}\inprod{\varepsilon_i^{(0)}}{TF_i^{(0)}}_V\Biggr\rvert,\\
&R_{\Sigma,\eta}^{(j)} =\norm{ \{\Sigma^{(j)}+\eta I_W\}^{-1/2} \{\widehat\Sigma^{(j)}-\Sigma^{(j)}\} \{\Sigma^{(j)}+\eta I_W\}^{-1/2} }_{\mathrm{op}}.
\end{aligned}
\end{equation}
For these proofs, let $Q^{(j)}(\eta):W\to W$ be the unitary polar factor such that
\begin{equation}\label{eq:polar_alignment}
\begin{aligned}
&Q^{(j)}(\eta)
\{\widehat\Sigma^{(j)}+\eta I_W\}^{-1/2}
\{\Sigma^{(j)}+\eta I_W\}^{1/2}\\
&\qquad=
\bigl[
\{\Sigma^{(j)}+\eta I_W\}^{1/2}
\{\widehat\Sigma^{(j)}+\eta I_W\}^{-1}
\{\Sigma^{(j)}+\eta I_W\}^{1/2}
\bigr]^{1/2}.
\end{aligned}
\end{equation}
Set $Q_V^{(j)}(\eta)=I_V\otimes Q^{(j)}(\eta)$ on $V\otimes W$. Define $R_{M,\eta}^{(j)} =\norm{\widehat M_\eta^{(j)}(A^{(0)},\theta^{(j)})}_{V\otimes W}$,
\begin{equation}
\begin{aligned}
&R_{O,\Gamma,\eta}^{(j)} =\sup_{\norm{T}_\Gamma\le1} \norm{ \{Q_V^{(j)}(\eta)\widehat O^{(j)}(\eta) -O^{(j)}(\eta)\}T}_{V\otimes W},\\
&R_{\Phi,\Gamma,\eta}^{(j)} =\sup_{\norm{T}_\Gamma\le1} \max_{1\le l\le p} \norm{ \{Q_V^{(j)}(\eta)\widehat\Phi_l^{(j)}(\eta) -\Phi_l^{(j)}(\eta)\}T}_{V\otimes W}.
\end{aligned}
\end{equation}

Denote
\begin{equation}
\begin{aligned}
&R_{\Phi A,\eta}^{(j)} =\max_{1\le l\le p} \norm{Q_V^{(j)}(\eta) \widehat\Phi_l^{(j)}(\eta)\widetilde A -\Phi_l^{(j)}(\eta)A^{(0)}}_{V\otimes W},\\
&R_{\theta\Phi,\Gamma,\eta}^{(j)} =\sup_{\norm{T}_{\Gamma}\le1} \norm{\{Q_V^{(j)}(\eta) \widetilde\theta^{(j)}\cdot\widehat\Phi^{(j)}(\eta) -\theta^{(j)}\cdot\Phi^{(j)}(\eta)\}T}_{V\otimes W}.
\end{aligned}
\end{equation}

\subsection{Upper bounds}\label{app:upper_proofs}\label{app:structural_conditions}
\begin{proof}\textbf{of Lemma~\ref{lem:resolvent}.}
Since $L^{(j)}=L^{(0)}+\theta^{(j)}\cdot D$, we have ${L^{(0)}A^{(j)}=I_W-\theta^{(j)}\cdot DA^{(j)}}$. Left multiplication by $A^{(0)}$ proves \eqref{eq:resolvent}. Substitution of \eqref{eq:reg_model} into the left-hand side of \eqref{eq:transformed_regression}, followed by \eqref{eq:resolvent}, proves the second assertion.
\end{proof}

\begin{proof}\textbf{of Theorem~\ref{thm:upper_bound}.}
We first verify that \eqref{eq:upper_admissibility} implies the conditions needed for calibration and curvature control. By the uniform bounds in Section~\ref{sec:upper_bounds}, $s_j$ is uniformly bounded and $\norm{A^{(0)}}_{\mathcal G_0}\le (C_\Gamma+1)^{1/2}\norm{A^{(0)}}_H\le C$. Also, $r_\Sigma^{(j)}(\eta)\le d_j$. Thus, for a sufficiently small constant in \eqref{eq:upper_admissibility}, $C r_\Sigma^{(0)}(\lambda_0)+C\max_j r_\Sigma^{(j)}(\eta)\le1/2$, and, for each nonempty support,
\begin{equation}\label{eq:calibration_admissibility}
Cd_j\le
\min\Bigl[
1,\frac{\kappa_j}{2s_j^{1/2}},
c\frac{\alpha_j\kappa_j^2}
{s_j^{3/2}\{1+\phi_j\norm{A^{(0)}}_{\mathcal G_0}\}}
\Bigr],
\qquad
\lambda_j\le c\frac{\tau_j\kappa_j^2}{s_j^{1/2}}.
\end{equation}
Empty supports follow the convention in Appendix~\ref{app:empty_supports}. The same condition also gives
\begin{equation}\label{eq:curvature_smallness}
C\varepsilon_{\mathrm{curv}}\le\frac18.
\end{equation}
Write $\widehat\Delta=\widehat A-A^{(0)}$. Note that \eqref{eq:reg_model} and \eqref{eq:emp_moment} imply
\begin{equation}
\begin{aligned}
&\frac{1}{n_0}\sum_{i=1}^{n_0}
\bigl\{\norm{U_i^{(0)}-\widehat{A}F_i^{(0)}}_V^2
-\norm{U_i^{(0)}-A^{(0)}F_i^{(0)}}_V^2\bigr\}\\
&\quad=\frac{1}{n_0}\sum_{i=1}^{n_0}
\norm{\widehat{\Delta}F_i^{(0)}}_V^2
-\frac{2}{n_0}\sum_{i=1}^{n_0}
\inprod{\varepsilon_i^{(0)}}{\widehat{\Delta}F_i^{(0)}}_V,
\end{aligned}
\end{equation}
and \eqref{eq:emp_moment_expansion} implies
\begin{equation}
\widetilde\Pi_\perp^{(j)}(\eta)
\widehat M_\eta^{(j)}(\widehat{A},\widetilde\theta^{(j)})
-\widetilde\Pi_\perp^{(j)}(\eta)
\widehat M_\eta^{(j)}(A^{(0)},\widetilde\theta^{(j)})
=-\widetilde\Psi^{(j)}(\eta)\widehat\Delta,
\end{equation}
where
\begin{equation}
\widetilde\Psi^{(j)}(\eta)
=\widetilde\Pi_\perp^{(j)}(\eta)
\{\widehat O^{(j)}(\eta)
-\widetilde\theta^{(j)}\cdot\widehat\Phi^{(j)}(\eta)\}.
\end{equation}
By optimality of $\widehat{A}$ and feasibility of $A^{(0)}$,
\begin{equation}
\begin{aligned}
&-\frac{1}{n_0}\sum_{i=1}^{n_0} \norm{\widehat{\Delta}F_i^{(0)}}_V^2 +\frac{2}{n_0}\sum_{i=1}^{n_0} \inprod{\varepsilon_i^{(0)}}{\widehat{\Delta}F_i^{(0)}}_V \ge \sum_{j=1}^m\rho_j \biggl[ \norm{\widetilde\Pi_\perp^{(j)}(\eta) \widehat M_\eta^{(j)}(\widehat{A},\widetilde\theta^{(j)})}_{V\otimes W}^2 \\
&\quad -\norm{\widetilde\Pi_\perp^{(j)}(\eta) \widehat M_\eta^{(j)}(A^{(0)},\widetilde\theta^{(j)})}_{V\otimes W}^2 \biggr] +\lambda \bigl\{\norm{\widehat{A}}_H^2-\norm{A^{(0)}}_H^2\bigr\}\\
&\ge \frac{1}{2}\sum_{j=1}^m\rho_j \norm{\widetilde\Psi^{(j)}(\eta)\widehat\Delta}_{V\otimes W}^2 -2\sum_{j=1}^m\rho_j \norm{\widetilde\Pi_\perp^{(j)}(\eta) \widehat M_\eta^{(j)}(A^{(0)},\widetilde\theta^{(j)})}_{V\otimes W}^2 \\
&\quad +\lambda \Bigl\{\frac{1}{2}\norm{\widehat\Delta}_H^2 -2\norm{A^{(0)}}_H^2\Bigr\},
\end{aligned}
\end{equation}
where the last step uses $\norm{a+b}^2 \ge \frac{1}{2}\norm{a}^2 - \norm{b}^2$. Rearranging gives
\begin{equation}
\begin{aligned}
& \frac{1}{n_0}\sum_{i=1}^{n_0}\norm{\widehat{\Delta}F_i^{(0)}}_V^2
+ \frac{1}{2}\sum_{j=1}^m\rho_j
\norm{\widetilde\Psi^{(j)}(\eta)\widehat\Delta}_{V\otimes W}^2
+ \frac{\lambda}{2}\norm{\widehat\Delta}_H^2 \\
&\le \frac{2}{n_0}\sum_{i=1}^{n_0}
\inprod{\varepsilon_i^{(0)}}{\widehat{\Delta}F_i^{(0)}}_V
+2\sum_{j=1}^m\rho_j
\norm{\widetilde\Pi_\perp^{(j)}(\eta)
\widehat M_\eta^{(j)}(A^{(0)},\widetilde\theta^{(j)})}_{V\otimes W}^2
+2\lambda\norm{A^{(0)}}_H^2 .
\end{aligned}
\end{equation}
We have
\begin{equation}
\frac{1}{n_0}\sum_{i=1}^{n_0}\inprod{\varepsilon_i^{(0)}}{\widehat{\Delta}F_i^{(0)}}_V \le R_{M,\mathcal G}^{(0)} \norm{\widehat\Delta}_{\mathcal G},
\qquad
\frac{1}{n_0}\sum_{i=1}^{n_0}\norm{\widehat{\Delta}F_i^{(0)}}_V^2 \ge \norm{\widehat\Delta}_{\Gamma^{(0)}}^2 - R_{\Sigma,\mathcal G}^{(0)} \norm{\widehat\Delta}_{\mathcal G}^2,
\end{equation}
and, by Lemma~\ref{lem:projection},
\begin{equation}
\begin{aligned}
&\abs{ \norm{\widetilde\Psi^{(j)}(\eta)\widehat\Delta}_{V\otimes W} -\norm{\widehat\Delta}_{\Gamma^{(j)}(\eta)}}\\
&\le C\biggl[ \frac{s_j^{1/2}}{\kappa_j} R_{\Phi A,\eta}^{(j)} \biggl\{b_j\Bigl(\frac{n_0}{N}\Bigr)^{1/2} +\rho_j^{-1/2}\biggr\} +R_{O,\mathcal G,\eta}^{(j)} +R_{\theta\Phi,\mathcal G,\eta}^{(j)} \biggr] \norm{\widehat\Delta}_{\mathcal G} =: R_{\Psi,\mathcal G}^{(j)} \norm{\widehat\Delta}_{\mathcal G} .
\end{aligned}
\end{equation}
Combining the above equations,
\begin{equation}
\begin{aligned}
& 2R_{M,\mathcal G}^{(0)} \norm{\widehat\Delta}_{\mathcal G}
+2\sum_{j=1}^m\rho_j
\norm{\widetilde\Pi_\perp^{(j)}(\eta)
\widehat M_\eta^{(j)}(A^{(0)},\widetilde\theta^{(j)})}_{V\otimes W}^2
+2\lambda\norm{A^{(0)}}_H^2 \\
&\ge \norm{\widehat\Delta}_{\Gamma^{(0)}}^2
-R_{\Sigma,\mathcal G}^{(0)} \norm{\widehat\Delta}_{\mathcal G}^2
+\frac{1}{2}\sum_{j=1}^m\rho_j
\Bigl[
\frac{1}{2}\norm{\widehat\Delta}_{\Gamma^{(j)}(\eta)}^2
-\{R_{\Psi,\mathcal G}^{(j)}\}^2
\norm{\widehat\Delta}_{\mathcal G}^2
\Bigr]
+\frac{\lambda}{2}\norm{\widehat\Delta}_H^2 \\
&\ge \biggl[\frac{1}{4} - R_{\Sigma,\mathcal G}^{(0)} - \frac{1}{2}\sum_{j=1}^m\rho_j \{R_{\Psi,\mathcal G}^{(j)}\}^2\biggr] \norm{\widehat\Delta}_{\mathcal G}^2 .
\end{aligned}
\end{equation}
By Lemma~\ref{lem:emp_concentration}, Propositions~\ref{prop:pilot_rate} and \ref{prop:calibration_rate}, and $\rho_j^{-1/2}=(n_0/n_j)^{1/2}$, the expression subtracted from $1/4$ in the last display is at most $C\varepsilon_{\mathrm{curv}}$. Equation~\eqref{eq:curvature_smallness} therefore leaves a coefficient of at least $1/8$ on $\norm{\widehat\Delta}_{\mathcal G}^2$. Using $2R_{M,\mathcal G}^{(0)}\norm{\widehat\Delta}_{\mathcal G}\le\norm{\widehat\Delta}_{\mathcal G}^2/16+16\{R_{M,\mathcal G}^{(0)}\}^2$ gives
\begin{equation}
\norm{\widehat\Delta}_{\mathcal G}^2
\le \{16 R_{M,\mathcal G}^{(0)}\}^2
+64\sum_{j=1}^m\rho_j
\norm{\widetilde\Pi_\perp^{(j)}(\eta)
\widehat M_\eta^{(j)}(A^{(0)},\widetilde\theta^{(j)})}_{V\otimes W}^2
+64\lambda\norm{A^{(0)}}_H^2.
\end{equation}
Note that \eqref{eq:emp_moment_expansion} gives
\begin{equation}
\widehat M_\eta^{(j)}(A^{(0)},\widetilde\theta^{(j)})
-\widehat M_\eta^{(j)}(A^{(0)},\theta^{(j)})
=\{\widetilde\theta^{(j)}-\theta^{(j)}\}
\cdot\widehat\Phi^{(j)}(\eta)A^{(0)}.
\end{equation}
On the event $\{\supp(\widetilde\theta^{(j)})=S_j\}$,
\begin{equation}
\widetilde\Pi_\perp^{(j)}(\eta)
[\widehat M_\eta^{(j)}(A^{(0)},\widetilde\theta^{(j)})
-\widehat M_\eta^{(j)}(A^{(0)},\theta^{(j)})]
=\widetilde\Pi_\perp^{(j)}(\eta)
\{\widetilde\theta^{(j)}-\theta^{(j)}\}
\cdot\widehat\Phi^{(j)}(\eta)
\{A^{(0)}-\widetilde{A}\},
\end{equation}
and thus
\begin{equation}
\norm{\widetilde\Pi_\perp^{(j)}(\eta) \widehat M_\eta^{(j)}(A^{(0)},\widetilde\theta^{(j)})}_{V\otimes W} \le R_{M,\eta}^{(j)} +C\bigl\{\phi_j+R_{\Phi,\mathcal G_0,\eta}^{(j)}\bigr\} \norm{\widetilde\theta^{(j)}-\theta^{(j)}}_{\ell^1} \norm{\widetilde{A}-A^{(0)}}_{\mathcal G_0}.
\end{equation}

The bound $Cr_\Sigma^{(0)}(\lambda_0)\le1/2$, \eqref{eq:calibration_admissibility}, and \eqref{eq:curvature_smallness}, together with Lemma~\ref{lem:emp_concentration}, Proposition~\ref{prop:pilot_rate}, Proposition~\ref{prop:calibration_rate}, and Lemma~\ref{lem:projection}, imply all random inequalities used above on an event of probability at least $1-C(m+1)\exp(-t)$.  On this event, substituting \eqref{eq:target_cross_error}, \eqref{eq:source_true_moment_bound}, \eqref{eq:pilot_source_Phi_error}, \eqref{eq:pilot_rate_explicit}, and \eqref{eq:calibration_rate_general} into the preceding two displays gives
\begin{equation}
\begin{aligned}
&\norm{\widehat\Delta}_{\mathcal G}^2 \le C\biggl( \frac{n_0}{N}r_M^{(0)}(\bar\lambda)^2 +\lambda\norm{A^{(0)}}_H^2\\
&\quad +\sum_{j=1}^m\rho_j \biggl[ r_M^{(j)}(\eta) +\{1+r_{\op}^{(j)}(\eta,\lambda_0)\} \frac{s_j\lambda_j}{\kappa_j^2} \times \{r_M^{(0)}(\lambda_0) +\lambda_0^{1/2}\norm{A^{(0)}}_H\} \biggr]^2 \biggr).
\end{aligned}
\end{equation}
Since $\mathcal G\succeq c(N/n_0)\Gamma^{(0)}$, multiplying by $Cn_0/N$ and using $\lambda=(N/n_0)\bar\lambda$ and $1+r_{\op}^{(j)}(\eta,\lambda_0)\le1+d_j\le C$ proves \eqref{eq:upper_rate_explicit}.
\end{proof}

\begin{proof}\textbf{of Corollary~\ref{cor:transparent_rate}.}
The choice \eqref{eq:calibration_rate_simplified} satisfies \eqref{eq:lambda_j_choice} by the pilot bound. Since $0<\alpha_j\le1$ in the irrepresentability condition (taking $\alpha_j=1$ for empty supports), it also gives
\begin{equation}
q_j\le\frac{C s_j}{\alpha_j\kappa_j^2}
\biggl\{r_M^{(j)}(\bar\lambda)
+\Bigl(\frac{t}{n_0}\Bigr)^{\frac{\beta}{2\beta+1}}\biggr\}.
\end{equation}
Theorem~\ref{thm:upper_bound} and Proposition~\ref{prop:pilot_rate} therefore give
\begin{equation}
\begin{aligned}
\norm{\widehat A-A^{(0)}}_{\Gamma^{(0)}}^2
\le C\biggl[&
\frac{tn_0\mathcal N(\bar\lambda)}{N^2}
+\frac{t^2}{N^2\bar\lambda}
+\bar\lambda\norm{A^{(0)}}_H^2
+\sum_{j=1}^mw_jr_M^{(j)}(\bar\lambda)^2\\
&+\sum_{j=1}^mw_j
\frac{s_j^2}{\alpha_j^2\kappa_j^4}
\biggl\{r_M^{(j)}(\bar\lambda)
+\Bigl(\frac{t}{n_0}\Bigr)^{\frac{\beta}{2\beta+1}}
\biggr\}^2
\Bigl(\frac{t}{n_0}\Bigr)^{\frac{2\beta}{2\beta+1}}
\biggr].
\end{aligned}
\end{equation}
By Assumption~\ref{asm:spectrum}, $\mathcal N(a)\asymp a^{-1/(2\beta)}$.  Moreover,
\begin{equation}
\begin{aligned}
\sum_{j=1}^mw_jr_M^{(j)}(\bar\lambda)^2
&\le C\biggl\{
\frac{mt}{N}\bar\lambda^{-1/(2\beta)}
+\frac{t^2}{N\bar\lambda}
\sum_{j=1}^m\frac1{n_j}\biggr\}.
\end{aligned}
\end{equation}
By the uniform bounds in Section~\ref{sec:upper_bounds}, $s_j$, $\alpha_j^{-1}$, and $\kappa_j^{-1}$ are uniformly bounded. Since $\lambda_0\le1$, we have
\begin{equation}
\max_{1\le j\le m}\frac{s_j^2}{\alpha_j^2\kappa_j^4}
\Bigl(\frac{t}{n_0}\Bigr)^{\frac{2\beta}{2\beta+1}}\le C.
\end{equation}
Consequently, the part of the calibration term containing $r_M^{(j)}(\bar\lambda)^2$ is at most $C\sum_jw_jr_M^{(j)}(\bar\lambda)^2$. The remaining part is bounded by $Cc_\theta(t/n_0)^{4\beta/(2\beta+1)}$. Adding these bounds proves \eqref{eq:readable_upper_rate}.
\end{proof}

\begin{proof}\textbf{of Corollary~\ref{cor:transfer_improvement}.}
Under the stated choices,
\begin{equation}
\mathcal N(\lambda_0)\asymp
\Bigl(\frac{n_0}{t}\Bigr)^{\frac1{2\beta+1}},
\qquad
\mathcal N(\eta)\asymp
\Bigl(\frac{N}{t}\Bigr)^{\frac1{2\beta+1}}.
\end{equation}
The growth condition on $p$ gives $r_{\op}^{(j)}(\eta,\bar\lambda)=o(1)$.  Indeed,
\begin{equation}
r_{\op}^{(j)}(\eta,\bar\lambda) \le r_\Sigma^{(j)}(\eta)+C\biggl[ \Bigl\{ \frac{(t+\log p)(N/t)^{2/(2\beta+1)}}{N} \Bigr\}^{1/2} +\frac{t+\log p} {N(t/N)^{2\beta/(2\beta+1)}}\biggr]=o(1).
\end{equation}
Since $\lambda_0\ge\eta$ for all sufficiently large samples, $r_{\op}^{(j)}(\eta,\lambda_0)=o(1)$. Also, $r_\Sigma^{(0)}(\lambda_0)=o(1)$ and $\max_jr_\Sigma^{(j)}(\eta)=o(1)$. For the refit, $\mathcal N(\bar\lambda)=o(N)$ and $t=o(N)$ give $(n_0/N)r_\Sigma^{(0)}(\bar\lambda)=o(1)$, even if the unweighted target radius does not vanish. The score radii satisfy
\begin{equation}
r_M^{(j)}(\eta)
\le C\Bigl(\frac{t}{N}\Bigr)^{\frac{\beta}{2\beta+1}},
\qquad
r_M^{(0)}(\lambda_0)+\lambda_0^{1/2}\norm{A^{(0)}}_H
\le C\Bigl(\frac{t}{n_0}\Bigr)^{\frac{\beta}{2\beta+1}}.
\end{equation}
Together with the uniform bounds in Section~\ref{sec:upper_bounds}, these bounds give $\max_j(d_j+\lambda_j+q_j)=o(1)$. Since $m$ is fixed and $n_j\asymp N$, the definition of $\delta_j$ yields $\delta_j=o\{(n_0/N)^{1/2}\}$ and hence $\varepsilon_{\mathrm{curv}}=o(1)$. Thus \eqref{eq:upper_admissibility} holds for all sufficiently large samples. Since $N\ge n_0$, the displayed bounds give $r_M^{(j)}(\eta)+e_0\le C(t/n_0)^{\beta/(2\beta+1)}$. Together with $r_{\op}^{(j)}(\eta,\lambda_0)=o(1)$ and the uniform bounds in Section~\ref{sec:upper_bounds}, this shows that the chosen $\lambda_j$, with a sufficiently large constant, satisfies \eqref{eq:lambda_j_choice}. Moreover, $q_j\le (Cs_j/\kappa_j^2)(t/n_0)^{\beta/(2\beta+1)}$, so the argument in the proof of Corollary~\ref{cor:transparent_rate} gives \eqref{eq:readable_upper_rate} for these penalties. To obtain the rate directly, put $a_\beta=2\beta/(2\beta+1)$. Since $m$ is fixed and $n_j\asymp N$, the definitions give $v(t)\le Ct/N$, $u(t)\le C(t/N)^2$, and $c_\theta\le C$. For $\bar\lambda=(t/N)^{a_\beta}$, the first three terms in \eqref{eq:readable_upper_rate} satisfy
\begin{equation}
\bar\lambda\|A^{(0)}\|_H^2\le C(t/N)^{a_\beta},\quad v(t)\bar\lambda^{-1/(2\beta)}\le C(t/N)^{a_\beta},\quad u(t)\bar\lambda^{-1}\le C(t/N)^{2-a_\beta} \le C(t/N)^{a_\beta}.
\end{equation}
The last inequality uses $0<a_\beta<1$ and $t/N\le1$ for sufficiently large samples. The fourth term is bounded by $C(t/n_0)^{2a_\beta}$, which proves \eqref{eq:transfer_improvement}. Finally, $(t/n_0)^{2a_\beta}\le C(t/N)^{a_\beta}$ when $tN\le Cn_0^2$.
\end{proof}

\subsection{Refinement bounds}\label{app:iterative_proof}\label{app:coverage_proof}
\begin{proof}\textbf{of Theorem~\ref{thm:iterated_refit}.}
For $B\in H$, write, within this proof only, $e(B)=\Bigl(\frac{n_0}{N}\Bigr)^{1/2} \norm{B-A^{(0)}}_{\mathcal G}$. The equivalence in Lemma~\ref{lem:emp_concentration} implies
\begin{equation}\label{eq:iteration_geometry}
\norm{B-A^{(0)}}_{\Gamma^{(0)}}
+\bar\lambda^{1/2}\norm{B-A^{(0)}}_H
\le Ce(B).
\end{equation}
Consider the common event in Lemma~\ref{lem:emp_concentration}, Propositions~\ref{prop:pilot_rate} and \ref{prop:calibration_rate}, and Lemma~\ref{lem:projection}.  Its probability is at least $1-C\mathrm{e}^{-t}$ under the fixed-$m$ conditions.  No additional random event is introduced below.

Let $B$ satisfy $e(B)\le c$.  Rescaling the $\mathcal G$-unit ball in \eqref{eq:source_Phi_error}, using \eqref{eq:iteration_geometry} and Assumption~\ref{asm:nonp}, gives
\begin{equation}
\max_{1\le l\le p} \norm{Q_V^{(j)}(\eta) \widehat\Phi_l^{(j)}(\eta)B -\Phi_l^{(j)}(\eta)A^{(0)}}_{V\otimes W} \le C\{r_{\op}^{(j)} (\eta,\bar\lambda)+e(B)\}.\label{eq:iterated_generator}
\end{equation}
Indeed, the centered operator applied to $B-A^{(0)}$ is bounded by $R_{\Phi,\mathcal G,\eta}^{(j)} \norm{B-A^{(0)}}_{\mathcal G} \le Cr_{\op}^{(j)}(\eta,\bar\lambda)e(B)$, whereas the population part is at most $C\phi_je(B)$.  The same operator bound applied to $A^{(0)}$, together with $(n_0/N)^{1/2}\norm{A^{(0)}}_{\mathcal G}\le C$, contributes the first term on the right of \eqref{eq:iterated_generator}.

Since the operator radii vanish under Corollary~\ref{cor:transfer_improvement}, \eqref{eq:iterated_generator} and Assumption~\ref{asm:para} imply, whenever $e(B)\le c$, that the restricted empirical Gram matrix has minimum eigenvalue at least $\kappa_j^2/2$.  The empirical moment expansion gives
\begin{equation}
\widehat M_\eta^{(j)}(B,\theta^{(j)}) ={}\widehat M_\eta^{(j)}(A^{(0)},\theta^{(j)}) -\{\widehat O^{(j)}(\eta) -\theta^{(j)}\cdot\widehat\Phi^{(j)}(\eta)\} \{B-A^{(0)}\}.
\end{equation}
Equations \eqref{eq:source_true_moment_bound}, \eqref{eq:source_O_error}, and \eqref{eq:source_Phi_error}, followed by \eqref{eq:iteration_geometry}, therefore yield
\begin{equation}\label{eq:iterated_score}
\norm{\widehat M_\eta^{(j)}(B,\theta^{(j)})}_{V\otimes W}
\le C\{r_M^{(j)}(\eta)+e(B)\}.
\end{equation}

Suppose now that $\widehat S_j=S_j$.  The normal equations for \eqref{eq:iterated_calibration}, the empirical eigenvalue bound, and \eqref{eq:iterated_score} give
\begin{equation}\label{eq:iterated_theta}
\norm{\widehat\theta_B^{(j)}-\theta^{(j)}}_{\ell^1}
\le
C\frac{s_j}{\kappa_j^2}
\{r_M^{(j)}(\eta)+e(B)\},
\end{equation}
where $\widehat\theta_B^{(j)}$ denotes the restricted recalibration based on $B$.  To see this directly, write
\begin{equation}
\widehat G_j(B)
=\bigl\{
\inprod{\widehat\Phi_k^{(j)}(\eta)B}
{\widehat\Phi_l^{(j)}(\eta)B}_{V\otimes W}
\bigr\}_{1\le k,l\le p}.
\end{equation}
The restricted normal equations are
\begin{equation}
\widehat G_j(B)[S_j,S_j]
\{\widehat\theta_{B,S_j}^{(j)}-\theta_{S_j}^{(j)}\}
=-
\bigl[
\inprod{\widehat\Phi_l^{(j)}(\eta)B}
{\widehat M_\eta^{(j)}(B,\theta^{(j)})}_{V\otimes W}
\bigr]_{l\in S_j}.
\end{equation}
The generator norms are bounded by \eqref{eq:iterated_generator}; inversion of the Gram matrix first gives the corresponding $\ell^2$ bound, and multiplication by $s_j^{1/2}$ gives \eqref{eq:iterated_theta}.

Let $\widehat\Pi_B^{(j)}(\eta)$ denote the projection generated by $\widehat\Phi_l^{(j)}(\eta)B$, $l\in S_j$. Then $\widehat\Pi_{B,\perp}^{(j)}(\eta)[\{\widehat\theta_B^{(j)}-\theta^{(j)}\}\cdot\widehat\Phi^{(j)}(\eta)B]=0$, so
\begin{equation}\label{eq:iterated_residual}
\begin{aligned}
&\norm{\widehat\Pi_{B,\perp}^{(j)}(\eta) \widehat M_\eta^{(j)}(A^{(0)},\widehat\theta_B^{(j)}) }_{V\otimes W}\\
&\le \norm{\widehat M_\eta^{(j)}(A^{(0)},\theta^{(j)})}_{V\otimes W} + \norm{\widehat\theta_B^{(j)}-\theta^{(j)}}_{\ell^1} \max_l\norm{\widehat\Phi_l^{(j)}(\eta) \{A^{(0)}-B\}}_{V\otimes W}\\
&\le C\{r_M^{(j)}(\eta)+e(B)^2\}.
\end{aligned}
\end{equation}
The last line follows from \eqref{eq:iterated_theta}, \eqref{eq:iteration_geometry}, and the fixed bounds on $s_j$ and $\kappa_j^{-1}$; the product satisfies $r_M^{(j)}(\eta)e(B)\le c r_M^{(j)}(\eta)$ because $e(B)\le c$.

The proof of Lemma~\ref{lem:projection}, with $B$ in place of $\widetilde A$, shows that the projection error is bounded by $C\frac{s_j^{1/2}}{\kappa_j} \{r_{\op}^{(j)}(\eta,\bar\lambda)+e(B)\}$. Moreover, \eqref{eq:iterated_theta}, \eqref{eq:source_Phi_error}, and Assumption~\ref{asm:nonp} give
\begin{equation}
\begin{aligned}
&\sup_{\norm{T}_{\mathcal G}\le1}
\norm{\{Q_V^{(j)}(\eta)
\widehat\theta_B^{(j)}\cdot\widehat\Phi^{(j)}(\eta)
-\theta^{(j)}\cdot\Phi^{(j)}(\eta)\}T}_{V\otimes W}\\
&\quad\le
C\Bigl(\frac{n_0}{N}\Bigr)^{1/2}
\{r_{\op}^{(j)}(\eta,\bar\lambda)
+r_M^{(j)}(\eta)+e(B)\}.
\end{aligned}
\end{equation}
Consequently, the complete source-curvature perturbation on the $\mathcal G$-unit ball is at most $C\Bigl(\frac{n_0}{N}\Bigr)^{1/2} \{r_{\op}^{(j)}(\eta,\bar\lambda) +r_M^{(j)}(\eta)+e(B)\}$. Since $n_j\asymp N$ and $m$ is fixed, the sum of the squared perturbations weighted by $\rho_j$ is bounded by $C[\max_j\{r_{\op}^{(j)}(\eta,\bar\lambda)+r_M^{(j)}(\eta)\}^2+e(B)^2]$. Together with the target quadratic deviation, this is at most $1/8$ for all sufficiently large samples if $c$ in $e(B)\le c$ is chosen sufficiently small. The basic inequality in the proof of Theorem~\ref{thm:upper_bound}, with \eqref{eq:iterated_residual} as the source remainder, therefore gives
\begin{equation}
e(\widehat A^{[k+1]})^2 \le C\biggl( \Bigl(\frac{n_0}{N}\Bigr)^2 r_M^{(0)}(\bar\lambda)^2 +\bar\lambda\norm{A^{(0)}}_H^2 +\sum_{j=1}^m\frac{n_j}{N} [r_M^{(j)}(\eta)+e(\widehat A^{[k]})^2]^2 \biggr).
\end{equation}
The preceding $1/8$ bound leaves a coefficient of at least $1/8$ on $\|\widehat A^{[k+1]}-A^{(0)}\|_{\mathcal G}^2$ in that argument. Under the tuning choices of Corollary~\ref{cor:transfer_improvement}, the target score, penalty bias, and source scores in the preceding display are all bounded by $\Bigl(\frac{t}{N}\Bigr)^{\frac{2\beta}{2\beta+1}}$. Taking square roots therefore gives
\begin{equation}\label{eq:iteration_recursion}
e(\widehat A^{[k+1]})
\le C\biggl[
\Bigl(\frac{t}{N}\Bigr)^{\frac{\beta}{2\beta+1}}
+e(\widehat A^{[k]})^2
\biggr].
\end{equation}

For the initial iterate, Lemma~\ref{lem:emp_concentration}, Proposition~\ref{prop:pilot_rate}, and $\bar\lambda\le\lambda_0$ give $e(\widehat A^{[0]}) \le C\Bigl(\frac{t}{n_0}\Bigr)^{\frac{\beta}{2\beta+1}}$. The original Lasso calibration recovers $\widehat S_j=S_j$ and the proof of Corollary~\ref{cor:transfer_improvement} gives \eqref{eq:iteration_recursion} for $k=0$.  Induction in \eqref{eq:iteration_recursion} yields, for every fixed $K$,
\begin{equation}
e(\widehat A^{[K]})
\le C_K\biggl\{
\Bigl(\frac{t}{N}\Bigr)^{\frac{\beta}{2\beta+1}}
+
\Bigl(\frac{t}{n_0}\Bigr)^{
\frac{2^K\beta}{2\beta+1}}
\biggr\}.
\end{equation}
Both terms tend to zero.  Hence, for every fixed $K$, the same induction also keeps all preceding iterates inside the neighborhood $e(B)\le c$ for sufficiently large samples; this justifies each use of the local Gram and projection bounds. Since $\norm{T}_{\Gamma^{(0)}}\le C(n_0/N)^{1/2}\norm{T}_{\mathcal G}$, squaring the last display proves \eqref{eq:iterated_rate}.  Comparing its two terms proves the final assertion.
\end{proof}

\begin{proof}\textbf{of Proposition~\ref{prop:revision-coverage}.}
The hypothesis $\{e_j\cdot\widehat\Phi^{(j)}\}\widetilde A\in\operatorname{ran}(\Pi_j)$ gives $u_{\Pi,j}=0$. The identity $r_{\Pi,j}=z_{\Pi,j}+u_{\Pi,j}+p_{\Pi,j}$ in Section~\ref{sec:iterative_theory} therefore reduces to $r_{\Pi,j}=z_{\Pi,j}+p_{\Pi,j}$. Contractivity of $\Pi_{\perp,j}$ and the triangle inequality give
\begin{equation}
\|p_{\Pi,j}\|_{V\otimes W}\le\sum_{l=1}^p|e_{jl}|\|\widehat\Phi_l^{(j)}\{A^{(0)}-\widetilde A\}\|_{V\otimes W}\le\|e_j\|_{\ell^1}\max_{1\le l\le p}\|\widehat\Phi_l^{(j)}\{A^{(0)}-\widetilde A\}\|_{V\otimes W}.
\end{equation}
This proves \eqref{eq:revision-product-residual}.
\end{proof}

\subsection{Lower bounds}\label{app:minimax_proofs}\label{app:all_source_lower_proof}\label{app:common_shift_lower_proof}
The block models used below are verified in Appendix~\ref{app:anchored_model_verification}.

\begin{proof}\textbf{of Theorem~\ref{thm:minimax_source}.}
Let $K=\lfloor c_1N^{1/(2\beta+1)}\rfloor$ and $d^2=c_2/(N\nu_K)$, where $c_1,c_2>0$ are fixed constants to be chosen. For $\omega\in\{-1,1\}^K$, set $(a_\omega)_k=d\omega_{k-K}$ for $K<k\le2K$ and zero otherwise. Since
\begin{equation}
\begin{aligned}
\|a_\omega\|_{\ell^2}^2=Kd^2
=\frac{c_2K^{2\beta+1}}{c_\beta N}
\le\frac{c_2c_1^{2\beta+1}}{c_\beta},
\end{aligned}
\end{equation}
we have $\|a_\omega\|_{\ell^2}\le R$ whenever $c_2c_1^{2\beta+1}\le c_\beta R^2$.

Let $P_a$ denote the joint law of all $N$ observations under \eqref{eq:block_observation}, and write $F=(0,F_2)$. The forcing law is independent of $a$, and, conditional on $F$, only the first coordinate of $U$ depends on $a$. It is Gaussian with variance $\sigma^2$. Independence across observations therefore gives
\begin{equation}\label{eq:block_KL}
\begin{aligned}
\operatorname{KL}(P_a,P_{a'})
&=\frac{N}{2\sigma^2}\E(|\langle a-a',F_2\rangle|^2)
=\frac{N}{2\sigma^2}\sum_{k\ge1}\nu_k(a_k-a_k')^2.
\end{aligned}
\end{equation}
For sufficiently large $N$, we have $K\ge8$. The Varshamov--Gilbert bound \citep[Lemma~2.9]{tsybakov2009} supplies a set $\Omega\subset\{-1,1\}^K$ with $\log|\Omega|\ge cK$ whose distinct elements differ in at least $K/8$ coordinates. By \eqref{eq:final-block-prediction}, for $\omega\ne\omega'$ in $\Omega$,
\begin{equation}\label{eq:block-packing-separation}
\begin{aligned}
\|A_{a_\omega}-A_{a_{\omega'}}\|_{\Gamma^{(0)}}^2
&=\sum_{k=K+1}^{2K}\nu_k\{(a_\omega)_k-(a_{\omega'})_k\}^2\\
&\ge\frac{K}{2}d^2\nu_{2K}
\ge c c_2\frac KN,
\end{aligned}
\end{equation}
where we used $\nu_{2K}=2^{-2\beta}\nu_K$. On the other hand, \eqref{eq:block_KL} gives
\begin{equation}\label{eq:block-packing-KL}
\begin{aligned}
\max_{\omega,\omega'\in\Omega}
\operatorname{KL}(P_{a_\omega},P_{a_{\omega'}})
\le\frac{2NKd^2\nu_K}{\sigma^2}
=\frac{2c_2K}{\sigma^2}.
\end{aligned}
\end{equation}
Since $\log|\Omega|\ge cK$, choose $c_2>0$ so that $2c_2K/\sigma^2\le\frac18\log(|\Omega|-1)$ for all sufficiently large $N$, and then choose $c_1>0$ with $c_2c_1^{2\beta+1}\le c_\beta R^2$. Applying Fano's inequality \citep[Corollary~2.6]{tsybakov2009} with \eqref{eq:block-packing-KL} gives, for all sufficiently large $N$ and some fixed $\gamma_0\in(0,1/2)$,
\begin{equation}\label{eq:block-fano-bound}
\begin{aligned}
\inf_{\widehat\omega}\max_{\omega\in\Omega}
\pr_{P_{a_\omega}}(\widehat\omega\ne\omega)\ge2\gamma_0,
\end{aligned}
\end{equation}
where the infimum is over all decision rules taking values in $\Omega$.

For any operator estimator $\widehat A$, choose
$\widehat\omega\in\argmin_{\omega'\in\Omega}\|\widehat A-A_{a_{\omega'}}\|_{\Gamma^{(0)}}$, with ties resolved by a fixed ordering. If $\widehat\omega\ne\omega$, the triangle inequality gives
$\|\widehat A-A_{a_\omega}\|_{\Gamma^{(0)}}\ge\frac12\|A_{a_{\widehat\omega}}-A_{a_\omega}\|_{\Gamma^{(0)}}$.
Thus \eqref{eq:block-packing-separation} and \eqref{eq:block-fano-bound} give, for a sufficiently small fixed $c>0$,
\begin{equation}\label{eq:block-loss-probability}
\begin{aligned}
\max_{\omega\in\Omega}\pr_{P_{a_\omega}}
\bigl\{\ell_{P_{a_\omega}}(\widehat A)>cK/N\bigr\}
\ge2\gamma_0>\gamma_0.
\end{aligned}
\end{equation}
Since $\{P_{a_\omega}:\omega\in\Omega\}\subset\mathcal P_{\rm anc}(R)$, \eqref{eq:block-loss-probability} proves $R_{\gamma_0}^*(\mathcal P_{\rm anc}(R))\ge cK/N$. Multiplying the loss threshold $cK/N$ by the probability $2\gamma_0$ in \eqref{eq:block-loss-probability} also gives
$\sup_{\omega\in\Omega}\E_{P_{a_\omega}}\{\ell_{P_{a_\omega}}(\widehat A)\}\ge2\gamma_0cK/N$ for every $\widehat A$.
Taking the infimum over $\widehat A$ yields $R_\mathrm{e}^*(\mathcal P_{\rm anc}(R))\ge2\gamma_0cK/N$. Finally, $K/N\asymp N^{-2\beta/(2\beta+1)}$ proves Theorem~\ref{thm:minimax_source}.
\end{proof}

\begin{proof}\textbf{of Theorem~\ref{thm:minimax_shift}.}
For every $1\le j\le m$, $L_s^{(0)}+\theta_s^{(j)}\cdot D =L^{(0)}+\theta^{(j)}\cdot D=L^{(j)}$. Because $\supp(h)\subseteq S_j$, no inactive coefficient changes. For each fixed path, taking $|s|<\min_{j,l:h_l\ne0}|\theta_l^{(j)}|/(2|h_l|)$ also prevents any active coefficient from vanishing, so $\supp(\theta_s^{(j)})=S_j$. Hence $A^{(j)}$, and therefore the entire source distribution, is constant along the path.  Since $(h\cdot D)A^{(0)}$ is bounded, the Neumann expansion gives
\begin{equation}
A_s^{(0)} =A^{(0)}\{I_W+s(h\cdot D)A^{(0)}\}^{-1} =A^{(0)}-sT_h+O(s^2)
\end{equation}
in operator norm, and hence also in prediction norm because $\E\{\|F^{(0)}\|_W^2\}<\infty$.  Consequently, $\norm{A_s^{(0)}-A^{(0)}}_{\Gamma^{(0)}} \ge\frac{|s|}{2}\norm{T_h}_{\Gamma^{(0)}}$ for all sufficiently small $|s|$.

Take $s=c_0n_0^{-1/2}$. Since the source laws are fixed and observations are independent, the assumed bound $\operatorname{KL}(P_s^{(0)},P_0^{(0)})\le Cs^2$ gives a joint divergence of at most $Cn_0s^2=Cc_0^2$. Choose $c_0$ so that Pinsker's inequality bounds the total variation distance by $1/2$. Write $d=\|A_s^{(0)}-A^{(0)}\|_{\Gamma^{(0)}}$. The open balls of radius $d/2$ around the two parameters are disjoint, so the two-point testing inequality gives $\max_{u\in\{0,s\}}\pr_{P_u}\{\|\widehat A-A_u^{(0)}\|_{\Gamma^{(0)}}^2\ge d^2/4\}\ge1/4$ for every estimator. In particular, the probability of loss strictly greater than $d^2/8$ is at least $1/4$ for one of the two laws. Since $d^2\ge c/n_0$, this proves the fixed-confidence assertion for $\gamma_1<1/4$ and the expected-risk assertion.
\end{proof}

\section{Auxiliary results}\label{sec:supporting_bounds}
\subsection{Concentration}\label{app:basic_concentration}
\begin{lemma}\label{lem:concentration_mean}
Let $\xi_1,\dots,\xi_n$ be independent mean-zero random elements in a separable Hilbert space equipped with norm $\norm{\cdot}$. Suppose that there exist constants $v,b>0$ such that for every integer $q\ge2$, $\sum_{i=1}^n\E(\norm{\xi_i}^q) \le \frac{q!}{2}v^2b^{q-2}$. Then there is a universal constant $C>0$ such that for every $t\ge1$, with probability at least $1-\exp(-t)$, $\norm{\sum_{i=1}^n\xi_i} \le C\bigl(vt^{1/2}+bt\bigr)$.
\end{lemma}
\begin{proof}
For $0<u<b^{-1}$, the moment assumption gives $\sum_i\E(\mathrm{e}^{u\|\xi_i\|}-1-u\|\xi_i\|)\le v^2u^2/\{2(1-bu)\}$. The Hilbert-space case of \citet[Theorem~3.1]{pinelis1994optimum}, applied to the partial sums, therefore gives
\begin{equation}
\begin{aligned}
\pr\biggl(\biggl\|\sum_i\xi_i\biggr\|\ge r\biggr)&\le2\exp\biggl\{-ur+\frac{v^2u^2}{2(1-bu)}\biggr\}\\
&=2\exp\biggl\{-\frac{r^2}{2(v^2+br)}\biggr\},\qquad u=\frac{r}{v^2+br},\quad r>0.
\end{aligned}
\end{equation}
Taking $r=C(vt^{1/2}+bt)$ with a sufficiently large numerical $C$ proves the assertion for $t\ge1$.
\end{proof}

\begin{lemma}
\label{lem:concentration_cov}
Let $X_1,\dots,X_n$ be independent copies of a mean-zero random element $X$ in a separable Hilbert space. Suppose that $\E(\|X\|^2)<\infty$ and that its covariance $\Omega$ satisfies $\norm{\inprod{X}{x}}_{\psi_2}\le C\inprod{\Omega x}{x}^{1/2}$ for every $x$ in that space. For $a>0$, set $d_\Omega(a)=\tr\{\Omega(\Omega+aI)^{-1}\}$. Then, for every $t\ge1$, with probability at least $1-\mathrm{e}^{-t}$,
\begin{equation}
\Biggl\lVert(\Omega+aI)^{-1/2} \biggl(\frac1n\sum_{i=1}^nX_i\otimes X_i-\Omega\biggr) (\Omega+aI)^{-1/2}\Biggr\rVert_{\op} \le C\biggl[\Bigl\{\frac{d_\Omega(a)+t}{n}\Bigr\}^{1/2} +\frac{d_\Omega(a)+t}{n}\biggr].
\end{equation}
\end{lemma}

\begin{proof}
Set $Y_i=(\Omega+aI)^{-1/2}X_i$ and $C_a=(\Omega+aI)^{-1/2}\Omega(\Omega+aI)^{-1/2}$. Then $\E(Y_i\otimes Y_i)=C_a$ and
\begin{equation}\label{eq:covariance-subgaussian}
\begin{aligned}
\|\langle Y_i,x\rangle\|_{\psi_2}
&\le C\langle\Omega(\Omega+aI)^{-1/2}x,(\Omega+aI)^{-1/2}x\rangle^{1/2}
=C\langle C_ax,x\rangle^{1/2}.
\end{aligned}
\end{equation}
Since $\tr(C_a)\le a^{-1}\E(\|X\|^2)<\infty$, there is a centered Gaussian random element $G$ with covariance $C_a$. Thus $G$ and $Y_i$ have the same covariance, which is the pre-Gaussian requirement in \citet[Theorem~9]{koltchinskii2017concentration}. If $C_a=0$, then $Y_i=0$ almost surely and $n^{-1}\sum_iY_i\otimes Y_i-C_a=0$. For $C_a\ne0$, \citet[Theorem~9]{koltchinskii2017concentration}, applied with \eqref{eq:covariance-subgaussian} and the effective-rank bound $\{\E(\|G\|)\}^2/\|C_a\|_{\op}\le\tr(C_a)/\|C_a\|_{\op}$, gives, with probability at least $1-\mathrm{e}^{-t}$,
\begin{equation}\label{eq:covariance-normalized-bound}
\begin{aligned}
\biggl\|\frac1n\sum_{i=1}^nY_i\otimes Y_i-C_a\biggr\|_{\op}
&\le C\biggl[
\Bigl\{\frac{\|C_a\|_{\op}\tr(C_a)}{n}\Bigr\}^{1/2}
+\frac{\tr(C_a)}{n}\\
&\qquad{}+\|C_a\|_{\op}\Bigl(\frac tn\Bigr)^{1/2}
+\|C_a\|_{\op}\frac tn\biggr].
\end{aligned}
\end{equation}
Substituting $Y_i=(\Omega+aI)^{-1/2}X_i$, $\|C_a\|_{\op}\le1$, and $\tr(C_a)=d_\Omega(a)$ into \eqref{eq:covariance-normalized-bound} proves Lemma~\ref{lem:concentration_cov}. The constant depends only on the constant in the assumed inequality $\|\langle X,x\rangle\|_{\psi_2}\le C\langle\Omega x,x\rangle^{1/2}$.
\end{proof}

\begin{lemma}\label{lem:target_quadratic}
Suppose that $F_1,\dots,F_n$ are independent copies of $F^{(0)}$ and \eqref{eq:subgaussian_forcing} holds.  Then, for every $0<\lambda\le1$ and $t\ge1$, with probability at least $1-\exp(-t)$,
\begin{equation}
\begin{aligned}
\sup_{\norm{T}_{\Gamma^{(0)}+\lambda I_H}\le1}
\Biggl\lvert\frac1n\sum_{i=1}^n\norm{TF_i}_V^2
-\E\{\norm{TF^{(0)}}_V^2\}\Biggr\rvert
\le C\biggl[
\Bigl\{\frac{\mathcal N(\lambda)+t}{n}\Bigr\}^{1/2}
+\frac{\mathcal N(\lambda)+t}{n}
\biggr].
\end{aligned}
\end{equation}
\end{lemma}

\begin{proof}
Let $\mathcal B_\lambda=\{T\in H:\|T\|_{\Gamma^{(0)}+\lambda I_H}\le1\}$ and $f_T(F)=\|TF\|_V$. Equip $\mathcal B_\lambda$ with the semimetric $d(T,S)=\|f_T(F^{(0)})-f_S(F^{(0)})\|_{\psi_2}$. The reverse triangle inequality and \eqref{eq:subgaussian_forcing} give
\begin{equation}\label{eq:quadratic-subgaussian-bounds}
\begin{aligned}
d(T,S)\le C\|T-S\|_{\Gamma^{(0)}},\qquad
\sup_{T\in\mathcal B_\lambda}\|f_T(F^{(0)})\|_{\psi_2}\le C.
\end{aligned}
\end{equation}
Write $\gamma_2(\mathcal B_\lambda,d)$ for the generic chaining functional defined in \citet[Section~2]{dirksen2015tail}. In the eigenbasis of $\Gamma^{(0)}$, the image of $\mathcal B_\lambda$ under $T\mapsto\{\Gamma^{(0)}\}^{1/2}T$ is an ellipsoid with semiaxes $\{\nu_k/(\nu_k+\lambda)\}^{1/2}$. Talagrand's majorizing measure theorem, in the form of \citet[Theorem~2]{koltchinskii2017concentration}, and $d(T,S)\le C\|T-S\|_{\Gamma^{(0)}}$ in \eqref{eq:quadratic-subgaussian-bounds} give
\begin{equation}\label{eq:final-quadratic-width}
\begin{aligned}
\gamma_2(\mathcal B_\lambda,d)
&\le C\E\Biggl\{\biggl(\sum_{k\ge1}\frac{\nu_k}{\nu_k+\lambda}g_k^2\biggr)^{1/2}\Biggr\}
\le C\mathcal N(\lambda)^{1/2},
\end{aligned}
\end{equation}
where the $g_k$ are independent standard Gaussian variables; the second inequality in \eqref{eq:final-quadratic-width} follows from Jensen's inequality and $\sum_k\nu_k/(\nu_k+\lambda)=\mathcal N(\lambda)$.

The bound $\sup_{T\in\mathcal B_\lambda}\|f_T(F^{(0)})\|_{\psi_2}\le C$ in \eqref{eq:quadratic-subgaussian-bounds} implies
\begin{equation}\label{eq:quadratic-square-moments}
\begin{aligned}
\sup_{T\in\mathcal B_\lambda}\E\biggl(\bigl|f_T(F^{(0)})^2-\E[f_T(F^{(0)})^2]\bigr|^q\biggr)
\le q!C^q,\qquad q=2,3,\ldots.
\end{aligned}
\end{equation}
Applying \citet[Theorem~5.5]{dirksen2015tail} with \eqref{eq:quadratic-subgaussian-bounds}, \eqref{eq:final-quadratic-width}, and \eqref{eq:quadratic-square-moments} gives, with probability at least $1-\mathrm{e}^{-t}$,
\begin{equation}\label{eq:quadratic-process-bound}
\begin{aligned}
\sup_{T\in\mathcal B_\lambda}
\biggl|\frac1n\sum_{i=1}^nf_T(F_i)^2-\E[f_T(F^{(0)})^2]\biggr|
&\le C\biggl[\frac{\mathcal N(\lambda)}n
+\Bigl\{\frac{\mathcal N(\lambda)}n\Bigr\}^{1/2}
+\Bigl(\frac tn\Bigr)^{1/2}+\frac tn\biggr].
\end{aligned}
\end{equation}
To justify \eqref{eq:quadratic-process-bound} on $\mathcal B_\lambda$, first apply \citet[Theorem~5.5]{dirksen2015tail} to nested finite subsets whose union is dense in $\mathcal B_\lambda$ in the $H$ norm. The right-hand side of \eqref{eq:quadratic-process-bound} is independent of the finite subset, so continuity from above of probability gives \eqref{eq:quadratic-process-bound} on their union. Continuity of $T\mapsto TF_i$ and $T\mapsto\|T\|_{\Gamma^{(0)}}^2$ extends \eqref{eq:quadratic-process-bound} to $\mathcal B_\lambda$. Finally, $\{\mathcal N(\lambda)/n\}^{1/2}+(t/n)^{1/2}\le[2\{\mathcal N(\lambda)+t\}/n]^{1/2}$ proves Lemma~\ref{lem:target_quadratic}.
\end{proof}

\subsection{Joint concentration}\label{app:joint_concentration}
\begin{lemma}\label{lem:emp_concentration}
Under Assumptions~\ref{asm:nonp}, \ref{asm:spectrum}, and \ref{asm:bernstein}, there is a constant $C>0$ such that, if $b_j$, $\phi_j$, and $\norm{\theta^{(j)}}_{\ell^1}$ are uniformly bounded, then, for every $t\ge1$, $0<\lambda_0\le1$, and $0<\eta\le\bar\lambda\le1$ satisfying $C\max_{1\le j\le m}r_\Sigma^{(j)}(\eta)\le\frac12$, with probability at least $1-C(m+1)\exp(-t)$,
\begin{equation}\label{eq:target_quad_error}
\begin{aligned}
R_{\Sigma,\mathcal G}^{(0)}
&\le C\frac{n_0}{N}r_\Sigma^{(0)}(\bar\lambda),
\end{aligned}
\end{equation}
\begin{equation}\label{eq:target_cross_error}
\begin{aligned}
R_{M,\mathcal G}^{(0)}
&\le C\Bigl(\frac{n_0}{N}\Bigr)^{1/2}
r_M^{(0)}(\bar\lambda),
\end{aligned}
\end{equation}
\begin{equation}\label{eq:source_cov_error}
\begin{aligned}
R_{\Sigma,\eta}^{(j)}
&\le C r_\Sigma^{(j)}(\eta),
\end{aligned}
\end{equation}
\begin{equation}\label{eq:source_true_moment_bound}
\begin{aligned}
R_{M,\eta}^{(j)}
&\le C r_M^{(j)}(\eta),
\qquad 1\le j\le m,
\end{aligned}
\end{equation}
\begin{equation}\label{eq:source_O_error}
\begin{aligned}
R_{O,\mathcal G,\eta}^{(j)}
&\le C\biggl\{
\Bigl(\frac{n_0}{N}\Bigr)^{1/2}
r_{\op}^{(j)}(\eta,\bar\lambda)
+\rho_j^{-1/2}r_\Sigma^{(j)}(\eta)
\biggr\},
\end{aligned}
\end{equation}
\begin{equation}\label{eq:source_Phi_error}
\begin{aligned}
R_{\Phi,\mathcal G,\eta}^{(j)}
&\le C\Bigl(\frac{n_0}{N}\Bigr)^{1/2}
r_{\op}^{(j)}(\eta,\bar\lambda).
\end{aligned}
\end{equation}
On the same event,
\begin{equation}\label{eq:pilot_quad_error}
\begin{aligned}
R_{\Sigma,\mathcal G_0}^{(0)}
&\le C r_\Sigma^{(0)}(\lambda_0),
\end{aligned}
\end{equation}
\begin{equation}\label{eq:pilot_cross_error}
\begin{aligned}
R_{M,\mathcal G_0}^{(0)}
&\le C r_M^{(0)}(\lambda_0),
\end{aligned}
\end{equation}
\begin{equation}\label{eq:pilot_O_error}
\begin{aligned}
R_{O,\mathcal G_0,\eta}^{(j)}
&\le C r_{\op}^{(j)}(\eta,\lambda_0),
\end{aligned}
\end{equation}
\begin{equation}\label{eq:pilot_source_Phi_error}
\begin{aligned}
R_{\Phi,\mathcal G_0,\eta}^{(j)}
&\le C r_{\op}^{(j)}(\eta,\lambda_0).
\end{aligned}
\end{equation}
\noeqref{eq:pilot_cross_error,eq:pilot_O_error}
The bounds \eqref{eq:source_cov_error}, \eqref{eq:source_true_moment_bound}, and \eqref{eq:pilot_quad_error}--\eqref{eq:pilot_source_Phi_error} remain valid for every $0<\lambda_0,\eta\le1$; they do not require $\eta\le\bar\lambda$. Moreover, $\mathcal N(\lambda)\le C\lambda^{-1/(2\beta)}$ for $0<\lambda\le1$.
\end{lemma}

\begin{proof}\textbf{of Lemma~\ref{lem:emp_concentration}.}
For $\nu,\eta,\bar\lambda>0$ with $\eta\le\bar\lambda$, $\nu+\rho\frac{\nu^2}{\nu+\eta} +(1+\rho)\bar\lambda \asymp (1+\rho)(\nu+\bar\lambda)$. Indeed, use $\nu^2/(\nu+\eta)\ge\nu/2$ when $\nu\ge\eta$; when $\nu<\eta$, the regularization term dominates $(1+\rho)\nu$.  Assumption~\ref{asm:spectrum} therefore gives $\mathcal G \asymp \frac{N}{n_0} \{\Gamma^{(0)}+\bar\lambda I_H\}$. Consequently, a $\mathcal G$-unit vector has $\Gamma^{(0)}+\bar\lambda I_H$ norm at most $C(n_0/N)^{1/2}$.  Lemma~\ref{lem:target_quadratic}, applied to the resulting rescaled ball, proves \eqref{eq:target_quad_error}; the quadratic deviation acquires the factor $n_0/N$.  Lemma~\ref{lem:concentration_mean} and the target-score moment condition similarly prove \eqref{eq:target_cross_error}, with factor $(n_0/N)^{1/2}$.

Lemma~\ref{lem:concentration_cov}, applied on $W$ with $\Omega=\Sigma^{(j)}$ and \eqref{eq:source_subgaussian_forcing}, gives \eqref{eq:source_cov_error}.  Its effective dimension is bounded because $\tr\bigl[\Sigma^{(j)} \{\Sigma^{(j)}+\eta I_W\}^{-1}\bigr] \le C\mathcal N(\eta)$. On the event $R_{\Sigma,\eta}^{(j)}\le1/2$, functional calculus gives
\begin{equation}
\norm{ [ \{\Sigma^{(j)}+\eta I_W\}^{1/2} \{\widehat\Sigma^{(j)}+\eta I_W\}^{-1} \{\Sigma^{(j)}+\eta I_W\}^{1/2} ]^{1/2} -I_W}_{\mathrm{op}} \le C R_{\Sigma,\eta}^{(j)}.\label{eq:whitening_alignment_error}
\end{equation}
The operator $Q_V^{(j)}(\eta)$ is unitary on $V\otimes W$. It preserves norms and Gram matrices and carries each empirical projection to the projection onto the rotated span.  After this alignment, \eqref{eq:whitening_alignment_error} shows that replacing the population whitening operator by its empirical version contributes $C r_\Sigma^{(j)}(\eta)$.  This is a pathwise argument, so the same observations may be used in $\widehat\Sigma^{(j)}$ and in the empirical moments; no independence or sample splitting is needed.

Within this proof, denote the square-root operator in \eqref{eq:whitening_alignment_error} before $-I_W$ by $B^{(j)}(\eta)$, and set $B_V^{(j)}(\eta)=I_V\otimes B^{(j)}(\eta)$ on $V\otimes W$. The resolvent identity and \eqref{eq:polar_alignment} give
\begin{equation}
Q_V^{(j)}(\eta)
\widehat M_\eta^{(j)}(A^{(0)},\theta^{(j)})
=B_V^{(j)}(\eta)
\frac1{n_j}\sum_{i=1}^{n_j}
\{I_V+A^{(0)}\theta^{(j)}\cdot D\}\varepsilon_i^{(j)}
\otimes Z_i^{(j)}(\eta),
\end{equation}
where $Z_i^{(j)}(\eta) =\{\Sigma^{(j)}+\eta I_W\}^{-1/2}F_i^{(j)}$. The sum on the right is an average of independent mean-zero elements. Since $\norm{B^{(j)}(\eta)}_{\mathrm{op}}\le C$ on the same event, Lemma~\ref{lem:concentration_mean} and the source-score moment condition give \eqref{eq:source_true_moment_bound}.  For the source operators, \eqref{eq:polar_alignment} gives, for every $T\in H$,
\begin{equation}
\begin{aligned}
&Q_V^{(j)}(\eta)\widehat O^{(j)}(\eta)T
-O^{(j)}(\eta)T\\
&\quad=
B_V^{(j)}(\eta)
\biggl\{\frac1{n_j}\sum_{i=1}^{n_j}
TF_i^{(j)}\otimes Z_i^{(j)}(\eta)
-O^{(j)}(\eta)T\biggr\}
+\{B_V^{(j)}(\eta)-I_{V\otimes W}\}O^{(j)}(\eta)T,
\end{aligned}
\end{equation}
and the identical decomposition holds with $\widehat O^{(j)}(\eta)$ and $O^{(j)}(\eta)$ replaced by $\widehat\Phi_l^{(j)}(\eta)$ and $\Phi_l^{(j)}(\eta)$. Lemma~\ref{lem:concentration_mean}, applied to the centered operators in Assumption~\ref{asm:bernstein}, controls the terms in braces; a union bound over $1\le l\le p$ gives the $t+\log p$ factor.  Rescaling the $\mathcal G$-unit ball gives the factor $(n_0/N)^{1/2}$. The remaining terms are bounded by \eqref{eq:whitening_alignment_error}. For $\widehat O^{(j)}(\eta)$, the whitening error also multiplies the population moment map.  Assumption~\ref{asm:nonp} gives
\begin{equation}
\sup_{\norm{T}_{\mathcal G}\le1}
\norm{O^{(j)}(\eta)T}_{V\otimes W}
\le C\biggl\{
\Bigl(\frac{n_0}{N}\Bigr)^{1/2}+\rho_j^{-1/2}
\biggr\},
\end{equation}
which yields the additional $\rho_j^{-1/2}r_\Sigma^{(j)}(\eta)$ term in \eqref{eq:source_O_error}. For $\widehat\Phi_l^{(j)}(\eta)$, Assumption~\ref{asm:nonp} gives a population bound of order $(n_0/N)^{1/2}$, and its whitening error is already included in the first term $r_\Sigma^{(j)}(\eta)$ of $r_{\op}^{(j)}(\eta,\bar\lambda)$.  This proves \eqref{eq:source_O_error} and \eqref{eq:source_Phi_error}. Repeating the argument with $\mathcal G_0=\Gamma^{(0)}+\lambda_0 I_H$ gives \eqref{eq:pilot_quad_error}--\eqref{eq:pilot_source_Phi_error}. No comparison between $\eta$ and $\lambda_0$ is needed here: Assumption~\ref{asm:spectrum} implies $\Gamma^{(j)}(\eta)\preceq C\Gamma^{(0)} \preceq C\mathcal G_0$, while Assumption~\ref{asm:nonp} controls the projected component and $\Phi_l^{(j)}(\eta)$ by the same target curvature.  Thus the population terms created by empirical whitening are bounded on the $\mathcal G_0$-unit ball and are covered by the $r_\Sigma^{(j)}(\eta)$ part of $r_{\op}^{(j)}(\eta,\lambda_0)$. Finally, $\mathcal N(\lambda) =\sum_{k\ge1}\frac{\nu_k}{\nu_k+\lambda} \le C\lambda^{-1/(2\beta)}$ by Assumption~\ref{asm:spectrum} and comparison with an integral.
\end{proof}

\subsection{Pilot}\label{app:pilot_proof}
\begin{proposition}\label{prop:pilot_rate}
Suppose that $A^{(0)}\in H$.  On the event $R_{\Sigma,\mathcal G_0}^{(0)}\le1/2$,
\begin{equation}\label{eq:pilot_rate_general}
\norm{\widetilde A-A^{(0)}}_{\mathcal G_0}
\le
4R_{M,\mathcal G_0}^{(0)}
+4\lambda_0^{1/2}\norm{A^{(0)}}_H.
\end{equation}
Consequently, under Assumptions~\ref{asm:spectrum} and \ref{asm:bernstein}, for $0<\lambda_0\le1$ and $t\ge1$ with $C r_\Sigma^{(0)}(\lambda_0)\le1/2$, with probability at least $1-C\exp(-t)$,
\begin{equation}
\norm{\widetilde A-A^{(0)}}_{\mathcal G_0}
\le C\bigl\{
r_M^{(0)}(\lambda_0)
+\lambda_0^{1/2}\norm{A^{(0)}}_H\bigr\}.
\label{eq:pilot_rate_explicit}
\end{equation}
If $\norm{A^{(0)}}_H$ is bounded and $\lambda_0=(t/n_0)^{2\beta/(2\beta+1)}$, then $\norm{\widetilde A-A^{(0)}}_{\mathcal G_0}\le C(t/n_0)^{\beta/(2\beta+1)}$ on the same event.
\end{proposition}

\begin{proof}
The optimality of $\widetilde A$, with $A^{(0)}$ used as a comparison point, gives
\begin{equation}
\begin{aligned}
&\frac1{n_0}\sum_{i=1}^{n_0}
\norm{\{\widetilde A-A^{(0)}\}F_i^{(0)}}_V^2
+\lambda_0\norm{\widetilde A-A^{(0)}}_H^2\\
&\quad\le
\frac2{n_0}\sum_{i=1}^{n_0}
\inprod{\varepsilon_i^{(0)}}
{\{\widetilde A-A^{(0)}\}F_i^{(0)}}_V
-2\lambda_0
\inprod{A^{(0)}}{\widetilde A-A^{(0)}}_H.
\end{aligned}
\end{equation}
The left-hand side is at least $\{1-R_{\Sigma,\mathcal G_0}^{(0)}\}\norm{\widetilde A-A^{(0)}}_{\mathcal G_0}^2$. The absolute value of the right-hand side is at most $2\{R_{M,\mathcal G_0}^{(0)}+\lambda_0^{1/2}\norm{A^{(0)}}_H\}\norm{\widetilde A-A^{(0)}}_{\mathcal G_0}$. This proves \eqref{eq:pilot_rate_general}. Lemma~\ref{lem:target_quadratic} bounds $R_{\Sigma,\mathcal G_0}^{(0)}$, and Lemma~\ref{lem:concentration_mean} applied to the target score bounds $R_{M,\mathcal G_0}^{(0)}$. A union bound proves \eqref{eq:pilot_rate_explicit}; substituting $\mathcal N(\lambda_0)\le C\lambda_0^{-1/(2\beta)}$ gives the final rate.
\end{proof}

\subsection{Calibration}\label{app:calibration_proofs}\label{app:empty_supports}
 
When $S_j=\varnothing$, we have $\theta^{(j)}=0$ and $s_j=0$; set $\Pi^{(j)}(\eta)=0$ and $\alpha_j=1$. The conditions in Assumption~\ref{asm:para} are then vacuous. In Lemma~\ref{lem:deterministic_calibration} and Proposition~\ref{prop:calibration_rate}, omit the restrictions involving $s_j^{-3/2}$ and $s_j^{-1/2}$, while retaining all other conditions. In particular, retain $R_{\Phi A,\eta}^{(j)}\le1$ and \eqref{eq:calibration_score_condition}. Since $\theta^{(j)}=0$, the latter gives
\begin{equation}
\begin{aligned}
\max_{1\le l\le p}
	\bigl|2\langle\widehat\Phi_l^{(j)}(\eta)\widetilde A,
	\widehat M_\eta^{(j)}(\widetilde A,0)\rangle_{V\otimes W}\bigr|
	\le \lambda_j/8<\lambda_j.
\end{aligned}
\end{equation}
Thus every partial derivative of the squared moment loss at $\theta=0$ has magnitude strictly below $\lambda_j$. Together with convexity, this implies that the unique minimizer in \eqref{eq:calibration} is $\widetilde\theta^{(j)}=0$. The calibration error is therefore zero, and terms proportional to $s_j$ or $s_j^{1/2}$ in the error bounds are interpreted as zero.

\begin{lemma}\label{lem:deterministic_calibration}
Under Assumptions~\ref{asm:nonp} and \ref{asm:para}, suppose that
\begin{equation}\label{eq:generator_smallness}
R_{\Phi A,\eta}^{(j)}\le1,
\qquad
R_{\Phi A,\eta}^{(j)}
\le
c\frac{\alpha_j\kappa_j^2}
{s_j^{3/2}\{1+\phi_j\norm{A^{(0)}}_{\mathcal G_0}\}}.
\end{equation}
If
\begin{equation}\label{eq:calibration_score_condition}
\lambda_j\ge
\frac8{\alpha_j}
\max_{1\le l\le p}
\abs{2\inprod{\widehat\Phi_l^{(j)}(\eta)\widetilde A}
{\widehat M_\eta^{(j)}(\widetilde A,\theta^{(j)})}_{V\otimes W}}
\end{equation}
and $\lambda_j\le c \tau_j\kappa_j^2/s_j^{1/2}$, then
\begin{equation}
\supp(\widetilde\theta^{(j)})=S_j,
\qquad
\norm{\widetilde\theta^{(j)}-\theta^{(j)}}_{\ell^1}
\le C\frac{s_j\lambda_j}{\kappa_j^2}.
\end{equation}
\end{lemma}

\begin{proof}
Write $\widehat g_l^{(j)}(\eta) =Q_V^{(j)}(\eta) \widehat\Phi_l^{(j)}(\eta)\widetilde A$ within this proof.  Since $\norm{g_l^{(j)}(\eta)}_{V\otimes W} \le\phi_j\norm{A^{(0)}}_{\Gamma^{(0)}}$, every entry of the empirical Gram matrix differs from the corresponding entry of $G_j(\eta)$ by at most $C R_{\Phi A,\eta}^{(j)} \{1+\phi_j\norm{A^{(0)}}_{\mathcal G_0}\}$. Condition \eqref{eq:generator_smallness}, Weyl's inequality, and the Neumann-series identity give
\begin{equation}
\begin{aligned}
\lambda_{\min}
\bigl(\{\inprod{\widehat g_k^{(j)}(\eta)}{\widehat g_l^{(j)}(\eta)}_{V\otimes W}
\}_{k,l\in S_j}\bigr)
&\ge\kappa_j^2/2,\\
\norm{
\{\inprod{\widehat g_k^{(j)}(\eta)}{\widehat g_l^{(j)}(\eta)}_{V\otimes W}
\}_{k\in S_j^\complement,l\in S_j}
\{\inprod{\widehat g_k^{(j)}(\eta)}{\widehat g_l^{(j)}(\eta)}_{V\otimes W}
\}_{k,l\in S_j}^{-1}
}_\infty
&\le1-\alpha_j/2.
\end{aligned}
\end{equation}
Indeed, the two perturbation terms are bounded by $Cs_j^{3/2}R_{\Phi A,\eta}^{(j)} \{1+\phi_j\norm{A^{(0)}}_{\mathcal G_0}\}/\kappa_j^2$.

Let $w_l=2\inprod{\widehat g_l^{(j)}(\eta)}{Q_V^{(j)}(\eta)\widehat M_\eta^{(j)}(\widetilde A,\theta^{(j)})}_{V\otimes W}$.
The optimality equations on $S_j$ give
\begin{equation}
\widetilde\theta_{S_j}^{(j)}-\theta_{S_j}^{(j)}
=-\frac12\widehat G_j[S_j,S_j]^{-1}
(w_{S_j}+\lambda_j z_{S_j}),
\qquad \norm{z_{S_j}}_\infty\le1,
\end{equation}
where $\widehat G_j$ is the Gram matrix of $\widehat g_1^{(j)}(\eta),\dots,\widehat g_p^{(j)}(\eta)$.  Since $\norm{w}_{\ell^\infty}\le\alpha_j\lambda_j/8$, $\norm{\widetilde\theta_{S_j}^{(j)}-\theta_{S_j}^{(j)}}_{\ell^2} \le C\frac{s_j^{1/2}\lambda_j}{\kappa_j^2}$. For the inactive coordinates, substitution of the preceding display gives
\begin{equation}
\begin{aligned}
&w_{S_j^\complement}
-\widehat G_j[S_j^\complement,S_j]\widehat G_j[S_j,S_j]^{-1}
(w_{S_j}+\lambda_jz_{S_j}).
\end{aligned}
\end{equation}
Its $\ell^\infty$ norm is at most $(1-\alpha_j/2)\lambda_j+2\norm{w}_{\ell^\infty}\le(1-\alpha_j/4)\lambda_j<\lambda_j$. Extending the restricted solution by zeros therefore satisfies the full optimality conditions, with strict inequalities on $S_j^\complement$. The active Gram matrix is positive definite, so this minimizer is unique. Finally, $\lambda_j\le c\tau_j\kappa_j^2/s_j^{1/2}$ implies $\norm{\widetilde\theta_{S_j}^{(j)}-\theta_{S_j}^{(j)}}_{\ell^\infty}<\tau_j$. Hence its support is $S_j$, and Cauchy--Schwarz gives the $\ell^1$ bound.
\end{proof}

\begin{proposition}\label{prop:calibration_rate}\sloppy
Under Assumptions~\ref{asm:nonp}--\ref{asm:bernstein}, suppose that $b_j$, $\phi_j$, and $C_\Gamma$ are uniformly bounded. Suppose also that $\norm{A^{(0)}}_H$ and $\norm{\theta^{(j)}}_{\ell^1}$ are uniformly bounded.  Let $0<\lambda_0,\eta\le1$ and suppose that $C r_\Sigma^{(0)}(\lambda_0) +C\max_{1\le j\le m}r_\Sigma^{(j)}(\eta)\le\frac12$. Choose $\lambda_j$ according to \eqref{eq:lambda_j_choice}. Suppose that $r_{\op}^{(j)}(\eta,\lambda_0) +r_M^{(0)}(\lambda_0) +\lambda_0^{1/2}\norm{A^{(0)}}_H\le1$. With probability at least $1-C(m+1)\exp(-t)$,
\begin{equation}
R_{\Phi A,\eta}^{(j)}
\le C\bigl\{
r_{\op}^{(j)}(\eta,\lambda_0)
+r_M^{(0)}(\lambda_0)
+\lambda_0^{1/2}\norm{A^{(0)}}_H
\bigr\}.
\label{eq:generator_rate}
\end{equation}
If the right-hand side of \eqref{eq:generator_rate} is at most $\min\Bigl[1, c\frac{\alpha_j\kappa_j^2} {s_j^{3/2}\{1+\phi_j\norm{A^{(0)}}_{\mathcal G_0}\}} \Bigr]$ and $\lambda_j\le c \tau_j\kappa_j^2/s_j^{1/2}$, then, on the same event,
\begin{equation}\label{eq:calibration_rate_general}
\supp(\widetilde\theta^{(j)})=S_j,\qquad
\norm{\widetilde\theta^{(j)}-\theta^{(j)}}_{\ell^1}
\le C\frac{s_j\lambda_j}{\kappa_j^2}.
\end{equation}
If, in addition, $0<\eta\le\bar\lambda\le1$, then
\begin{equation}
R_{\theta\Phi,\mathcal G,\eta}^{(j)}
\le C\Bigl(\frac{n_0}{N}\Bigr)^{1/2}\biggl[
\Bigl\{\norm{\theta^{(j)}}_{\ell^1}
+\frac{s_j\lambda_j}{\kappa_j^2}\Bigr\}
r_{\op}^{(j)}(\eta,\bar\lambda)
+\frac{s_j\lambda_j}{\kappa_j^2}
\biggr].
\label{eq:thetaPhi_rate}
\end{equation}
\end{proposition}

\begin{proof}
On the event $R_{\Sigma,\eta}^{(j)}\le1/2$, add and subtract $\Phi_l^{(j)}(\eta)\widetilde A$ in the definition of $R_{\Phi A,\eta}^{(j)}$. The triangle inequality and Assumption~\ref{asm:nonp}(\ref{asm:nonp:perturb}) give
\begin{equation}
\begin{aligned}
R_{\Phi A,\eta}^{(j)}
&\le R_{\Phi,\mathcal G_0,\eta}^{(j)}
\bigl\{\norm{A^{(0)}}_{\mathcal G_0}
+\norm{\widetilde A-A^{(0)}}_{\mathcal G_0}\bigr\}\\
&\quad+C\phi_j\norm{\widetilde A-A^{(0)}}_{\Gamma^{(0)}}.
\end{aligned}
\end{equation}
Together with Lemma~\ref{lem:emp_concentration} and Proposition~\ref{prop:pilot_rate}, this proves \eqref{eq:generator_rate}.

For $h\in\mathbb R^p$, linearity gives $\widehat M_\eta^{(j)}(\widetilde A,\theta^{(j)}+h)=\widehat M_\eta^{(j)}(\widetilde A,\theta^{(j)})+h\cdot\widehat\Phi^{(j)}(\eta)\widetilde A$. Moreover, \eqref{eq:emp_moment_expansion} and Assumptions~\ref{asm:nonp} and \ref{asm:spectrum} imply
\begin{equation}
\norm{\widehat M_\eta^{(j)}(\widetilde A,\theta^{(j)})}_{V\otimes W} \le{}R_{M,\eta}^{(j)} +C\{1+R_{O,\mathcal G_0,\eta}^{(j)} +\norm{\theta^{(j)}}_{\ell^1} R_{\Phi,\mathcal G_0,\eta}^{(j)}\} \norm{\widetilde A-A^{(0)}}_{\mathcal G_0}.
\end{equation}
The assumed bound on the right-hand side of \eqref{eq:generator_rate} implies \eqref{eq:generator_smallness}. Together with $\max_l\norm{g_l^{(j)}(\eta)}_{V\otimes W}\le\phi_j\norm{A^{(0)}}_{\Gamma^{(0)}}\le C$, this gives $\max_l\norm{\widehat\Phi_l^{(j)}(\eta)\widetilde A}_{V\otimes W}\le C$. Lemma~\ref{lem:emp_concentration} and Proposition~\ref{prop:pilot_rate} therefore show that \eqref{eq:lambda_j_choice} implies \eqref{eq:calibration_score_condition}. Lemma~\ref{lem:deterministic_calibration} proves \eqref{eq:calibration_rate_general}.

For $R_{\theta\Phi,\Gamma,\eta}^{(j)}$, add and subtract $\widetilde\theta^{(j)}\cdot\Phi^{(j)}(\eta)$ in its definition. Assumption~\ref{asm:nonp}(\ref{asm:nonp:perturb}) and $\norm{\widetilde\theta^{(j)}}_{\ell^1}\le\norm{\theta^{(j)}}_{\ell^1}+\norm{\widetilde\theta^{(j)}-\theta^{(j)}}_{\ell^1}$ then give
\begin{equation}
\begin{aligned}
R_{\theta\Phi,\Gamma,\eta}^{(j)}
&\le\bigl\{\norm{\theta^{(j)}}_{\ell^1}
+\norm{\widetilde\theta^{(j)}-\theta^{(j)}}_{\ell^1}\bigr\}
R_{\Phi,\Gamma,\eta}^{(j)}\\
&\quad+C\phi_j\norm{\widetilde\theta^{(j)}-\theta^{(j)}}_{\ell^1}
\sup_{\norm{T}_{\Gamma}\le1}\norm{T}_{\Gamma^{(0)}}.
\end{aligned}
\end{equation}
When $\eta\le\bar\lambda$, taking $\Gamma=\mathcal G$ and using $\mathcal G\succeq c(N/n_0)\Gamma^{(0)}$, \eqref{eq:calibration_rate_general}, and \eqref{eq:source_Phi_error} proves \eqref{eq:thetaPhi_rate}.
\end{proof}

\subsection{Projections}\label{app:projection_proofs}
\begin{lemma}\label{lem:projection}
Under Assumption~\ref{asm:para}\ref{asm:para:eigen}, suppose that $\supp(\widetilde\theta^{(j)})=S_j$, $s_j^{1/2}R_{\Phi A,\eta}^{(j)}\le\kappa_j/2$.  Then, for every $x\in V\otimes W$,
\begin{equation}\label{eq:projection_error}
\begin{aligned}
&\norm{
Q_V^{(j)}(\eta)\widetilde\Pi^{(j)}(\eta)
Q_V^{(j)}(\eta)^*
-\Pi^{(j)}(\eta)}_{\mathrm{op}}
\le
C\frac{s_j^{1/2}}{\kappa_j}R_{\Phi A,\eta}^{(j)},
\\
&\abs{
\norm{\widetilde\Pi_\perp^{(j)}(\eta)x}_{V\otimes W}
-\norm{\Pi_\perp^{(j)}(\eta)
Q_V^{(j)}(\eta)x}_{V\otimes W}}
\le
C\frac{s_j^{1/2}}{\kappa_j}
R_{\Phi A,\eta}^{(j)}\norm{x}_{V\otimes W}.
\end{aligned}
\end{equation}
\end{lemma}

\begin{proof}\textbf{of Lemma~\ref{lem:projection}.}
For any $\theta$ supported on $S_j$,
\begin{equation}
\norm{\theta\cdot\Phi^{(j)}(\eta)A^{(0)}}_{V\otimes W}
=\{\theta^\top G_j(\eta)[S_j,S_j]\theta\}^{1/2}
\ge\kappa_j\norm{\theta}_{\ell^2}.
\end{equation}
Moreover,
\begin{equation}
\norm{\theta\cdot\Phi^{(j)}(\eta)A^{(0)}
-Q_V^{(j)}(\eta)
\theta\cdot\widehat\Phi^{(j)}(\eta)\widetilde A}_{V\otimes W}
\le s_j^{1/2}\norm{\theta}_{\ell^2}R_{\Phi A,\eta}^{(j)}.
\end{equation}
The two generating families therefore have rank $s_j$, and their directed gaps in $V\otimes W$ are bounded by $Cs_j^{1/2}R_{\Phi A,\eta}^{(j)}/\kappa_j$.  Indeed, the population lower bound controls the coefficient vector for the gap from the population span to the empirical span; the perturbation bound and $s_j^{1/2}R_{\Phi A,\eta}^{(j)}\le\kappa_j/2$ give an empirical lower bound $\kappa_j/2$, which controls the reverse gap.  The projection identity in \citet[Theorem~I-6.34]{kato1995perturbation} states, for closed subspaces with orthogonal projections $\Pi_1$ and $\Pi_2$, that
\begin{equation}
\norm{\Pi_1-\Pi_2}_{\mathrm{op}}
=\max\bigl\{
\norm{(I-\Pi_2)\Pi_1}_{\mathrm{op}},
\norm{(I-\Pi_1)\Pi_2}_{\mathrm{op}}
\bigr\}.
\end{equation}
The two terms on the right are precisely the two directed gaps, which proves \eqref{eq:projection_error}.  The second inequality follows because $Q_V^{(j)}(\eta)$ is unitary and $Q_V^{(j)}(\eta)\widetilde\Pi^{(j)}(\eta) Q_V^{(j)}(\eta)^*$ is the projection onto the rotated empirical span.
\end{proof}

\section{Verification of examples}\label{sec:additional_examples}

\subsection{Elliptic model}
\label{app:elliptic_verification}
We verify the claims of Example~\ref{ex:elliptic_theory} for the model in Example~\ref{ex:elliptic}. Constants may depend on $\kappa,s,c_s$, the fixed noise envelope, the lower source-weight bound, and the fixed perturbation bounds, but not on $\eta\in(0,1]$ or the sample sizes.

Let $e_k$ denote the normalized complex Fourier functions on $\mathbb T^2$. Choose a set $\mathcal K_+$ containing one element from each pair $\{k,-k\}$ with $k\ne0$. For each task, take independent Rademacher signs $\xi_0,\xi_{k,c},\xi_{k,s}$ and put $\xi_k=(\xi_{k,c}-i\xi_{k,s})/2^{1/2}$ and $\xi_{-k}=\overline{\xi_k}$ for $k\in\mathcal K_+$. Then $F=\sum_kq_k^{1/2}\xi_ke_k$ is real valued and is precisely the independent-sign construction in the normalized real sine--cosine basis. It satisfies $|F_k|^2=q_k$ and has covariance eigenvalues $q_k$. Use an independent copy of these signs for the noise, with even amplitudes $\sigma_{-k}=\sigma_k$. Operators in $H$ have $a_{-k}=\overline{a_k}$ and $a_0\in\mathbb R$; $H$ is regarded as a real Hilbert space. The complex notation below represents this real model.

For bounded $\theta^{(j)}$, the symbols in Example~\ref{ex:elliptic} obey
\begin{equation}
\begin{aligned}
\ell_0(k)\le|\ell_j(k)|&\le C\ell_0(k),&
\sup_k\frac{1+|k|^2}{|\ell_j(k)|}&<\infty,&
\sup_k\frac{|k_l|}{|\ell_j(k)|}&<\infty.
\end{aligned}
\end{equation}
Thus $L^{(j)}:H^2_{\rm per}\to L^2$ is bijective and its graph norm is equivalent to the $H^2$ norm, which proves closedness on $L^2$. The inverse maps $L^2$ boundedly into $H^2_{\rm per}$, and $D_lA^{(j)}$ is bounded on $L^2$. The space $H$ embeds isometrically into the Hilbert--Schmidt operators, so evaluations are continuous. Also, $\|A^{(0)}\|_H^2=\sum_k\ell_0(k)^{-2}<\infty$ in dimension two. Right composition by $D_lA^{(j)}$ multiplies $a_k$ by $ik_l/\ell_j(k)$ and therefore preserves $H$ boundedly.

Since $s>1$, $\sum_kq_k<\infty$, so the forcing is in $L^2$ almost surely. For every $T_a\in H$, $\|T_aF\|_V^2=\sum_kq_k|a_k|^2=\|T_a\|_{\Gamma^{(0)}}^2$ almost surely. Ordering the lattice modes by radius gives $\nu_r\asymp r^{-s}$, because the number of modes in a radius-$R$ disk is of order $R^2$. Consequently $\beta=s/2>1/2$ and $\mathcal N(a)\asymp a^{-1/s}$ for $0<a\le1$. Identical forcing covariances give the forcing-comparability condition and the source trace bound in Assumption~\ref{asm:spectrum}.

Write $B_j=O^{(j)}(\eta)-\theta^{(j)}\cdot\Phi^{(j)}(\eta)$. Conditional centering of the noise and its derivatives gives the following diagonal coefficients of $B_jT_a$ and $g_l^{(j)}=\Phi_l^{(j)}(\eta)A^{(0)}$:
\begin{equation}\label{eq:elliptic_multipliers}
a_k\frac{q_k}{(q_k+\eta)^{1/2}}\frac{\ell_0(k)}{\ell_j(k)},
\qquad
\frac{q_k}{(q_k+\eta)^{1/2}}
\frac{ik_l}{\ell_0(k)\ell_j(k)}.
\end{equation}
The corresponding coefficient of $\Phi_l^{(j)}(\eta)T_a$ is $a_kq_kik_l/\{(q_k+\eta)^{1/2}\,\ell_j(k)\}$. The multiplier bounds above imply
\begin{equation}
\begin{aligned}
\|B_jT_a\|_{V\otimes W}^2
&\asymp\sum_k\frac{q_k^2}{q_k+\eta}|a_k|^2
\le C\|T_a\|_{\Gamma^{(0)}}^2,\\
\max_l\|\Phi_l^{(j)}(\eta)T_a\|_{V\otimes W}^2
&\le C\sum_k\frac{q_k^2}{q_k+\eta}|a_k|^2
\le C\|T_a\|_{\Gamma^{(0)}}^2.
\end{aligned}
\end{equation}
Contractivity of the projections proves Assumption~\ref{asm:nonp} and the upper source-curvature bound in Assumption~\ref{asm:spectrum}.

Let $M_\eta$ be the full map at zero perturbation, so $\|M_\eta T_a\|^2=\sum_kq_k^2|a_k|^2/(q_k+\eta)$. Let $g_l^*$ be the second multiplier in \eqref{eq:elliptic_multipliers} at $\theta=0$, and let $\Pi_{l,\mathrm{ref}}$ project onto its real span. The real Hilbert--Schmidt inner product of $g_1^*$ and $g_2^*$ is zero: Its summand is a radial weight times $k_1k_2$, and the absolutely convergent sum vanishes by sign symmetry. The norms of these generators are bounded above and away from zero uniformly over $0<\eta\le1$; the modes $(1,0)$ and $(0,1)$ give the lower bounds. Hence
\begin{equation}
\|(I-\Pi_{1,\mathrm{ref}})M_\eta T\|^2+\|(I-\Pi_{2,\mathrm{ref}})M_\eta T\|^2 =2\|M_\eta T\|^2-\|\Pi_{1,\mathrm{ref}}M_\eta T\|^2-\|\Pi_{2,\mathrm{ref}}M_\eta T\|^2 \ge\|M_\eta T\|^2.
\end{equation}

For the two complementary sources in Example~\ref{ex:elliptic_theory}, the bound $\sup_k|\ell_0(k)/\ell_j(k)-1|\le C\epsilon$ gives $\|(B_j-M_\eta)T\|\le C\epsilon\|M_\eta T\|$ and $\|g_l^{(j)}-g_l^*\|\le C\epsilon$.
For nonzero vectors with norms bounded below, their rank-one orthogonal projections vary by at most a constant times the vector difference. Indeed, normalize the vectors, use $\|u\otimes u-v\otimes v\|_{\op}\le2\|u-v\|$ for unit vectors, and bound the change in normalization by the reverse triangle inequality. Thus $\|\Pi^{(1)}-\Pi_{1,\mathrm{ref}}\|_{\op}+\|\Pi^{(2)}-\Pi_{2,\mathrm{ref}}\|_{\op} \le C\epsilon$ uniformly in $\eta$. The triangle inequality in the direct sum of the two moment spaces now gives
\begin{equation}
\begin{aligned}
\bigl\{\|\Pi_\perp^{(1)}B_1T\|^2+
\|\Pi_\perp^{(2)}B_2T\|^2\bigr\}^{1/2}
&\ge(1-C\epsilon)\|M_\eta T\|.
\end{aligned}
\end{equation}
Choose $\epsilon$ so that $C\epsilon\le1/2$. If $n_1/n_S,n_2/n_S\ge\omega_*>0$, positivity of all remaining source curvatures yields
\begin{equation}
\sum_{j=1}^m\rho_j\|\Pi_\perp^{(j)}B_jT_a\|^2
\ge\frac{\omega_*\rho}{4}
\sum_k\frac{q_k^2}{q_k+\eta}|a_k|^2.
\end{equation}
This proves the required aggregate lower bound.

For each fixed bounded source coefficient vector, the modes $(1,0)$ and $(0,1)$ give $\|v_1g_1^{(j)}+v_2g_2^{(j)}\|^2\ge c(v_1^2+v_2^2)$ uniformly in $\eta$. Every active Gram matrix therefore has its minimum eigenvalue bounded away from zero. If $S_j=\{1\}$, then $\theta_2^{(j)}=0$, and reflection of $k_2$ in \eqref{eq:elliptic_multipliers} gives $\langle g_1^{(j)},g_2^{(j)}\rangle=0$; the case $S_j=\{2\}$ follows by reflecting $k_1$. Thus $\alpha_j=1$ is valid for singleton supports. When both coordinates are active there are no inactive coordinates, and empty supports use the convention in Appendix~\ref{app:empty_supports}. For each fixed nonempty coefficient vector, take $\tau_j=\min_{l\in S_j}|\theta_l^{(j)}|>0$; the two complementary sources in Example~\ref{ex:elliptic_theory} satisfy $\tau_j\ge\tau$. This verifies Assumption~\ref{asm:para}.

The noise envelope gives $\sum_k(1+|k|^2)^2\sigma_k^2\le C\sum_k(1+|k|^2)^{-(s+1)}<\infty$. Thus $\varepsilon^{(j)}\in H^2_{\rm per}$ almost surely. Independence and centering of its real Fourier signs imply both conditions in \eqref{eq:conditional_centering}. Moreover, $|k_l\sigma_k|\le Cq_k^{1/2}$ and $|k_l/\ell_j(k)|\le C$, so $|\{D_lU^{(j)}\}_k|\le Cq_k^{1/2}$ almost surely. The transformed noise has multiplier $\ell_j(k)/\ell_0(k)$; its $L^2$ norm and the target noise norm are bounded uniformly.

For $X=F^{(j)}$ or $X=D_lU^{(j)}$, evaluation in the Fourier-diagonal space therefore satisfies
\begin{equation}
\begin{aligned}
\|(T\mapsto TX)\{\Gamma^{(0)}+aI_H\}^{-1/2}\|_{\mathrm{HS}}^2
&=\sum_k\frac{|X_k|^2}{q_k+a}\le C\mathcal N(a),\\
\|Z^{(j)}(\eta)\|_W^2
&=\sum_k\frac{q_k}{q_k+\eta}=\mathcal N(\eta)
\le\eta^{-1}\sum_kq_k.
\end{aligned}
\end{equation}
The tensor-product norm identity and the noise bounds therefore give second-moment bounds $C\mathcal N(a)$, $C\mathcal N(\eta)\mathcal N(a)$, and $C\mathcal N(\eta)$ for the target score, the centered source moment operators, and the transformed source score, respectively. Since $\mathcal N(a)\le C/a$ and $\mathcal N(\eta)\le C/\eta$, their almost-sure norms are bounded by $C/a^{1/2}$, $C/(\eta a)^{1/2}$, and $C/\eta^{1/2}$, respectively. For any Hilbert-space random element $\xi$ satisfying $\E(\norm{\xi}^2)\le v^2$ and $\norm{\xi}\le b$ almost surely, $\E(\norm{\xi}^q)\le b^{q-2}\E(\norm{\xi}^2)\le(q!/2)v^2b^{q-2}$ for $q\ge2$.
Applying this inequality to the three quantities proves the Bernstein moment bounds in Assumption~\ref{asm:bernstein}. The target sub-Gaussian condition is immediate from the deterministic identity for $\|T_aF\|_V$. For each real linear functional of the forcing, independence of the signs and $\cosh(x)\le\exp(x^2/2)$ bound its moment generating function by that of a Gaussian with its forcing variance. This proves the source sub-Gaussian condition. All parts of Assumption~\ref{asm:bernstein} are therefore verified separately.

\subsection{Block model}
\label{app:anchored_model_verification}\label{app:block_shift_verification}
\begin{proposition}
\label{prop:anchored_submodel}
For every fixed $R$, the class $\mathcal P_{\rm anc}(R)$ consists of valid functional observations from closed operators and satisfies Assumptions~\ref{asm:nonp}--\ref{asm:bernstein} uniformly, with the fixed, nonzero operator $D$ in Example~\ref{exp:3}. Thus it is a subclass of the model used for the upper bounds, with $\theta^{(j)}=0$ for every source.
\end{proposition}

\begin{proof}\textbf{of Proposition~\ref{prop:anchored_submodel}.}
The operators $A_a$ and $L_a$ in \eqref{eq:block_operator} are mutually inverse, and $V_0=\R\oplus\operatorname{ran}(\Lambda)$ is dense in $V$ because $\mu_k>0$. To check that $L_a$ is closed, suppose $(y_r,g_r)\in V_0$ converges to $(y,g)$ in $V$ and $L_a(y_r,g_r)\to(x,f)$ in $W$. Then $\Lambda^{-1}g_r\to f$, so $g=\Lambda f$ and $x=y-\langle a,f\rangle$. Thus $(y,g)\in V_0$ and $L_a(y,g)=(x,f)$. Its inverse $A_a:W\to V$ is bounded uniformly for $\|a\|_{\ell^2}\le R$. Moreover, \eqref{eq:exp-3-H-def} gives
\begin{equation}
\begin{aligned}
\|T_{c,b}\|_{V\otimes W}^2
=c^2(1+\|\Lambda\|_{\HS}^2)+\|b\|_{\ell^2}^2
\le C\|T_{c,b}\|_H^2.
\end{aligned}
\end{equation}
Hence $H$ is separable and continuously embedded in the Hilbert--Schmidt operators, and $T\mapsto Tw$ is continuous for every $w\in W$. Put $h_\Lambda=\Lambda h_0$. Since $D(y,g)=(-\langle h_0,g\rangle,0)$ is bounded, $DA_a(x,f)=(-\langle h_\Lambda,f\rangle,0)$, and $T_{c,b}DA_a=T_{0,-c h_\Lambda}$, the map $T\mapsto TDA_a$ is bounded on $H$, as required in Section~\ref{sec:assumptions}.

All tasks have the observation law \eqref{eq:block_observation}; we omit the task index when writing a generic observation. Let $F_2=(\nu_k^{1/2}\xi_k)_{k\ge1}$ and $\gamma_c=\sum_k\mu_k^2\nu_k>0$. Then $\|F\|_W^2=\sum_k\nu_k<\infty$ and $\varepsilon=(\sigma\zeta,0)\in V_0$. Since $\zeta$ is centered and independent of $F$, and $D\varepsilon=0$, both conditions in \eqref{eq:conditional_centering} hold. Independence of the Rademacher coordinates gives
\begin{equation}\label{eq:final-block-prediction}
\begin{aligned}
\|T_{c,b}F\|_V^2&=|\langle b,F_2\rangle|^2+c^2\gamma_c,\qquad
\|T_{c,b}\|_{\Gamma^{(0)}}^2=\sum_k\nu_k b_k^2+c^2\gamma_c.
\end{aligned}
\end{equation}
By \eqref{eq:final-block-prediction}, $\Gamma^{(0)}$ is diagonal in the coordinates $(c,b)$, with eigenvalues $\gamma_c$ and $(\nu_k)_{k\ge1}$. Since $\nu_k=c_\beta k^{-2\beta}$, adding the fixed eigenvalue $\gamma_c>0$ preserves the eigenvalue bounds in Assumption~\ref{asm:spectrum}.

Since $\theta^{(j)}=0$, we have $S_j=\varnothing$, $\Pi^{(j)}(\eta)=0$, and $\Psi^{(j)}(\eta)=O^{(j)}(\eta)$. Assumption~\ref{asm:para} is therefore vacuous. Direct calculation from \eqref{eq:block_observation} gives
\begin{equation}\label{eq:block-source-curvature}
\begin{aligned}
\|\Psi^{(j)}(\eta)T_{c,b}\|_{V\otimes W}^2
&=\sum_k\frac{\nu_k^2b_k^2}{\nu_k+\eta}
+c^2\sum_k\frac{\mu_k^2\nu_k^2}{\nu_k+\eta}.
\end{aligned}
\end{equation}
The ratio $\{(\gamma_c+\eta)/\gamma_c^2\}\sum_k\mu_k^2\nu_k^2/(\nu_k+\eta)$ is positive and continuous on $[0,1]$, with value one at zero, and is therefore bounded above and away from zero. Comparing the coefficients in \eqref{eq:block-source-curvature} and \eqref{eq:final-block-prediction} gives
\begin{equation}\label{eq:block-source-comparison}
\begin{aligned}
c_\Gamma\{\Gamma^{(0)}\}^2\{\Gamma^{(0)}+\eta I_H\}^{-1}
\preceq\Gamma^{(j)}(\eta)
\preceq C_\Gamma\{\Gamma^{(0)}\}^2\{\Gamma^{(0)}+\eta I_H\}^{-1}.
\end{aligned}
\end{equation}
Multiplying the lower inequality in \eqref{eq:block-source-comparison} by $\rho_j$ and summing over $j$ gives $\sum_j\rho_j\Gamma^{(j)}(\eta)\succeq c_\Gamma\rho\{\Gamma^{(0)}\}^2\{\Gamma^{(0)}+\eta I_H\}^{-1}$. Moreover, the common forcing law gives $\E\{\|TF^{(j)}\|_V^2\}=\|T\|_{\Gamma^{(0)}}^2$ and
\begin{equation}
\begin{aligned}
\tr\bigl[\Sigma^{(j)}\{\Sigma^{(j)}+\eta I_W\}^{-1}\bigr]
=\sum_k\frac{\nu_k}{\nu_k+\eta}\le\mathcal N(\eta).
\end{aligned}
\end{equation}
This completes the verification of Assumption~\ref{asm:spectrum}. Part~(\ref{asm:nonp:feat}) of Assumption~\ref{asm:nonp} follows from $\Pi^{(j)}(\eta)=0$. For part~(\ref{asm:nonp:perturb}), $p=1$ and
\begin{equation}
\begin{aligned}
\|\Phi_1^{(j)}(\eta)T_{c,b}\|_{V\otimes W}^2
=c^2\sum_k\frac{(h_\Lambda)_k^2\nu_k^2}{\nu_k+\eta}
\le Cc^2\le C\|T_{c,b}\|_{\Gamma^{(0)}}^2,
\end{aligned}
\end{equation}
where the last inequality uses \eqref{eq:final-block-prediction} and $\gamma_c>0$.

We next verify Assumption~\ref{asm:bernstein}. Independence of the signs and $\cosh(x)\le\exp(x^2/2)$ give
\begin{equation}\label{eq:block-rademacher-bound}
\begin{aligned}
\|\langle b,F_2\rangle\|_{\psi_2}\le C\biggl(\sum_k\nu_kb_k^2\biggr)^{1/2}.
\end{aligned}
\end{equation}
Taking $b$ to be the $\ell^2$ component of $w\in W$ in \eqref{eq:block-rademacher-bound} proves \eqref{eq:source_subgaussian_forcing}. Also, \eqref{eq:final-block-prediction} and \eqref{eq:block-rademacher-bound} imply
$\|\|T_{c,b}F\|_V\|_{\psi_2}\le C\{(\sum_k\nu_kb_k^2)^{1/2}+|c|\gamma_c^{1/2}\}\le C\|T_{c,b}\|_{\Gamma^{(0)}}$, proving \eqref{eq:subgaussian_forcing}.
For the moment bounds in Assumption~\ref{asm:bernstein}, write $r$ for its regularization parameter $a$, to distinguish it from the block parameter $a\in\ell^2$. For $Y=F$ or $Y=DU$, let $\mathcal E_Y:H\to V$ be the evaluation map $T\mapsto TY$. In the coordinates of \eqref{eq:exp-3-H-def},
\begin{equation}\label{eq:block-evaluation-bounds}
\begin{aligned}
\|\mathcal E_F\{\Gamma^{(0)}+rI_H\}^{-1/2}\|_{\HS}^2
&=\frac{\gamma_c}{\gamma_c+r}+\sum_k\frac{\nu_k}{\nu_k+r}
=\mathcal N(r),\\
\|\mathcal E_{DU}\{\Gamma^{(0)}+rI_H\}^{-1/2}\|_{\HS}^2
&=\frac{|\langle h_\Lambda,F_2\rangle|^2}{\gamma_c+r}
\le C\mathcal N(r),\\
\|Z(\eta)\|_W^2&=\sum_k\frac{\nu_k}{\nu_k+\eta}\le\mathcal N(\eta).
\end{aligned}
\end{equation}
The second line of \eqref{eq:block-evaluation-bounds} uses $DU=(-\langle h_\Lambda,F_2\rangle,0)$, $|\langle h_\Lambda,F_2\rangle|^2\le\|h_\Lambda\|_{\ell^2}^2\sum_k\nu_k$ by Cauchy--Schwarz, and $\mathcal N(r)\ge\gamma_c/(\gamma_c+r)$.

For $Y\in\{F,DU\}$, \eqref{eq:block-evaluation-bounds} gives
\begin{equation}
\begin{aligned}
\bigl\|\{T\mapsto TY\otimes Z(\eta)\}\{\Gamma^{(0)}+rI_H\}^{-1/2}\bigr\|_{\HS}^2
\le C\mathcal N(r)\mathcal N(\eta)\le\frac{C}{r\eta},
\end{aligned}
\end{equation}
where we used $\mathcal N(u)\le\tr(\Gamma^{(0)})/u$ for $u>0$. For $G$ equal to the centered Hilbert--Schmidt norm in the third row of Assumption~\ref{asm:bernstein}, centering gives $G\le C/(r\eta)^{1/2}$ almost surely and $\E(G^2)\le C\mathcal N(r)\mathcal N(\eta)$. Applying $\E(G^q)\le B^{q-2}\E(G^2)$ with $B=C/(r\eta)^{1/2}$ and increasing $C$ yields
\begin{equation}
\begin{aligned}
\E(G^q)\le\frac{q!}{2}C^2\mathcal N(r)\mathcal N(\eta)
\biggl\{\frac{C}{(r\eta)^{1/2}}\biggr\}^{q-2},\qquad q=2,3,\ldots,
\end{aligned}
\end{equation}
which is the third Bernstein inequality in Assumption~\ref{asm:bernstein}.

For the first two rows of Assumption~\ref{asm:bernstein}, respectively, the random quantities are
\begin{equation}\label{eq:block-noise-quantities}
\begin{aligned}
\sup_{\|T\|_H\le1}
\bigl|\langle\varepsilon,[\{\Gamma^{(0)}+rI_H\}^{-1/2}T]F\rangle_V\bigr|
&=\sigma|\zeta|\biggl(\sum_k\frac{\nu_k}{\nu_k+r}\biggr)^{1/2},\\
\|\varepsilon\otimes Z(\eta)\|_{V\otimes W}
&=\sigma|\zeta|\biggl(\sum_k\frac{\nu_k}{\nu_k+\eta}\biggr)^{1/2},
\end{aligned}
\end{equation}
where $\{I_V+A^{(0)}\theta^{(j)}\cdot D\}\varepsilon=\varepsilon$ because $\theta^{(j)}=0$. For $G$ equal to the first or second quantity in \eqref{eq:block-noise-quantities}, take $u=r$ or $u=\eta$, respectively. Using $\E(|\zeta|^q)\le q!C^q$ and $\sum_k\nu_k/(\nu_k+u)\le\mathcal N(u)\le\tr(\Gamma^{(0)})/u$, and increasing $C$, we obtain
\begin{equation}
\begin{aligned}
\E(G^q)\le\frac{q!}{2}C^2\mathcal N(u)
\biggl(\frac{C}{u^{1/2}}\biggr)^{q-2},\qquad q=2,3,\ldots.
\end{aligned}
\end{equation}
This verifies the first two rows of Assumption~\ref{asm:bernstein}. All constants are uniform over $\|a\|_{\ell^2}\le R$, $r,\eta\in(0,1]$, and the sample sizes, with dependence only on the fixed $h_0,\Lambda,\beta,\sigma,R$.
\end{proof}

We next verify the common shift in Theorem~\ref{thm:minimax_shift} for the same block model. Use $D$ from Example~\ref{exp:3}, fix $a\in\ell^2$, and let $L^{(j)}=L_a+\theta^{(j)}D$, where every $|\theta^{(j)}|$ is bounded away from zero. The definitions in \eqref{eq:block_operator} give $L_a+sD=L_{a+s\Lambda h_0}$ and $A_{a+s\Lambda h_0}-A_a=sT_{0,\Lambda h_0}$. Hence the replacements $a\mapsto a+s\Lambda h_0$ and $\theta^{(j)}\mapsto\theta^{(j)}-s$ leave each $L^{(j)}$ unchanged. For small enough $|s|$, every source coefficient remains nonzero.

For $h=1$, $T_h=A_aDA_a=-T_{0,\Lambda h_0}\in H$. Its squared prediction norm is $\norm{T_h}_{\Gamma^{(0)}}^2 =\sum_{k\ge1}\nu_k\mu_k^2(h_0)_k^2>0$. The sum is finite because $h_0\in\ell^2$ and $(\nu_k)$ and $(\mu_k)$ are bounded; it is positive because $h_0\ne0$ and $\nu_k,\mu_k>0$. Also, $DA_a(x,f)=(-\langle\Lambda h_0,f\rangle,0)$ is bounded. The formulas for $L_a$ and $A_a$ remain valid with $a+s\Lambda h_0$ in place of $a$.

The forcing and noise laws are fixed. Conditional on $F$, only the first response coordinate changes, and its Gaussian variance is $\sigma^2$. Therefore $\operatorname{KL}(P_s^{(0)},P_0^{(0)})=\frac{s^2}{2\sigma^2}\E(\langle\Lambda h_0,F_2\rangle^2)=\frac{s^2}{2\sigma^2}\norm{T_{0,\Lambda h_0}}_{\Gamma^{(0)}}^2$. This verifies the target KL inequality and completes the verification of the conditions in Theorem~\ref{thm:minimax_shift}. The two block constructions use the same spaces, operator $D$, forcing, and noise. Their source coefficients differ: $\theta^{(j)}=0$ in $\mathcal P_{\rm anc}(R)$, whereas the common-shift construction uses nonzero $\theta^{(j)}$ and fails the aggregate lower bound in Assumption~\ref{asm:spectrum}.

\section{Experimental details}
\label{app:experiments}

\subsection{Adaptation and tuning of literature baselines}
\label{app:literature-baselines}

\subsubsection{Lifting scalar-response methods to operator spaces}

We detail the implementation of the three functional linear regression (FLR) transfer baselines—Jiao--McKeague (JM; \citealt{jiao2025coefficient}), Lin--Reimherr (LR; \citealt{lin2024hypothesis}), and Hu--Lin (HL; \citealt{hu2025functional})—introduced in Section~\ref{sec:baselines}. Although these methods were developed for scalar-response FLR, their estimation, source-ranking, and aggregation rules interact with the data strictly through the empirical normal equations and principal-component directions of the squared prediction loss. In our setting, each candidate operator $A\in H$ is parameterized by its coordinates in a finite-dimensional basis of $H$. Because the task-$j$ empirical squared loss is quadratic in these coordinates, it is completely characterized by a task-specific empirical normal (Gram) matrix and predictor--response cross-covariance vector on $H$. We therefore implement JM, LR, and HL by identifying the operator coordinates in $H$ with the regression slope vector, extracting empirical principal directions from the corresponding normal matrices, and applying each method's shape-alignment, ridge-offset, or $\ell^1$-offset transfer mechanism in $H$. All empirical moments are uncentered with no additional intercept, and none of the methods is given prior knowledge of the informative source sets.

Specifically, the coordinate representation and regularization geometry in each study are configured as follows:
\begin{itemize}
 \item \emph{Functional regression (Section~\ref{sec:functional-per-n0}):} With $c=1$ fixed by the shared noise-free output block $\Lambda f$, the operator $T_{1,a}\in H$ is parameterized directly by the scalar-response slope vector $a\in\mathbb R^K$ ($K=512$), equipped with the Euclidean inner product and standard ridge penalty.
 \item \emph{Advection--diffusion (Section~\ref{sec:exp-diffusion}):} The space $H$ is parameterized by the 625 real Fourier-multiplier coordinates equipped with the Hilbert--Schmidt inner product. Because the empirical normal matrix is block-diagonal across conjugate frequency pairs, spectral truncation retains the constant mode and complete conjugate frequency pairs, yielding 313 distinct spectral units.
 \item \emph{Linear time-varying filter (Section~\ref{sec:real_ltv}):} Operators are parameterized by the complex B-spline harmonic-transfer coefficients in \eqref{eq:real-ltv-class}, and the quadratic loss measures output prediction error as in \eqref{eq:real-ltv-output}. JM and HL use the Euclidean coordinate geometry, whereas LR uses the second-difference spline smoothing penalty (including its numerical stabilization ridge) of the selected operator class.
\end{itemize}

\subsubsection{Estimator constructions and source selection}

\paragraph{Jiao--McKeague (JM).}
The target empirical normal matrix defines a common principal-component basis. Given a candidate spectral truncation level and a subset of sources, least-squares coefficients are fitted separately on the target and on each selected source within the retained basis. Each source coefficient vector is sign-aligned with the target fit according to the sign of their real inner product, and the aligned source fits are averaged with the target fit using weights proportional to task sample sizes. The resulting pooled shape vector is then multiplied by a scalar amplitude fitted via target least squares (reducing to truncated target least squares when the selected source set is empty).

In both simulation studies, the truncation level and source subset (searching over all four subsets of the two sources) are selected jointly by minimizing mean prediction risk on independent development datasets. In the real-data study, candidate sources are ranked by the absolute sine of the angle between their fitted coefficient vectors and the target fit, corresponding to the shape-misalignment criterion of \citet{jiao2025coefficient}. The target validation record then selects the truncation level and nested source subset (including the empty set) directly, omitting concave-penalty fusion, residual-bootstrap thresholding, and cumulative-variance pre-screening.

\paragraph{Lin--Reimherr (LR).}
For any candidate source subset, LR first computes a pooled source ridge estimator—normalizing each source loss by its sample size so that selected sources receive equal task weight, as in Algorithm~1 of \citet{lin2024hypothesis}—and then fits a ridge-penalized target offset (reducing to target ridge regression when the source set is empty).

To construct the adaptive ATL-FLR estimator, the target training sample is split into two halves (using the first and second halves of the observations in the simulations, and the single training and validation records in the real-data study). On the first half, sources are ranked by the full penalty-norm distance between their individual ridge fits and the target ridge fit. Candidate estimators are then formed from the empty set and successive nested source sets, sharing a common target-offset penalty while scaling the source-fit penalty with the total selected source sample size to the power $-6/7$, as in the authors' implementation. On the second half, these candidates are combined via the star-shaped aggregation procedure of \citet{gaiffas2011hyper}: in the simulations, the first quarter of the second half identifies an empirical-loss leader and preselects candidates, and the remaining quarter selects the optimal convex combination along segments connecting the leader to retained candidates; in the real-data study, the single validation record performs this segment aggregation directly without preselection. Following the original procedure, the aggregated estimator is not refitted on the combined target sample.

\paragraph{Hu--Lin (HL).}
For each nonempty candidate source set, empirical principal components are extracted from the sample-size-weighted average of the source normal matrices. A pooled least-squares fit in the retained principal coordinates yields the source estimator, after which a target offset is fitted in the same coordinates under an $\ell^1$ lasso penalty parameterized as a fraction of the zero-solution threshold (which simplifies to coordinatewise soft thresholding in the diagonal Fourier representation of Section~\ref{sec:exp-diffusion}, and penalizes real and imaginary parts separately in Section~\ref{sec:real_ltv} after fixing eigenvector phases so that each largest-magnitude entry is real and positive).

Similar to LR, the adaptive HL estimator splits the target data into two subsets. On the first subset, sources are ranked by the squared norm of the discrepancy between their predictor--response cross-moments and the target cross-moment. Candidates consist of target principal-component least squares and the $\ell^1$-offset estimators over successive nested source sets, with each candidate's truncation level and lasso fraction tuned via development risk (in the simulations) or validation error (in the real-data study). Sparse aggregation on the second target subset selects the best convex combination between the empirical-loss leader and one other candidate, without refitting on the full target sample.

\subsubsection{Hyperparameter grids and evaluation protocol}

In the functional regression study, JM and HL search over the candidate truncation grid
\begin{equation}
 \{1,2,3,4,5,6,8,10,12,16,20,24,32,40,48,64,96,128,192,256\},
\end{equation}
omitting dimensions that exceed the number of fitting observations. For LR, the target-offset and source-fit ridge penalties range over half-decade increments from $10^{-8.5}$ to $10^2$ and from $10^{-7.5}$ to $10^{1.5}$, respectively, scaled by the largest eigenvalue of the corresponding target or summed-source normal matrix. HL searches over the 14 lasso threshold fractions
\begin{equation}
 \{1,0.5,0.25,0.1,0.05,0.02,0.01,0.005,0.002,0.001,0.0005,0.0002,0.0001,0.00005\}.
\end{equation}
In the advection--diffusion study, the truncation grid additionally includes all $313$ mode groups (one constant mode and $312$ conjugate pairs), corresponding to $625$ real parameters; the LR target-offset multiplier ranges from $10^{-8.5}$ to $10^{0.5}$ in half-decades, and the source-fit multiplier grid augments half-decades from $10^{-8.5}$ to $10^{1.5}$ with $\{0,10^{-10},10^{-9}\}$. In the real-data study, truncation grids are
\begin{equation}\{8,16,32,64,128,256,512,768,1024,1152,1280,1408,1536,1600,1640\}\end{equation}
for JM and $\{64,256,512,1024,1280,1536,1640\}$ for HL (capped at the active class dimension $930$ or $1640$); LR searches over source-fit smoothing multipliers $\{10^{-8},10^{-6},10^{-4},10^{-2}\}$ and target-offset multipliers $\{10^{-8},10^{-6},10^{-4},10^{-2},1,100\}$; and HL extends the 14 lasso fractions above with $\{2\times10^{-5},10^{-5},5\times10^{-6},2\times10^{-6},10^{-6}\}$.

Across all experiments, hyperparameter selections are fixed prior to evaluating test performance (minimizing mean development risk in the simulations and full-band validation error in the real-data study), and bootstrap confidence intervals share identical resampling indices across all compared methods.

\subsection{Functional regression}
\label{app:functional-per-n0}

\subsubsection{Model specification and analytical risk}

We implement the finite-dimensional block model of Example~\ref{exp:3} on $V=W=\mathbb R\oplus\mathbb R^K$ with $K=512$, $\Lambda=\operatorname{diag}(\mu_1,\ldots,\mu_K)$, $\mu_k=k^{-1}$, and $\nu_k=k^{-2}$. The solution operators and their governing inverses take the block form
\begin{equation}
 \begin{aligned}
 A_a(x,f)&=(x+\langle a,f\rangle,\Lambda f),\\
 L_a(y,g)&=(y-\langle a,\Lambda^{-1}g\rangle,\Lambda^{-1}g),
 \end{aligned}
 \label{eq:block_operator-app}
\end{equation}
and the hypothesis space is the $(K+1)$-dimensional Hilbert space
\begin{equation}
 H=\{T_{c,a}:(x,f)\mapsto(cx+\langle a,f\rangle,c\Lambda f)
       \mid(c,a)\in\mathbb R\oplus\mathbb R^K\}
 \label{eq:exp-3-H-def-app}
\end{equation}
equipped with norm $\|T_{c,a}\|_H^2=c^2+\|a\|_2^2$. The target operator is $A^{(0)}=T_{1,a^{(0)}}$, where $a^{(0)}=(a_k)_{k=1}^K$ has unit Euclidean norm with coordinates
\begin{equation}
 a_k=\frac{k^{-3/4}}{\bigl(\sum_{r=1}^K r^{-3/2}\bigr)^{1/2}}.
\end{equation}

Source tasks perturb $L^{(0)}$ via two dictionary operators
\begin{equation}
 D_l(y,g)=(-\langle h_l,g\rangle,0),\qquad
 h_l=\frac{\widetilde h_l}{\|\widetilde h_l\|_2},\qquad l=1,2,
 \label{eq:fr-dictionary}
\end{equation}
where $\widetilde h_1$ and $\widetilde h_2$ have disjoint supports on the odd and even indices among the first 16 coordinates:
\begin{equation}
 \widetilde h_{1,k}=k^{-0.6}\mathbf 1_{\{k\le16,\ k\text{ odd}\}},
 \qquad
 \widetilde h_{2,k}=k^{-0.6}\mathbf 1_{\{k\le16,\ k\text{ even}\}}.
 \label{eq:fr-dictionary-vectors}
\end{equation}
Under $L^{(j)}=L^{(0)}+\theta^{(j)}\cdot D$, the source solution operators are
\begin{equation}
 A^{(j)}(x,f)=
 \biggl(x+\langle a^{(0)},f\rangle+
 \sum_{l=1}^2\theta_l^{(j)}\langle h_l,\Lambda f\rangle,
 \Lambda f\biggr),\qquad j=1,2.
 \label{eq:fr-source-operators}
\end{equation}

For each task $j$ and observation $i$, we draw independent Rademacher variables $\xi_{i,k}^{(j)}$ and standard Gaussian variables $\zeta_i^{(j)}$, setting
\begin{equation}
 \begin{aligned}
 f_{i,k}^{(j)}&=\nu_k^{1/2}\,\xi_{i,k}^{(j)},&
 F_i^{(j)}&=(0,f_i^{(j)}),\\
 \varepsilon_i^{(j)}&=(\sigma\zeta_i^{(j)},0),\quad\sigma=1,&
 U_i^{(j)}&=A^{(j)}F_i^{(j)}+\varepsilon_i^{(j)}.
 \end{aligned}
 \label{eq:fr-observations}
\end{equation}
Thus all tasks share the predictor covariance $\Sigma^{(j)}=\operatorname{diag}(0,\nu_1,\ldots,\nu_K)$, and observation noise enters exclusively in the scalar response block. Because $D_l\varepsilon_i^{(j)}=0$, this design cleanly isolates calibration and projection of the source perturbations from errors-in-variables noise amplification.

To set the perturbation magnitudes $\theta^{(j)}$, note that the population calibration generators $g_l^{(j)}(\eta)=\Phi_l^{(j)}(\eta)A^{(0)}$ are identical across sources and have squared norms
\begin{equation}
 \|g_l^{(j)}(\eta)\|_{V\otimes W}^2
 =\sum_{k=1}^K\frac{\nu_k^2\mu_k^2h_{l,k}^2}{\nu_k+\eta}.
 \label{eq:fr-generator-strengths}
\end{equation}
Because $\mu_k$ and $\nu_k$ decay rapidly, the second generator $g_2^{(j)}(\eta)$ (supported on even indices) has smaller norm than $g_1^{(j)}(\eta)$. We therefore set
\begin{equation}
 \theta^{(1)}=(1/2,0),\qquad
 \theta_1^{(2)}=0,\qquad
 \theta_2^{(2)}=
 \frac{\|g_1^{(1)}(10^{-3})\|_{V\otimes W}^2}
      {2\|g_2^{(2)}(10^{-3})\|_{V\otimes W}^2}
 \simeq10.58,
 \label{eq:fr-source-shifts}
\end{equation}
which equalizes the partial derivative magnitudes $2|\theta_l^{(j)}|\|g_l^{(j)}(\eta)\|_{V\otimes W}^2$ of the population calibration objective $\|M_\eta^{(j)}(A^{(0)},\theta)\|_{V\otimes W}^2$ at $\theta=0$ and the reference value $\eta=10^{-3}$. Consequently, although both sources have one-sparse governing perturbations, their induced regression-coefficient shifts $a^{(j)}-a^{(0)}=\theta_1^{(j)}\Lambda h_1+\theta_2^{(j)}\Lambda h_2$ have markedly different Euclidean norms ($\approx 0.39$ for Source~1 versus $\approx 3.73$ for Source~2).

In the coordinates $(c,a)$ of \eqref{eq:exp-3-H-def-app}, the target prediction curvature operator is diagonal:
\begin{equation}
 \Gamma^{(0)}=
 \operatorname{diag}\biggl(\sum_{k=1}^K\mu_k^2\nu_k,
                          \nu_1,\ldots,\nu_K\biggr)
 =\operatorname{diag}\biggl(\sum_{k=1}^K k^{-4},
                          1,2^{-2},\ldots,K^{-2}\biggr),
 \label{eq:fr-curvature-spectrum}
\end{equation}
corresponding to spectral decay exponent $\beta=1$, and the prediction risk of $\widehat A=T_{\widehat c,\widehat a}$ is computed analytically as
\begin{equation}
 \|\widehat A-A^{(0)}\|_{\Gamma^{(0)}}^2
 =\sum_{k=1}^K\nu_k(\widehat a_k-a_k)^2
  +(\widehat c-1)^2\sum_{k=1}^K\mu_k^2\nu_k.
 \label{eq:fr-risk}
\end{equation}

\subsubsection{Exact projection invariance in the block model}
\label{app:fr-estimation}

In this block model, the action of each empirical calibration generator on any candidate $T_{c,a}\in H$ is proportional to its action on the target ridge pilot $\widetilde A=T_{\widetilde c,\widetilde a}$ (where $\widetilde c>0$):
\begin{equation}
 \widehat\Phi_l^{(j)}(\eta)T_{c,a}
 =\frac{c}{\widetilde c}\,
   \widehat\Phi_l^{(j)}(\eta)\widetilde A,
 \label{eq:fr-generator-scalar-action}
\end{equation}
since $T_{c,a}D_l(y,g)=(-c\langle h_l,g\rangle,0)$ depends on $T_{c,a}$ only through the scalar $c$. Consequently, for every $A\in H$, the perturbation term in $\widehat M_\eta^{(j)}(A,\widetilde\theta^{(j)})$ lies entirely inside the active calibration span $\operatorname{ran}(\widetilde\Pi^{(j)}(\eta))$, yielding the exact cancellation identity
\begin{equation}
 \widetilde\Pi_\perp^{(j)}(\eta)
 \widehat M_\eta^{(j)}(A,\widetilde\theta^{(j)})
 =
 \widetilde\Pi_\perp^{(j)}(\eta)
 \widehat M_\eta^{(j)}(A,0),\qquad A\in H.
 \label{eq:fr-coefficient-cancellation}
\end{equation}
Thus, conditional on the selected supports $\widehat S_j=\operatorname{supp}(\widetilde\theta^{(j)})$ and tuning parameters $(\eta,\lambda)$, the projected refit $\widehat A_{\mathrm P}$ is completely invariant to the fitted perturbation magnitudes $\widetilde\theta^{(j)}$. Moreover, because the generator spans in \eqref{eq:fr-generator-scalar-action} depend only on $\widehat S_j$ and not on $a$ or the positive scalar $c$, fixed-support recalibration leaves both the projection operators and the strictly convex refit objective unchanged, so that
\begin{equation}
 \widehat A^{[k]}=\widehat A^{[1]},\qquad k\ge1.
 \label{eq:fr-fixed-support-iteration}
\end{equation}

\subsubsection{Regularization grids and Monte Carlo protocol}

Experiments are conducted across thirteen target sample sizes
\begin{equation}
 \begin{aligned}
 n_0&\in\{10,20,30,40,50,70,100,200,300,400,500,700,1000\},\\
 n_1&=n_2=9.5n_0,\qquad N=20n_0.
 \end{aligned}
 \label{eq:fr-sample-sizes}
\end{equation}
(Below the range $n_0\ge 50$ displayed in Figure~\ref{fig:functional-per-n0}, Pooling attains the lowest mean risk at $n_0=10$ and Unprojected at $n_0\in\{20,30\}$.) For Unprojected and Projected, the effective refit penalty is parameterized as $\bar\lambda=n_0\lambda/N$ with instrument regularization $\eta=\bar\lambda$. Regularization parameters are tuned over dimensionless multipliers of empirical covariance scales: for Target-only ($\lambda$) and the transfer pilot ($\lambda_0$), multipliers range from $10^{-8.5}$ to $10^{0.5}$ in half-decade steps times $\lambda_{\max}(\widehat\Sigma^{(0)})$; for Pooling ($\lambda$) and the transfer refit ($\bar\lambda$), multipliers range from $10^{-7.5}$ to $10^{1.5}$ in half-decade steps times the pooled source scale
\begin{equation}
 \lambda_{\max}\!\Bigl(
 \frac{n_1\widehat\Sigma^{(1)}+n_2\widehat\Sigma^{(2)}}{n_1+n_2}
 \Bigr).
 \label{eq:fr-source-covariance-scale}
\end{equation}
The calibration penalties $\lambda_j$ share a common multiplier chosen from 14 logarithmic grid points between $5\times 10^{-5}$ and $1$ times the source-specific zero-solution threshold
\begin{equation}
 2\max_{1\le l\le2}
 \Bigl|\bigl\langle
 \widehat\Phi_l^{(j)}(\eta)\widetilde A,
 \widehat M_\eta^{(j)}(\widetilde A,0)
 \bigr\rangle_{V\otimes W}\Bigr|,
 \label{eq:fr-calibration-scale}
\end{equation}
and the candidate grids for Unprojected and Projected also include the target-only fit.

At each $n_0$, all methods select their tuning multipliers by minimizing mean analytical risk \eqref{eq:fr-risk} over 500 independent development datasets. Using these fixed multipliers (with empirical covariance scales recomputed on each sample), we evaluate the estimators on 1,000 independent sample-size trajectories generated as nested prefixes of task sizes $(1000,9500,9500)$. Unprojected and Projected are tuned independently of one another; in a supplementary check where Unprojected is instead evaluated at the exact pilot, calibration, and refit parameters selected for Projected, projection still yields consistent risk reductions. Pointwise $95\%$ confidence intervals in Figure~\ref{fig:functional-per-n0} are obtained from the $2.5\%$ and $97.5\%$ quantiles of 5,000 paired bootstrap resamples of complete trajectories.

\subsection{Periodic advection--diffusion PDE}
\label{app:exp-diffusion}

\subsubsection{Fourier representation, spectral diagnostics, and prediction risk}

For the periodic advection--diffusion operators in Section~\ref{sec:exp-diffusion}, the Fourier symbols on $\mathbb Z^2$ are
\begin{equation}
 \ell_0(k)=1+|k|^2,\qquad
 \ell_j(k)=1+|k|^2+\mathrm{i}\theta^{(j)}\cdot k,
 \qquad j=1,2,
 \label{eq:exp-advection-symbols}
\end{equation}
with $\theta^{(1)}=(1,0)$ and $\theta^{(2)}=(0,1)$, so that $A^{(j)}=\{L^{(j)}\}^{-1}$ and $D_l=\partial_l$ have symbols $\ell_j(k)^{-1}$ and $\mathrm{i}k_l$, respectively. Because these multipliers preserve each Fourier mode, restricting to $\|k\|_\infty\le 12$ yields the exact solution operators on that $625$-dimensional subspace. The space $H$ of Hilbert--Schmidt Fourier multipliers with conjugate-symmetric symbols is represented in the real orthonormal basis consisting of the constant mode ($k=0$) and 312 sine--cosine pairs, with each nonzero pair contributing two real parameters (the real and imaginary symbol components).

In this real basis, the forcing and noise coordinates at mode $k$ are distributed as $\ell_0(k)^{-1}\xi$ and $\sigma\ell_0(k)^{-5/2}\zeta$, where $\xi$ and $\zeta$ are independent Rademacher signs across coordinates, observations, and tasks. The resulting predictor covariance $\Sigma^{(j)}$ and target curvature operator $\Gamma^{(0)}$ share eigenvalues $(1+|k|^2)^{-2}$ (corresponding to two-dimensional spectral decay $\beta=1$), with effective dimension
\begin{equation}
 \mathcal N(\eta)
 =\sum_{\|k\|_\infty\le12}
 \frac{(1+|k|^2)^{-2}}{(1+|k|^2)^{-2}+\eta}
 \label{eq:exp-advection-effective-dimension}
\end{equation}
and analytical prediction risk
\begin{equation}
 \|\widehat A-A^{(0)}\|_{\Gamma^{(0)}}^2
 =\E[\| \{\widehat A-A^{(0)}\}F^{(0)}\|_V^2].
 \label{eq:exp-advection-risk}
\end{equation}

Decomposing \eqref{eq:exp-advection-risk} across Fourier modes clarifies why direct parameter-space transfer baselines struggle in Figure~\ref{fig:exp-recalibration}. Because $\ell_j(0)=\ell_0(0)=1$, the constant mode $k=0$ is identical across the target and both sources, yet it accounts for $87$--$90\%$ of the full-sample target least-squares risk across the tested $n_0$ range. Conversely, the operator contrast $\|A^{(j)}-A^{(0)}\|_{\Gamma^{(0)}}^2\simeq 0.031$ vanishes at $k=0$ and concentrates $99.5\%$ of its energy in just six low-frequency units, producing an angle with sine $\approx 0.24$ between $A^{(j)}$ and $A^{(0)}$ in $H$. Consequently, $A^{(j)}$ is not a scalar rescaling of $A^{(0)}$ (causing JM to reject both sources), while the small offset penalties required by LR and HL to absorb the large low-frequency contrasts prevent them from borrowing strength on the shared constant mode.

\subsubsection{Regularization scaling, recalibration, and evaluation protocol}

In both the noise-level study ($\sigma\in\{0.05,0.1,0.25,0.5\}$ at $n_0\in\{256,1024\}$; Figure~\ref{fig:exp-projection}) and the sample-size study ($n_0\in\{16,32,64,128,256,512,1024,2048,4096\}$ at $\sigma=0.25$; Figure~\ref{fig:exp-recalibration}), the source sample sizes are fixed at $n_1=n_2=16384$ ($\rho_j=n_j/n_0$, $N=n_0+32768$). Whitening uses the full uncentered empirical source covariance matrix on the 625-dimensional state space without sample splitting. Because $p=2$, the Lasso calibration problem \eqref{eq:calibration} is solved exactly by enumerating active sets and signs, while the refit \eqref{eq:refit} solves the linear normal equations in $H$.

For Target-only, Pooling, Unprojected, and Projected, regularization parameters follow theoretical sample-size scaling laws with proportionality constants fixed in preliminary calibration checks prior to both studies: Target-only uses $\lambda\propto n_0^{-2/3}$, Pooling uses $\lambda=0$, and Unprojected and Projected (tuned separately) use
\begin{equation}\label{eq:exp-advection-ridge-scaling}
\begin{aligned}
\lambda_0&\propto n_0^{-2/3},\qquad
 \eta=\bar\lambda\propto N^{-2/3},\qquad
 \lambda=\frac{N}{n_0}\bar\lambda,
\end{aligned}
\end{equation}
\begin{equation}\label{eq:exp-advection-calibration-scaling}
\begin{aligned}
\lambda_j&\propto
 \Bigl\{\frac{\mathcal N(\eta)}{n_j}\Bigr\}^{1/2}
 +\frac{1}{n_j\eta^{1/2}}+n_0^{-1/3},\qquad j=1,2.
\end{aligned}
\end{equation}
By contrast, JM, LR, and HL are re-tuned specifically for each $(\sigma,n_0)$ setting over 300 independent development repetitions.

In Figure~\ref{fig:exp-recalibration}, iterative recalibration proceeds from $\widehat A^{[0]}=\widetilde A$ and $\widehat A^{[1]}=\widehat A_{\mathrm P}$ exactly as in Section~\ref{sec:recalibration}: at each step $k\ge 1$, source perturbations are re-estimated via unpenalized least squares on the fixed support $\widehat S_j=\operatorname{supp}(\widetilde\theta^{(j)})$ using $\widehat A^{[k]}$, and the generator spans and projections $\widehat\Pi_\perp^{(j)[k]}(\eta)$ are updated prior to computing $\widehat A^{[k+1]}$, keeping $(\eta,\lambda)$ unchanged. Each study is evaluated over 300 independent repetitions (retaining all runs, including any support-selection errors), with pointwise $95\%$ confidence intervals computed from 20,000 paired percentile-bootstrap resamples. Ratios in Figure~\ref{fig:exp-projection} and percentage changes $100\{\overline R_{\mathrm P}^{[1]}/\overline R_{\mathrm P}-1\}$ in Figure~\ref{fig:exp-recalibration} are computed from the Monte Carlo mean risks.

\subsection{Linear time-varying electronic filter}
\label{app:real_ltv}

\subsubsection{Benchmark data, preprocessing, and candidate source bank}

We use the periodic electronic band-pass filter dataset of \citet{lataire2015benchmark}. Each experimental record comprises three steady-state periods of $15640$ time-domain samples acquired at sampling frequency $f_s=156.25\text{ kHz}$. Following standard periodic frequency-domain preprocessing, we average the three periods into a single functional observation and apply the discrete Fourier transform normalized by $15640$, so that frequency line $k$ corresponds to angular frequency $\omega_k=2\pi k f_s/15640$. The evaluation band $\mathcal K_{\mathrm{full}}$ spans lines $k=51,\ldots,2001$ ($0.5$--$20\text{ kHz}$), partitioned into $73$ \emph{forced lines} (where the reference input spectrum exceeds $10^{-3}$ of its peak amplitude) and $1878$ unexcited \emph{sideband lines}. All estimators use the measured input spectrum $F_k$ for fitting and prediction; the reference input is used solely to label the forced lines.

The eight target regimes correspond to all combinations of two scheduling waveform types (\texttt{h1} with one harmonic and \texttt{h3} with three harmonics), two peak-to-peak scheduling amplitudes ($350\text{ mV}$ and $700\text{ mV}$), and two distinct scheduling realizations (groups $0$ and $1$), excited by random-phase multisine inputs of root-mean-square (RMS) amplitude $70\text{ mV}$. Each target regime contains three records with independent input realizations, which are rotated cyclically through training ($n_0=1$), validation, and testing to produce 24 evaluation runs.

From the benchmark's pool of 46 operating regimes (30 under constant scheduling voltages and 16 under periodic time-varying scheduling), we construct a 12-task candidate source bank for each target using only the input and scheduling channels (without inspecting output responses). After excluding the target's own scheduling trajectory, we measure inter-regime scheduling distance via the Euclidean distance of the constant component and first three Fourier harmonics of the scheduling voltage. We select the six nearest constant regimes, the four nearest time-varying regimes, and two additional time-varying regimes chosen greedily from the eight nearest to minimize the maximum cosine similarity of their scheduling contrasts relative to already selected sources. Each constant source contributes two records (one noise-excited and one multisine) and each time-varying source contributes three multisine records, yielding $m=12$ candidate source tasks and $30$ source records per target.

\subsubsection{Harmonic-transfer operator space and model selection}

At nonzero frequency $\omega_k$, the integrated output spectrum is $U_k=Y_k/(\mathrm{i}\omega_k)$. To capture harmonic coupling across neighbouring frequency lines, we parameterize $A\in H$ as a banded frequency-shift operator with smooth B-spline multipliers:
\begin{equation}
 (AF)_k=\frac{1}{\omega_c}
 \sum_{r=-R}^{R}\sum_{b=1}^{B}
 \beta_{r,b}B_b(\omega_k)F_{k-r},
 \qquad \omega_c=2\pi\cdot7500,
 \label{eq:real-ltv-class}
\end{equation}
where $\beta_{r,b}\in\mathbb C$, $B_b(\omega)$ are cubic B-splines with $B$ open uniform knots over the evaluation band, $R$ is the maximum harmonic shift radius, and $\omega_c$ is a numerical conditioning scale. The predicted output spectrum is accordingly
\begin{equation}
 \widehat Y_k
 =\mathrm{i}\omega_k(\widehat A F)_k
 =\frac{\mathrm{i}\omega_k}{\omega_c}
 \sum_{r=-R}^{R}\sum_{b=1}^{B}
 \widehat\beta_{r,b}B_b(\omega_k)F_{k-r}.
 \label{eq:real-ltv-output}
\end{equation}
All methods fit $\beta_{r,b}$ by minimizing squared output prediction error on the evaluation band. For Target-only, Pooling, Unprojected, Projected, and LR, regularization penalizes squared second differences of $(\beta_{r,b})_{b=1}^B$ along each shift band $r$, scaled by the mean diagonal entry of the target normal matrix (plus a $10^{-12}$ numerical stabilization ridge).

\begin{table}[t]
 \centering
 \caption{Target regimes and selected B-spline harmonic-transfer dimensions $(R,B)$ on the electronic filter benchmark. Scheduling types \texttt{h1} and \texttt{h3} contain one and three harmonics, respectively, and groups $0$ and $1$ denote distinct scheduling realizations.}
 \label{tab:real-ltv-classes}
 \begin{tabular}{ccccc}
 \toprule
 Scheduling & Peak-to-peak (mV) & Group & $R$ & $B$\\
 \midrule
 h1 & 350 & 0 & 20 & 40 \\
 h1 & 350 & 1 & 15 & 30 \\
 h1 & 700 & 0 & 20 & 40 \\
 h1 & 700 & 1 & 20 & 40 \\
 h3 & 350 & 0 & 20 & 40 \\
 h3 & 350 & 1 & 20 & 40 \\
 h3 & 700 & 0 & 15 & 30 \\
 h3 & 700 & 1 & 15 & 30 \\
 \bottomrule
 \end{tabular}
\end{table}

Before accessing source outputs, the representation dimensions $R\in\{7,10,15,20\}$, $B\in\{20,30,40\}$, and baseline smoothing multiplier $\lambda\in\{10^{-8},10^{-6},10^{-4}\}$ are selected via leave-one-target-out cross-validation. For each target regime, Target-only is fitted on the training record and evaluated on the validation record in each of the 21 rotations of the other seven target regimes. The configuration with the lowest mean full-band validation error is selected (breaking ties toward smaller $B$, smaller $R$, and larger $\lambda$) and shared by the target's three rotations; none of the target's records are used for this selection. Table~\ref{tab:real-ltv-classes} lists the resulting dimensions, which yield $(2R+1)B=1640$ or $930$ complex parameters and are shared by all seven estimators. Target-only, Unprojected, and Projected also retain the selected smoothing multiplier $\lambda=10^{-4}$ throughout, whereas Pooling and LR re-tune their smoothing multipliers on the current target validation record.

\subsubsection{Instrumental calibration and forward source selection}

Using the Target-only fit on the single target training record as the pilot $\widetilde A$, we construct the perturbation dictionary $D$ from the constant component and first three harmonics of the scheduling-voltage difference between source $j$ and the target (retaining components with amplitude at least $1\,\mu\mathrm{V}$ and, for harmonics, at least $1\%$ of the dominant harmonic). Multiplying the state $U$ and its first two time derivatives $U',U''$ by each retained constant, cosine, and sine scheduling component yields at most $p=21$ dictionary operators (rescaled by powers of $\omega_c$ for conditioning).

At each frequency line $k$, the instrument vector collects measured input lines within a local window of radius $R+3$, whitened with a ridge penalty of $10^{-4}$ times its mean diagonal covariance entry. Real perturbation coordinates $\widetilde\theta^{(j)}$ are fitted to the resulting whitened output moments via a column-normalized elastic net (unit-norm columns) with a small ridge penalty of $10^{-6}$; the $\ell^1$ penalty is a multiple of the zero-solution threshold. For time-varying sources, each of the three two-record calibration folds selects this multiple from $\{0,10^{-3},10^{-2},5\times 10^{-2}\}$ by leave-one-record-out validation on its two fitting records (ties toward the larger value); the median of the three selected values is used for the final refit. For constant-scheduling sources, which are calibrated on the noise-excited record and checked on the multisine record, the multiple is fixed at $10^{-3}$. Every candidate source is required to pass a held-out calibration check (achieving a whitened residual norm at most $0.03$ of the output norm and at least a fivefold residual reduction relative to $\theta=0$), which all 12 candidates satisfy in every rotation. Passing sources are then recalibrated on all of their records, and Projected projects out the real span of the active calibration generators evaluated at $\widetilde A$.

To guard against uninformative sources and select task weights, Unprojected and Projected perform greedy forward source selection on the single target validation record. Starting from Target-only, each step evaluates adding one remaining candidate source while jointly searching for a common source-weight multiplier in $\{0,0.01,0.03,0.1,0.3,1,3\}$ (where $\rho_j$ equals this multiplier times $n_j$). A source addition is accepted only if it reduces full-band validation error by at least $2\%$, up to a maximum of four sources, without refitting on the validation record.

\subsubsection{Target-specific performance and diagnostics}

Tables~\ref{tab:real-ltv-targets} and \ref{tab:real-ltv-targets-h3} report the mean normalized prediction errors $E_{\mathcal K}$ for each of the eight target regimes averaged over its three rotations. Unprojected and Projected reduce both full-band and sideband prediction errors relative to Target-only on every single target regime, whereas forced-line rankings vary across regimes. Across the 24 rotations, Unprojected and Projected select identical source sets and weight multipliers in 17 runs, typically retaining between one and three sources (specifically, $(1,2,3,4)$ sources are selected in $(8,11,4,1)$ runs by Unprojected and $(9,11,3,1)$ runs by Projected, with weight multipliers $(0.1,0.3,1,3)$ chosen in $(5,5,11,3)$ and $(5,6,9,4)$ runs, respectively).

Finally, spectral diagnostics explain the performance gap between spline-smoothed estimators (Target-only, LR, Unprojected, Projected) and hard spectral truncation methods (JM, HL) in this single-record regime. For a single multisine training record (e.g., target \texttt{h3}, $350\text{ mV}$, group $1$), 13 of the $1640$ eigenvalues of the empirical normal matrix fall below $10^{-12}$ times the largest eigenvalue. Second-difference spline smoothing regularizes these poorly excited frequency-shift spline modes smoothly without discarding the remaining signal, whereas hard principal-component truncation either drops informative directions or admits severe variance from ill-conditioned directions. Supplementary checks confirm that expanding LR's source-smoothing grid to $\{1,100\}$ alters its mean full-band and sideband errors by less than $1\%$, while removing HL's $\ell^1$ offset penalty substantially degrades validation accuracy in the $1640$-parameter class.

\begin{table}[p]
 \centering
 \caption{Target-specific mean normalized prediction errors $E_{\mathcal K}$ over the three rotations for the one-harmonic (\texttt{h1}) scheduling targets. Boldface and underlining mark the best and second-best means within each target regime and frequency band (evaluated prior to rounding).}
 \label{tab:real-ltv-targets}
 \begingroup
 \small
 \setlength{\tabcolsep}{6pt}
 \renewcommand{\arraystretch}{1.05}
 \begin{tabular}{lccc}
 \toprule
 Method & Full band & Forced lines & Sidebands\\
 \midrule
 \multicolumn{4}{l}{h1, 350 mV, group 0}\\
Target-only & $2.68\times10^{-5}$ & $1.06\times10^{-6}$ & $2.85\times10^{-4}$ \\
Pooling & $3.43\times10^{-2}$ & $1.37\times10^{-3}$ & $3.65\times10^{-1}$ \\
Jiao--McKeague & $3.85\times10^{-3}$ & $2.94\times10^{-3}$ & $1.30\times10^{-2}$ \\
Lin--Reimherr & $2.64\times10^{-5}$ & $6.74\times10^{-7}$ & $2.84\times10^{-4}$ \\
Hu--Lin & $3.63\times10^{-5}$ & $9.15\times10^{-6}$ & $3.09\times10^{-4}$ \\
Unprojected & $\boldsymbol{1.39\times10^{-5}}$ & $\underline{3.97\times10^{-7}}$ & $\boldsymbol{1.50\times10^{-4}}$ \\
Projected & $\underline{1.39\times10^{-5}}$ & $\boldsymbol{3.37\times10^{-7}}$ & $\underline{1.50\times10^{-4}}$ \\
 \midrule
 \multicolumn{4}{l}{h1, 350 mV, group 1}\\
Target-only & $3.12\times10^{-5}$ & $\underline{1.21\times10^{-6}}$ & $3.33\times10^{-4}$ \\
Pooling & $3.32\times10^{-2}$ & $4.89\times10^{-4}$ & $3.62\times10^{-1}$ \\
Jiao--McKeague & $3.15\times10^{-5}$ & $\boldsymbol{1.21\times10^{-6}}$ & $3.36\times10^{-4}$ \\
Lin--Reimherr & $3.12\times10^{-5}$ & $1.21\times10^{-6}$ & $3.33\times10^{-4}$ \\
Hu--Lin & $2.97\times10^{-5}$ & $1.22\times10^{-6}$ & $3.16\times10^{-4}$ \\
Unprojected & $\underline{2.45\times10^{-5}}$ & $1.24\times10^{-6}$ & $\underline{2.58\times10^{-4}}$ \\
Projected & $\boldsymbol{2.42\times10^{-5}}$ & $1.27\times10^{-6}$ & $\boldsymbol{2.54\times10^{-4}}$ \\
 \midrule
 \multicolumn{4}{l}{h1, 700 mV, group 0}\\
Target-only & $5.21\times10^{-5}$ & $3.56\times10^{-7}$ & $2.05\times10^{-4}$ \\
Pooling & $2.17\times10^{-1}$ & $5.99\times10^{-2}$ & $6.81\times10^{-1}$ \\
Jiao--McKeague & $2.38\times10^{-3}$ & $2.27\times10^{-3}$ & $2.68\times10^{-3}$ \\
Lin--Reimherr & $5.20\times10^{-5}$ & $\underline{3.45\times10^{-7}}$ & $2.05\times10^{-4}$ \\
Hu--Lin & $1.10\times10^{-4}$ & $4.99\times10^{-5}$ & $2.88\times10^{-4}$ \\
Unprojected & $\underline{2.88\times10^{-5}}$ & $3.49\times10^{-7}$ & $\underline{1.13\times10^{-4}}$ \\
Projected & $\boldsymbol{2.87\times10^{-5}}$ & $\boldsymbol{3.10\times10^{-7}}$ & $\boldsymbol{1.13\times10^{-4}}$ \\
 \midrule
 \multicolumn{4}{l}{h1, 700 mV, group 1}\\
Target-only & $5.46\times10^{-5}$ & $6.10\times10^{-7}$ & $2.14\times10^{-4}$ \\
Pooling & $2.09\times10^{-1}$ & $5.86\times10^{-2}$ & $6.55\times10^{-1}$ \\
Jiao--McKeague & $3.22\times10^{-3}$ & $3.45\times10^{-3}$ & $2.54\times10^{-3}$ \\
Lin--Reimherr & $5.44\times10^{-5}$ & $\underline{3.84\times10^{-7}}$ & $2.14\times10^{-4}$ \\
Hu--Lin & $5.33\times10^{-5}$ & $\boldsymbol{2.58\times10^{-7}}$ & $2.10\times10^{-4}$ \\
Unprojected & $\underline{2.87\times10^{-5}}$ & $4.32\times10^{-7}$ & $\underline{1.12\times10^{-4}}$ \\
Projected & $\boldsymbol{2.80\times10^{-5}}$ & $4.78\times10^{-7}$ & $\boldsymbol{1.10\times10^{-4}}$ \\
 \bottomrule
 \end{tabular}
 \endgroup
\end{table}

\begin{table}[p]
 \centering
 \caption{Target-specific mean normalized prediction errors $E_{\mathcal K}$ over the three rotations for the three-harmonic (\texttt{h3}) scheduling targets. Boldface and underlining mark the best and second-best means within each target regime and frequency band (evaluated prior to rounding).}
 \label{tab:real-ltv-targets-h3}
 \begingroup
 \small
 \setlength{\tabcolsep}{6pt}
 \renewcommand{\arraystretch}{1.05}
 \begin{tabular}{lccc}
 \toprule
 Method & Full band & Forced lines & Sidebands\\
 \midrule
 \multicolumn{4}{l}{h3, 350 mV, group 0}\\
Target-only & $2.59\times10^{-5}$ & $5.34\times10^{-7}$ & $4.11\times10^{-4}$ \\
Pooling & $7.66\times10^{-2}$ & $2.11\times10^{-4}$ & $1.24\times10^{0}$ \\
Jiao--McKeague & $2.49\times10^{-3}$ & $2.05\times10^{-3}$ & $9.12\times10^{-3}$ \\
Lin--Reimherr & $2.59\times10^{-5}$ & $5.18\times10^{-7}$ & $4.11\times10^{-4}$ \\
Hu--Lin & $4.41\times10^{-5}$ & $1.60\times10^{-5}$ & $4.72\times10^{-4}$ \\
Unprojected & $\underline{1.33\times10^{-5}}$ & $\underline{1.46\times10^{-7}}$ & $\underline{2.12\times10^{-4}}$ \\
Projected & $\boldsymbol{1.30\times10^{-5}}$ & $\boldsymbol{9.97\times10^{-8}}$ & $\boldsymbol{2.10\times10^{-4}}$ \\
 \midrule
 \multicolumn{4}{l}{h3, 350 mV, group 1}\\
Target-only & $4.17\times10^{-5}$ & $2.48\times10^{-6}$ & $7.20\times10^{-4}$ \\
Pooling & $4.31\times10^{-2}$ & $4.10\times10^{-3}$ & $7.18\times10^{-1}$ \\
Jiao--McKeague & $3.36\times10^{-3}$ & $2.96\times10^{-3}$ & $1.03\times10^{-2}$ \\
Lin--Reimherr & $4.13\times10^{-5}$ & $1.36\times10^{-6}$ & $7.32\times10^{-4}$ \\
Hu--Lin & $6.87\times10^{-5}$ & $2.29\times10^{-5}$ & $8.62\times10^{-4}$ \\
Unprojected & $\boldsymbol{1.53\times10^{-5}}$ & $\underline{5.69\times10^{-7}}$ & $\boldsymbol{2.70\times10^{-4}}$ \\
Projected & $\underline{1.53\times10^{-5}}$ & $\boldsymbol{2.79\times10^{-7}}$ & $\underline{2.76\times10^{-4}}$ \\
 \midrule
 \multicolumn{4}{l}{h3, 700 mV, group 0}\\
Target-only & $4.38\times10^{-5}$ & $\underline{7.09\times10^{-7}}$ & $2.62\times10^{-4}$ \\
Pooling & $1.12\times10^{-1}$ & $1.30\times10^{-2}$ & $6.18\times10^{-1}$ \\
Jiao--McKeague & $4.76\times10^{-5}$ & $\boldsymbol{7.09\times10^{-7}}$ & $2.85\times10^{-4}$ \\
Lin--Reimherr & $4.38\times10^{-5}$ & $7.22\times10^{-7}$ & $2.62\times10^{-4}$ \\
Hu--Lin & $4.35\times10^{-5}$ & $7.32\times10^{-7}$ & $2.61\times10^{-4}$ \\
Unprojected & $\boldsymbol{3.99\times10^{-5}}$ & $7.51\times10^{-7}$ & $\boldsymbol{2.39\times10^{-4}}$ \\
Projected & $\underline{3.99\times10^{-5}}$ & $7.51\times10^{-7}$ & $\underline{2.39\times10^{-4}}$ \\
 \midrule
 \multicolumn{4}{l}{h3, 700 mV, group 1}\\
Target-only & $5.72\times10^{-5}$ & $\boldsymbol{1.22\times10^{-6}}$ & $3.45\times10^{-4}$ \\
Pooling & $1.69\times10^{-1}$ & $1.21\times10^{-2}$ & $9.76\times10^{-1}$ \\
Jiao--McKeague & $6.10\times10^{-5}$ & $1.22\times10^{-6}$ & $3.69\times10^{-4}$ \\
Lin--Reimherr & $5.71\times10^{-5}$ & $1.23\times10^{-6}$ & $3.45\times10^{-4}$ \\
Hu--Lin & $5.70\times10^{-5}$ & $\underline{1.22\times10^{-6}}$ & $3.44\times10^{-4}$ \\
Unprojected & $\boldsymbol{4.90\times10^{-5}}$ & $1.25\times10^{-6}$ & $\boldsymbol{2.94\times10^{-4}}$ \\
Projected & $\underline{4.91\times10^{-5}}$ & $1.26\times10^{-6}$ & $\underline{2.95\times10^{-4}}$ \\
 \bottomrule
 \end{tabular}
 \endgroup
\end{table}

\clearpage
\bibliography{ref}
\end{document}